\documentclass[twocolumn,tighten]{aastex63}
\usepackage{amsmath}
\usepackage{amsfonts}
\usepackage{amssymb}
\usepackage{natbib}
\usepackage{graphicx}
\usepackage{subfigure}
\usepackage{latexsym, longtable, epsf}
\usepackage{fancyref}
\usepackage{lipsum}
\usepackage{gensymb}
\usepackage{hyperref}
\hypersetup{linkcolor=red,citecolor=blue,filecolor=cyan,urlcolor=magenta}

\newenvironment{rcases}
  {\left.\begin{aligned}}
  {\end{aligned}\right\rbrace}

\begin{document}

\title{Galactic Gamma Ray Background from Interactions of Cosmic Rays}

\correspondingauthor{Sayan Biswas }
\email{sayan@rri.res.in}

\author{Sayan Biswas}
\affiliation{Raman Research Institute,
C.V. Raman Avenue, Sadashivanagar, Bangalore 560080, India}

\author{Nayantara Gupta}
\affiliation{Raman Research Institute,
C.V. Raman Avenue, Sadashivanagar, Bangalore 560080, India}

\begin{abstract}
Various studies firmly establish the fact that gamma-ray observations  can act as a unique probe to detect the possible cosmic ray (CR) sources, study the CR density distribution and explore the average properties of interstellar medium (ISM) such as the gas density profile of ISM. We use the DRAGON code to study different propagation models by incorporating realistic source distribution, Galactic magnetic field (GMF) and gas density profile, and finally obtain the proton distribution (both spatial and energy) in the Galaxy by fitting the locally observed CR spectra. The uncertainties in the model parameters used to study the CR propagation are also shown here. Our obtained proton distribution is then used to calculate the diffuse gamma-ray flux  produced by proton-proton interactions in the Galactic halo. We compare our diffuse gamma-ray fluxes with the previous results and the isotropic gamma-ray background (IGRB) at sub-TeV  energy regime measured by \textit{Fermi}-LAT.  It is found to be much less than IGRB, which suggests IGRB is mostly of extragalactic origin. Finally, we use our obtained proton distribution to calculate gamma-ray fluxes from individual Galactic Molecular Clouds (GMCs) and those fluxes are compared with the \textit{Fermi}-LAT observations of GMCs.

\end{abstract}


\section{Introduction} \label{sec:intro}
The origin of cosmic rays (CRs) is still one of the enigmas of the high energy astrophysics.  The primary picture of the production and subsequent propagation of Galactic CRs is based on the pioneering studies in the early 1960s by \citet{ginz1964} which predict the confinement of a diffuse sea of high energy particles in a sizable diffusive halo. The motion of Galactic CRs is generally treated as a random walk of the particles in the magnetized, turbulent interstellar medium (ISM) and thus can be modeled in terms of a homogeneous, isotropic, diffusion equation with the addition of advection and loss terms.  Up to date, several theoretical and experimental advancements enrich the field in various ways;  see the excellent monographs, e.g. \citet{berez1990,gaiss2016}. However, the  issues related to the acceleration mechanism of CRs as well as the physical process, characterizing the interaction between CRs and the magnetized, turbulent interstellar plasma, which is supposed to be the reason behind the random walk and ultimately for the confinement of CRs are still under debate.  

Gamma-ray astronomy is undoubtedly considered as a unique probe for the investigation of acceleration and propagation of  CRs. While the acceleration sites of CRs can be revealed by the proper detection and identification of gamma-ray sources, the diffuse gamma-ray emission from the Galactic disk can trace out the distribution (both spatial and energy) of CRs. During their propagation, CRs randomly roam in different regions of the Milky Way Galaxy and produce diffuse gamma-ray emission by interacting with ambient gas via hadro-nuclear interactions or proton-proton interactions (hereafter p-p interactions). Indeed, this diffuse emission has been widely used in CR research as a tracer of 
 CR distribution in the Galactic plane  \citep[e.g.][]{abdo2009}. 

The gas density profile of Milky Way Galaxy plays a significant role in the production of diffuse gamma-ray emission via decay of neutral pions produced in p-p interactions. The diffuse gamma-ray flux is proportional to the total density of gas in our Galaxy. However, the gas density profile of our Galaxy is still uncertain. Gamma-ray observations can be used to map the density profile \citep[e.g.][]{dela2011,feld2013}. Moreover, various recent observations such as ion absorption lines against background quasars \citep{nica2002, ras2003, mill2013, fang2015, zheng2017} and emission lines \citep{hen2012, hen2013, mill2015} along different line of sights at high Galactic latitudes indicate the existence of a hot baryonic gas halo around the Galaxy known as circumgalactic medium (CGM). Such a claim from direct observations is further strengthened by various indirect observations \citep[e.g.][]{stan2002, fox2005, grc2009, put2011}. The protons present in the CGM can act as targets for  propagating CR protons,  which can  produce diffuse gamma-rays by p-p interactions. Gamma-rays produced at such high Galactic latitudes may contribute to the isotropic gamma-ray background (IGRB). Until now, IGRB has been measured  by various instruments such as \textit{SAS-2} satellite \citep{fich1975, fich1978}, EGRET on board the \textit{Compton Observatory} \citep{sreek1998, strong2004}, and the \textit{Fermi} Large Area Telescope (\textit{Fermi}-LAT) \citep{abdo2010, ackermann2013, ackermann2015}. Gamma-ray production from p-p interaction has also been studied by several authors \citep{steck1977, depaoli2000, cholis2012, feld2013, ahl2014, taylor2014, liu2019, zhao2019}  in their works. More recently, \textit{Fermi}-LAT extended their previous measurements on IGRB up to 820 GeV \citep{ackermann2015}. 
 
 In a more recent work \citep{kala2016}  the diffuse gamma-ray and neutrino fluxes have been calculated from CR interactions with circumgalactic gas. They have shown that the secondary gamma-ray flux produced in CR interactions may contribute non-negligibly to the diffuse gamma-ray background. These calculations depend on the values of many parameters which determine propagation and secondary production of CRs.

In this paper, we want to study the diffuse gamma-ray emission originated in interactions of CR protons, which require detailed modeling of CR propagation and interactions. For such purpose, we obtain CR proton distributions in our Galaxy from benchmark propagation models by fitting locally observed CR spectra.
The CR flux depends upon various uncertain parameters, namely the turbulent Galactic magnetic field (GMF), the halo size of Galaxy, hydrogen gas distribution in the Milky Way Galaxy and source distribution of CRs. The nature of the turbulent GMF is primarily modeled in our work with the Faraday rotation measurements \citep{han2009, pshir2011, janss2012} of Galactic and extragalactic radio sources. Alongside, synchrotron emission of Galactic CR electrons in the radio frequency range is also taken into account to model the turbulent component of GMF which relates the halo height ($z_{t}$) and GMF \citep{dibernardo2013}. Other local observables such as stable (${\rm{B/C}}$; $\rm{B}$ and $\rm{C}$ are boron and carbon respectively) and unstable ($^{10}{\rm{Be}}/^{9}{\rm{Be }}$; $\rm{Be}$ is beryllium) secondary to primary ratios can be used to disentangle energy dependent diffusion coefficient ($D({E_{k}})$) and halo height as $^{10}{\rm{Be}}/^{9}{\rm{Be }}\propto \sqrt{D(E_{k})}/z_{t}$  and  ${\rm{B/C}} \propto z_{t}/D(E_{k})$ with $E_{k}$ being the kinetic energy \citep{strong2011}. Realistic gas density profile can be modeled from the hydrodynamical simulations and observations in the radio, X-ray and gamma-ray wavebands. The source distribution of Galactic CRs has a less significant effect on the gamma-ray spectrum if we exclude the gamma-ray data of $|b|<10^{\degree}$ ($b$ is Galactic latitude) \citep{cholis2012}.  

 The obtained CR proton distribution is then used to calculate diffuse gamma-ray flux over the energy range of 1-1000 GeV. We have developed a code to calculate the diffuse gamma-ray emission from p-p interactions. In our code, we have used the framework of production of neutral pions in p-p interactions and  their decay to gamma-rays as provided by \citet{kelner2006}. To check the consistency of our code, we have compared our gamma-ray fluxes with the results obtained by \citet{kelner2006, cholis2012}.  In the present gamma-ray flux calculation, we exclude the gamma-ray emission of the inner Galaxy as well as of $|b|\leq 20^{\degree}$. We, thus, ensure that gamma-ray contribution from the CR sources is insignificant on the diffuse gamma-ray flux. Our result is compared with the IGRB measured by the \textit{Fermi}-LAT instrument.

 In section 2, we discuss the modeling of CR propagation with the Diffusion of cosmic RAys in Galaxy modelizatiON \texttt{DRAGON}\footnote{\url{https://github.com/cosmicrays/DRAGON}} \citep{evoli2008, dibernardo2010}. In this section, we also describe the methodology adopted for fitting the observed CR data. Section 3 contains the framework for the calculation of gamma-ray flux and section 4 is devoted to results obtained by us. We discuss and summarize our findings in section 5. The conclusion is presented in section 6.  

\section{Modeling of CR propagation in the Milky Way Galaxy}\label{sec:model}
Galactic CRs are generally believed to be accelerated by astrophysical sources such as supernova remnants (SNRs) \citep{bell1978a,bell1978b,bland1987}. Those accelerated CRs are, then, injected into the ISM where they propagate through the stochastic magnetic field to reach the Earth. The observed CR energy spectra distributed in a wide energy range of sub-GeV to multi-TeV are considered to be a combined effect of both acceleration and propagation mechanisms in our Galaxy.  In our case, we focus only on the propagation scenario. The propagation of CRs at energies below $10^{17}$~eV can be  described by diffusive transport equation \citep[e.g.][]{berez1990,feng2016}. CR density at any position of the ISM is obtained by solving such diffusive  transport equation following either semi-analytic \citep{putze2010}  or numerical procedures \citep{dibernardo2010, trotta2011}. In the present work, we use \texttt{DRAGON} code \citep{dibernardo2010,dibernardo2013} to solve the transport equation and study the propagation of CRs in the Galaxy.

\texttt{DRAGON} code numerically solves the diffusive transport equation, assuming the cylindrical symmetry and steady state approximation, in a 2+1D grid \footnote{We use 3D version of the \texttt{DRAGON} code which is available for download at \url{https://github.com/cosmicrays/DRAGON} } where each grid point is  described by its galactocentric radius, $r \in (0, 40~ \rm{kpc})$, vertical distance, $z \in (-L, +L)$  with $L = 3z_{t}$ and momentum $\tilde{p}$. In our Galactic geometry, we assume that the $z$-axis passes right through the Galactic center with Cartesian co-ordinates $(x=0, y=0, z=0)$. The Earth's position is represented as $(x = 8.5~\rm{kpc}, y=0, z=0)$. We also consider $r_{E}$ to be the galactocentric distance of the Earth and the relation, $r = \sqrt{x^{2} + y^{2}}$. In the following, we will discuss other necessary components needed for the modeling of the propagation of CRs.

\subsection{Primary sources and the injection spectra of CRs} \label{sec:source}

For our simulations, we consider the source distribution presented in the \citet{ferriere2001} which is constructed on the basis of progenitor stars and pulsar surveys. 

In our present work, we particularly pay attention to the proton spectrum in our Galaxy as proton plays the dominant role in the production of diffuse gamma-ray spectrum in the whole energy range, 0.1 GeV to $10^{5}$~GeV, considered here. In the publicly available version of the \texttt{DRAGON} code, the protons and all the other heavier nuclei are considered to be injected in the ISM with identical injection spectrum. So, proton injection spectrum is the representative of injection spectra of other heavier nuclei. We, here, describe the injection spectrum of protons as broken power law,

\begin{equation}\label{eq:inj} 
\frac{dN_{p}}{d\rho} \propto \Big( \frac{\rho}{\rho^{p}_{0,k}}  \Big)^{-\alpha^{p}_{k}},
\end{equation}   
 where, $\rho$, $\alpha^p$ and k denote the rigidity, spectral index and an integer number  respectively. For our present simulations, we consider two breaks in the injection spectrum at $\rho = \rho^{p}_{0,1} \sim 1-15$~GV and $\rho = \rho^{p}_{0,2} \sim 330$~GV with $\alpha^{p}_{1} \sim 1.80 - 2.10$ at low rigidities, $\alpha^{p}_{2}$ in the range $\sim 2.25-2.50$ at intermediate rigidities and $\alpha^{p}_{3}$ in the range $\sim 2.10 -2.45$ at high rigidities. The parameter choice is motivated a posteriori by fitting the proton flux with the observed data measured by Voyager \citep{stone2013,cum2016}, PAMELA \citep{adriani2011, adriani2013proton},  AMS-02 \citep{aguilar2015} and CREAM \citep{yoon2011}.

\subsection{Galactic magnetic field and Diffusion} \label{sec:gmfdf} 

To model GMF, we opt the geometry provided by \citet{pshir2011}. Our chosen GMF geometry contains three components,  namely the  disc, halo and turbulent, and their corresponding normalizations are denoted as $B_0^{\rm{disc}}$,  $B_0^{\rm{halo}}$ and $B_0^{\rm{turbulent}}$, respectively. Turbulent component is the more important component than the other two components as turbulent component largely affects the CR propagation. The z-dependence of the turbulent GMF is defined as \citep{dibernardo2013}

\begin{equation}\label{eq:turbgmf} 
B^{\rm{turbulent}}(z) \propto {\rm{exp} }\Big(- |z|/z_{t} \Big).
\end{equation} 

For our present simulations, we consider the diffusion in the form \citep{evoli2008, dibernardo2013, biswas2018}

\begin{equation}\label{eq:diff} 
  D(\rho,z) = \beta^{\eta} D_{0}\Big( \frac{\rho}{\rho_{0}} \Big)^{\delta} {\rm{exp} }\Big( \frac{|z|}{z_{t}} \Big), 
\end{equation} 
where, $\beta$, $\rho_{0}$, $\delta$ and $D_{0}$ are  the particle speed, reference rigidity, diffusion spectral index and normalization respectively. The other index $\eta$ accounts for uncertainties at low energy arises due to low energy  CR propagation. We can relate the turbulent component of GMF and the diffusion coefficient  from the equations \ref{eq:turbgmf} and \ref{eq:diff}, i.e.,

\begin{equation}\label{eq:gmfd} 
D(z)^{-1} \propto B^{\rm{turbulent}}(z) \propto {\rm{exp} }\Big(- |z|/z_{t} \Big).
\end{equation} 

The quasi-linear theory and the numerical simulations of particle propagation in turbulent magnetic fields \citep{marco2007} also support the above relation (see equation \ref{eq:gmfd}). 

In addition to spatial diffusion, we also incorporate the stochastic acceleration in modeling of CR propagation.  The diffusion in momentum space ($D_{\tilde{p}\tilde{p}}$, $\tilde{p}$ denotes momentum), connecting stochastic acceleration with the scattering of CRs on randomly moving magnetohydro-dynamical (MHD) waves, is the cause behind the stochastic acceleration. The diffusion coefficient in physical space ($D_{xx}$) is connected with $D_{\tilde{p}\tilde{p}}$ by a relation $D_{\tilde{p}\tilde{p}} \propto \tilde{p}^{2} v_{Alf}/D_{xx}$, where $v_{Alf}$ is the Alfv\'{e}n velocity of the propagation MHD waves \citep{berez1990}. Moreover, we also include wind speed ($V_{w}$) in one of our diffusion model by considering $V_{w} = |z|\times\frac{dv_{w}}{dz}$ ($\frac{dv_{w}}{dz}$ in units of $\rm{km~s^{-1}~kpc^{-1}}$) in the \texttt{DRAGON} code.

CRs having energies below 10 GeV are  largely affected due to solar activity. So, we  need to take into account the solar modulation effect for fitting the observed CR spectra.  We model the solar modulation with a potential ($\phi$) by following the prescription given in \citet{uso2005}.

\subsection{Hydrogen gas density profile in Milky Way Galaxy } \label{sec:gasden} 

In the \texttt{DRAGON} code, we use molecular hydrogen density ($n_{{\rm{H_{2}}}}$), atomic or neutral hydrogen density ($n_{{\rm{HI}}}$)  and ionized hydrogen density ($n_{{\rm{HII}}}$) profiles, where the subscripts $\rm{H_{2}}$, $\rm{HI}$ and $\rm{HII}$ denote the molecular, neutral and ionized hydrogen. The total gas density in our Galaxy is considered as
\begin{equation}
n_{g}(r,z) = 2n_{{\rm{H_{2}}}} + n_{{\rm{HI}}} + n_{{\rm{HII}}}  
\end{equation} 
 We, here, describe the three components of the total gas density in two regions; one region belongs to  $r\lesssim 3$~kpc (or, Galactic bulge (GB)) and the other one is $3~{\rm{kpc}} < r \leq 40~{\rm{kpc}} $. 
 
 \subsubsection{For $r\lesssim 3$~kpc}

\noindent \textbf{\Large{$n_{{\rm{H_{2}}}}$}:} The GB region can be divided into two parts, namely the central molecular zone (CMZ) and the GB disk.   CMZ, a layer of molecular hydrogen, exists in the core of GB and the average extension of CMZ is $r\sim 200$~pc. The rest part of GB is known as GB disk. Hence,  the total density profile of this region is the result of  combined contributions of both the CMZ and the GB disk. The density profile is based on the 2.6 mm CO emission line with the 18 cm OH absorption line which are complimented with theoretical and gas dynamical models. The density contribution of CMZ is denoted as \citep{ferriere2007} 

\begin{eqnarray}
n_{{\rm{H_{2}}}}^{\rm{CMZ}} =&& (150.0~{\rm{cm^{-3}}}) \nonumber \\
&& \times {\rm{exp}}\Bigg[- \Bigg(\frac{\sqrt{X^2 + (2.5Y)^2}-0.125~{\rm{kpc}}} {0.137~{\rm{kpc}}}\Bigg)^{4}\Bigg] \nonumber \\
&& \times {\rm{exp}}\Bigg[- \bigg(\frac{z}{0.018~{\rm{kpc}}}\bigg)^2\Bigg],
\end{eqnarray}
where,  CMZ coordinates ($X, Y$) and the Galactic coordinates ($x, y$) are related by the following equations \citep{ferriere2007}

 \begin{eqnarray}
 X = (x-x_{c})~{\rm{cos} } \theta_{c} + (y-y_{c})~{\rm{sin }} \theta_{c} \\
Y =- (x-x_{c})~{\rm{sin} } \theta_{c} + (y-y_{c})~{\rm{cos }} \theta_{c},
\end{eqnarray}

with $x_{c} =-50~{\rm{pc}},  y_{c} = 50~{\rm{pc}}$ and $\theta_{c} = 70^{\circ}$.

The GB disk, beyond CMZ region, is modeled as a tilted elliptical disk with a hole at the central region. The GB disk coordinates ($\cal{X, Y, Z}$) and the coordinates of our Galaxy ($x,y,z$)  are related  \citep{ferriere2007}

\begin{eqnarray}
{\cal{X}} =&&  x ~{\rm{cos} } \beta~{\rm{cos} } \theta_{d} \nonumber \\
 &&-y~ ( {\rm{sin} } \alpha ~{\rm{sin} } \beta ~{\rm{cos} } \theta_{d}  - {\rm{cos} } \alpha ~{\rm{sin} } \theta_{d} ) \nonumber \\
 && -z~({\rm{cos} } \alpha ~{\rm{sin} } \beta ~{\rm{cos} } \theta_{d}  + {\rm{sin} } \alpha~{\rm{sin} } \theta_{d})   \\
 {\cal{Y}} =&&  -x ~{\rm{cos} } \beta ~{\rm{sin} } \theta_{d} \nonumber \\
 &&+y~ ( {\rm{sin} } \alpha ~{\rm{sin} } \beta ~{\rm{sin} } \theta_{d}  + {\rm{cos} } \alpha ~{\rm{cos} } \theta_{d} ) \nonumber \\
 && +z~({\rm{cos} } \alpha ~{\rm{sin} } \beta~{\rm{sin} } \theta_{d}  - {\rm{sin} } \alpha ~{\rm{cos} } \theta_{d})   \\
 {\cal{Z}} =&&  x ~{\rm{sin} } \beta\nonumber \\
 &&+y~  {\rm{sin} } \alpha ~{\rm{cos} } \beta \nonumber \\
 && +z~{\rm{cos} } \alpha ~{\rm{cos} } \beta,
\end{eqnarray} 
  
 where, $\alpha = 13.5^{\circ}$, $\beta = 20 ^{\circ}$ and $\theta_{d} = 48.5 ^{\circ}$. The density contribution from holed GB disk can be expressed as \citep{ferriere2007} 
  
\begin{eqnarray}
n_{{\rm{H_{2}}}}^{\rm{disk}} =&& (4.8~{\rm{cm^{-3}}}) \nonumber \\
&& \times {\rm{exp}}\Bigg[- \Bigg(\frac{\sqrt{{\cal{X}}^2 + (3.1{\cal{Y}})^2}-1.2~{\rm{kpc}}} {0.438~{\rm{kpc}}}\Bigg)^{4}\Bigg] \nonumber \\
&& \times {\rm{exp}}\Bigg[- \bigg(\frac{{\cal{Z}}}{0.042~{\rm{kpc}}}\bigg)^2\Bigg].
\end{eqnarray}

The total density distribution of $\rm{H_{2}}$ is denoted as 

\begin{equation}
n_{{\rm{H_{2}}}} (r,z)= n_{{\rm{H_{2}}}}^{\rm{CMZ}}  + n_{{\rm{H_{2}}}}^{\rm{disk}}.     
\label{eq:h2fer07}
\end{equation}

\noindent \textbf{\Large{$n_{{\rm{HI}}}$}:}  Similar to $\rm{H_{2}}$,  $\rm{HI}$ density profile is also a sum of both the contributions of CMZ and holed GB disk.

Different surveys of CMZ indicate  that mass of $\rm{HI}$ is $8.8\%$ of  the mass of $\rm{H_{2}}$. The space-averaged density of ${\rm{HI}}$ is represented as \citep{ferriere2007}

\begin{eqnarray}
n_{{\rm{HI}}}^{\rm{CMZ}} =&& (8.8~{\rm{cm^{-3}}}) \nonumber \\
&&\times {\rm{exp}}\Bigg[- \Bigg(\frac{\sqrt{X^2 + (2.5Y)^2}-0.125~{\rm{kpc}}} {0.137~{\rm{kpc}}}\Bigg)^{4}\Bigg] \nonumber \\
&& \times {\rm{exp}}\Bigg[- \bigg(\frac{z}{0.054~{\rm{kpc}}}\bigg)^2\Bigg].
\end{eqnarray}
 
 Similarly, the space-averaged density of $\rm{HI}$ from the holed GB disk is expressed as \citep{ferriere2007}

\begin{eqnarray}
n_{{\rm{HI}}}^{\rm{disk}} =&& (0.34~{\rm{cm^{-3}}}) \nonumber \\
&& \times {\rm{exp}}\Bigg[- \Bigg(\frac{\sqrt{{\cal{X}}^2 + (3.1{\cal{Y}})^2}-1.2~{\rm{kpc}}} {0.438~{\rm{kpc}}}\Bigg)^{4}\Bigg] \nonumber \\
&& \times {\rm{exp}}\Bigg[- \bigg(\frac{{\cal{Z}}}{0.120~{\rm{kpc}}}\bigg)^2\Bigg].
\end{eqnarray}

So, the total density distribution of ${\rm{HI}}$ can be written as 

\begin{equation}
n_{{\rm{HI}}} (r,z)= n_{{\rm{HI}}}^{\rm{CMZ}}  + n_{{\rm{HI}}}^{\rm{disk}}.     
\label{eq:hifer07}
\end{equation}

\noindent \textbf{\Large{$n_{{\rm{HII}}}$}:} The density profile of ionized component is based  on non-axisymmetric spatial distribution of free electrons in our Galaxy, which is constructed from the data of dispersion, scattering and distance measurements of pulsars available till the end of 2001. 

In the present case, we only consider the contribution of weakly ionized medium (WIM). WIM contributes $83\%$ of the total mass of HII. Along with, we also assume that hydrogen gas is completely ionized whereas helium is completely neutral. The space-averaged density of HII is represented as \citep{ferriere2007}

\begin{eqnarray}
n_{{\rm{HII}}}(r,z) =&&  (8.0~{\rm{cm^{-3}}}) \nonumber \\
&& \times \Bigg\{ {\rm{exp}} \Bigg[  -  \frac{x^{2} + (y - y_{3})^{2}}{L_{3}^{2}}  \Bigg] \nonumber \\
&&\times {\rm{exp}} \Bigg[- \frac{(z-{z_{3})^{2}}}{H_{3}^{2}}   \Bigg] \nonumber \\
&&+ 0.009 \times {\rm{exp}}\Bigg[ - \Bigg(\frac{r - L_{2}}{L_{2}/2}\Bigg)^{2}  \Bigg]  {\rm{sech^{2}}} \bigg(\frac{z}{H_{2}}\bigg) \nonumber \\
&&+ 0.005\Bigg[ {\rm{cos}} \Bigg( \pi \frac{r}{2L_{1}}  \Bigg) u(L_{1} - r)  \Bigg]  \nonumber \\
&&  \times {\rm{sech^{2}}} \bigg(\frac{z}{H_{1}}\bigg)     \Bigg\},
\end{eqnarray}  

where, $u$ denotes the unit step function, $y_{3} =-10~{\rm{pc}}$, $z_{3} =-20~{\rm{pc}}$, $L_{3} =145~{\rm{pc}}$, $H_{3} =26~{\rm{pc}}$, $L_{2} =3.7~{\rm{kpc}}$, $H_{2} =140~{\rm{pc}}$,$L_{1} =17~{\rm{kpc}}$ and  $H_{1} =950~{\rm{pc}}$.  

\subsubsection{For $3~{\rm{kpc}}<r \leq 40$~kpc} 

\begin{figure}[h!]
\subfigure{}
\includegraphics[width=0.45\textwidth,clip,angle=0]{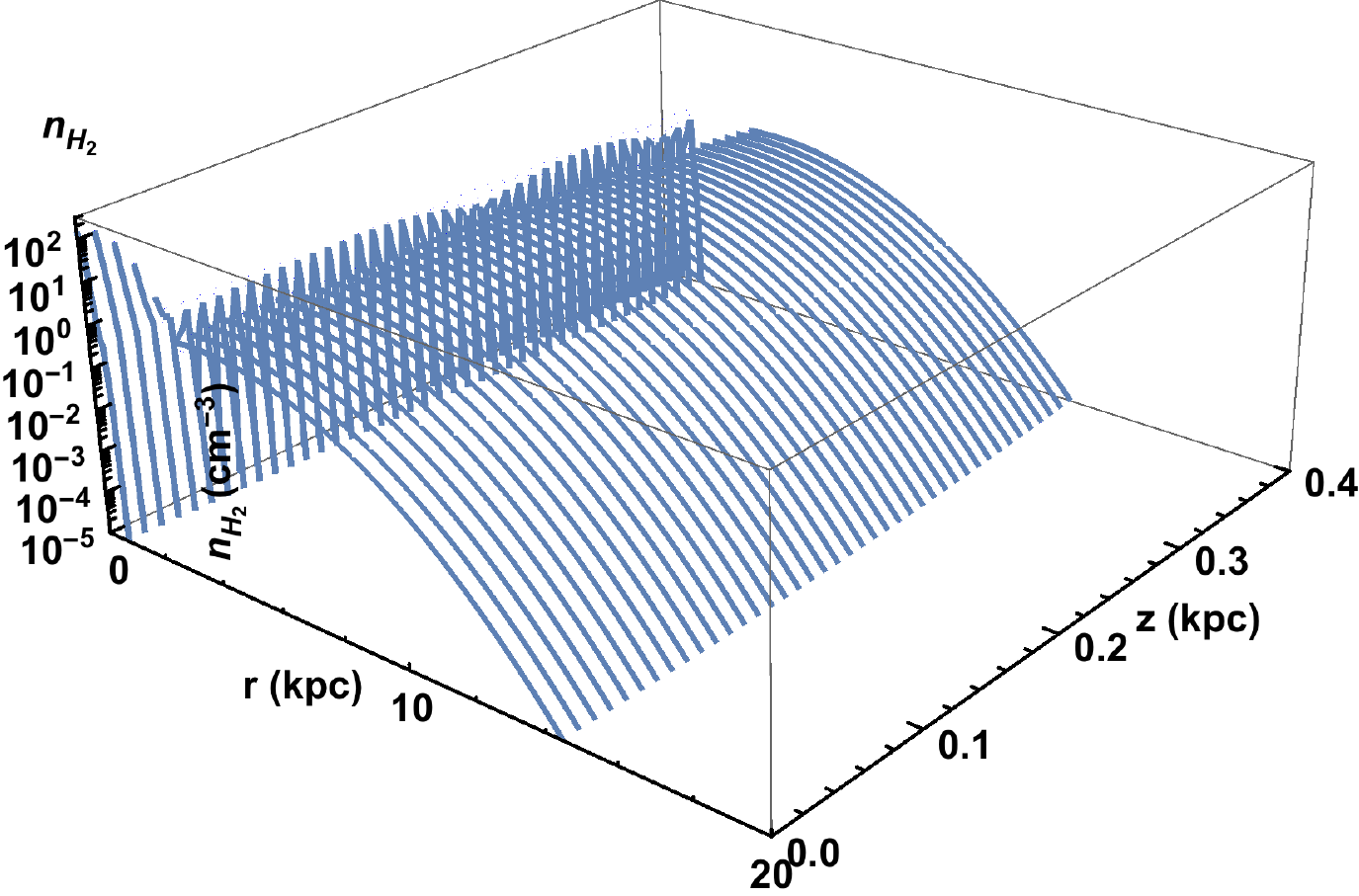}

\subfigure{}
\includegraphics[width=0.45\textwidth,clip,angle=0]{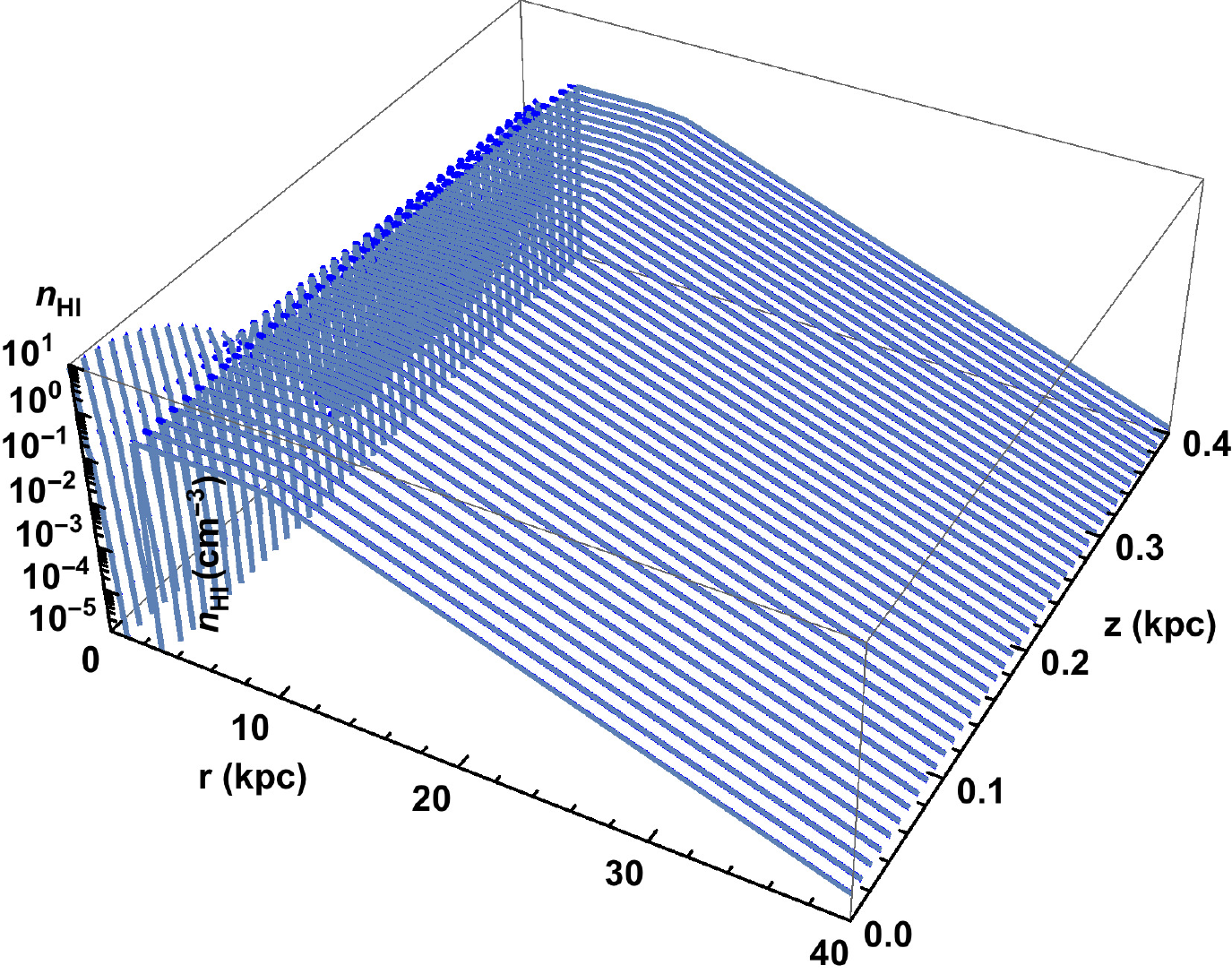}

\subfigure{}
\includegraphics[width=0.45\textwidth,clip,angle=0]{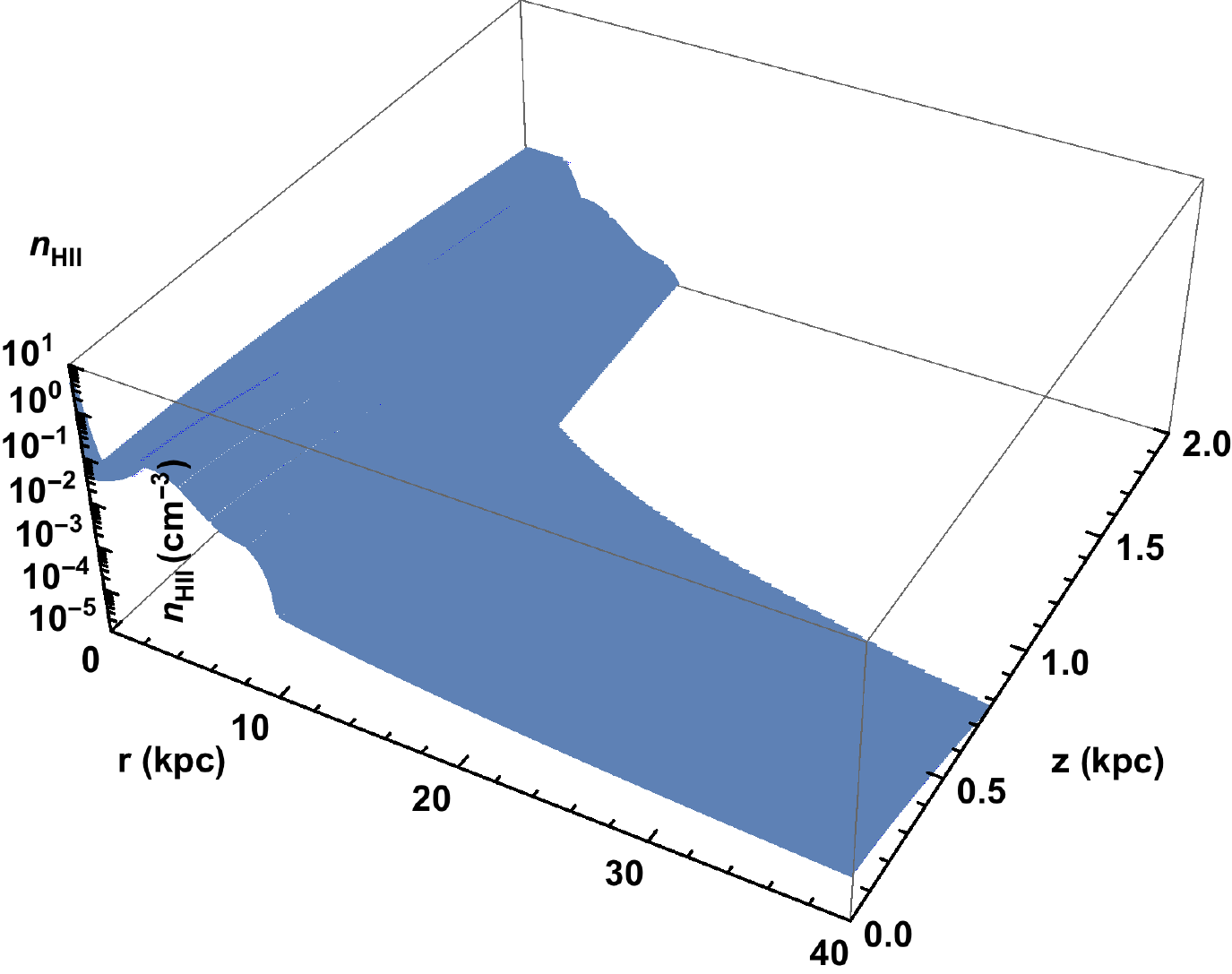}

\caption{\label{fig:gasdensity}  The 3D plots of  $n_{\rm{H_{2}}}$ (top), $n_{\rm{HI}}$ (middle) and $n_{\rm{HII}}$ (bottom) along with their radial and vertical dependence are shown. A gap in the range of $r \sim$1-3 kpc has been occurred in both $n_{\rm{H_{2}}}$ and $n_{\rm{HI}}$ which is considered as the Galactic bar effect \citep{ferriere2007}.}
\end{figure}

  In this region, we include radial and vertical distributions independently. The radial distributions are based on the measurements of CO emission, 21 cm line emission and absorption and dispersion measurements of pulsars. The vertical distributions are obtained by fitting the vertical density profiles of $\rm{H_{2}}$, $\rm{HI}$ and $\rm{HII}$ obtained from the gamma-ray observations and hydro-dynamical simulations \citep{feld2013}. We combine both the radial and vertical density distributions and construct a density profile as a function of  $r$ and $z$. We also normalize the combined density profile by following the prescription provided in \citet{biswas2018}. We follow the same procedure for all the three components and final forms are given below.

\noindent \textbf{\Large{$n_{{\rm{H_{2}}}}$}:}     The normalized density profile can be written as \citep{biswas2018}
\begin{eqnarray}\label{eq:moldisk}
n_{{\rm{H_{2}}}} (r, z)= && (0.5 \times 0.58~{\rm{cm^{-3}}}) \times \Bigg(  \frac{r}{8.5~{\rm{kpc}}}  \Bigg)^{-0.58} \nonumber \\
 && \times {\rm{exp}}\Bigg[  - \frac{(r - 4.5~{\rm{kpc}})^{2}  -  (4.0~{\rm{kpc}})^{2}   } {(2.9~{\rm{kpc}})^{2}}   \Bigg]   \nonumber \\
  &&  \times {\rm{exp}}\Bigg[   - \Big( \frac{|z|}{0.29}    \Big)^{1.96} \Bigg].
 \end{eqnarray}
 In equation \ref{eq:moldisk}, the radial part is obtained from the 2.6 mm CO emission line measurements \citep{ferriere1998}. 
 
 \noindent \textbf{\Large{$n_{{\rm{HI}}}$}:}  The normalized density profile of HI is 
\begin{eqnarray}
n_{{\rm{HI}}}(r,z) = &&\frac{(0.340~{\rm{cm^{-3}}} )}{(\alpha_{h} (r))^{2}}  \times {\rm{exp}}\Bigg[   - \Big( \frac{|z|}{0.38~{\rm{kpc}}}    \Big)^{1.76} \Bigg)      \Bigg] \nonumber \\
&&+ \frac{(0.226~{\rm{cm^{-3}}} )}{(\alpha_{h} (r))} \times {\rm{exp}}\Bigg[  - \Big( \frac{|z|}{0.38~{\rm{kpc}}}    \Big)^{1.76} \Bigg] \nonumber \\
&&  \times \Bigg\{\Bigg[  1.745 - \frac{1.289}{\alpha_{h} (r))} \Bigg] +  \Bigg[ 0.473 - \frac{0.070}{\alpha_{h} (r))} \Bigg] \nonumber \\
 && +  \Bigg[  0.283 - \frac{0.142}{\alpha_{h} (r))} \Bigg] \Bigg\}, \nonumber \\
 && {\rm{For} },~3~{\rm{kpc}}< r\leq 9~{\rm{kpc}}    \label{eq:atomlow}\\
 =&& \Big( 0.534~{\rm{cm^{-3}} }- 0.038~{\rm{cm^{-3}}}\times(r - 9.0)\Big) \nonumber \\
 && \times {\rm{exp}}\Bigg[  - \Big( \frac{|z|}{0.38~{\rm{kpc}}}    \Big)^{1.76} \Bigg] , \nonumber \\
 && {\rm{For} },~9~{\rm{kpc}}< r < 10.5~{\rm{kpc}} \label{eq:atommid}\\
 =&& \Bigg[0.9~{\rm{cm^{-3}}} \times {\rm{exp}}\Big(- \frac{r - 8.5~{\rm{kpc}}}{3.15~{\rm{kpc}}}\Big) \Bigg]    \nonumber \\
 &&\times {\rm{exp}}\Bigg[  - \Big( \frac{|z|}{0.38~{\rm{kpc}}} \Big)^{1.76} \Bigg], \nonumber \\
 && {\rm{For} },~10.5~{\rm{kpc}} \leq r \lesssim 40~{\rm{kpc}} \label{eq:atomup}
\end{eqnarray}
 where,
\begin{equation}
\begin{rcases}
 \alpha_{h} (r) &= 1.0,~ {\rm{For,~ r \leq 8.5~kpc}}  \\
                     &= \frac{r}{8.5 ~{\rm{kpc}}}, ~ {\rm{For,~ r > 8.5~kpc}}.\\
\end{rcases}
\end{equation}

\textbf{\Large{$n_{{\rm{HII}}}$}:}
In this case, the normalized density profile is considered as, 
\begin{widetext}
 \begin{eqnarray}
 n_{{\rm{HII}}}(r,z) =&& \Bigg( (0.0237~{\rm{cm^{-3}}} )~{\rm{exp}}\Bigg[-\frac{r^{2} - (8.5~{\rm{kpc}})^{2} }{(37.0~{\rm{kpc}})^{2}}\Bigg] \nonumber \\
&&+ (0.0013~{\rm{cm^{-3}}} )  \times  {\rm{exp}}\Bigg[-\frac{(r- 4.0~{\rm{kpc}})^{2} - (4.5~{\rm{kpc}})^{2}    }{(2.0~{\rm{kpc}})^{2}} \Bigg] \Bigg) \nonumber \\
&& \times \frac{1.0}{0.491} \Bigg( 0.49 \times {\rm{exp}}\Bigg[  - \bigg( \frac{|z|}{0.40~{\rm{kpc}}}    \bigg)^{1.36} \Bigg] + 7.05 \times 10^{-4} \times {\rm{exp}}\Bigg[  - \bigg( \frac{|z|}{9.17~{\rm{kpc}}}   \bigg) \Bigg]  \Bigg), \nonumber \\
&&  {\rm{For} },~3~{\rm{kpc}}< r\leq 9~{\rm{kpc}}    \label{eq:ionlow}\\
=&&\Big( 0.0239~{\rm{cm^{-3}} }- 0.0153~{\rm{cm^{-3}}}\times(r - 9.0)\Big) \nonumber \\ 
&& \times \frac{1.0}{0.491} \Bigg( 0.49 \times {\rm{exp}}\Bigg[  - \bigg( \frac{|z|}{0.40~{\rm{kpc}}}    \bigg)^{1.36} \Bigg] + 7.05 \times 10^{-4} \times {\rm{exp}}\Bigg[  - \bigg( \frac{|z|}{9.17~{\rm{kpc}}}   \bigg) \Bigg]  \Bigg), \nonumber \\
&&{\rm{For} },~9~{\rm{kpc}} < r  < 10.5~{\rm{kpc}} \label{eq:ionmid} \\ 
=&& \Bigg[0.045~{\rm{cm^{-3}}} \times \Big( \frac{r }{1.0~{\rm{kpc}}}\Big)^{-1.62} \Bigg]    \nonumber \\
&& \times \frac{1.0}{0.491} \Bigg( 0.49 \times {\rm{exp}}\Bigg[  - \bigg( \frac{|z|}{0.40~{\rm{kpc}}}    \bigg)^{1.36} \Bigg] + 7.05 \times 10^{-4} \times {\rm{exp}}\Bigg[  - \bigg( \frac{|z|}{9.17~{\rm{kpc}}}   \bigg) \Bigg]  \Bigg), \nonumber \\
&& {\rm{For} },~10.5~{\rm{kpc}} \leq r \lesssim~40~{\rm{kpc}} \label{eq:ionup}.
\end{eqnarray}
 \end{widetext}

Here, radial part in equation \ref{eq:atomlow} is derived on the basis of 21 cm emission and absorption line data \citep{ferriere1998}. The radial profile in  equation \ref{eq:atomup} is constructed from the 21 cm line survey along with the analysis of parameters for the warp or bending of the Galactic plane and rotation curve  of Milky Way Galaxy \citep{kalberla2008}. The equation in the middle, i.e. equation \ref{eq:atommid}, represents the interpolation between equations \ref{eq:atomlow} and  \ref{eq:atomup}.

 Similar to the previous one, the radial part in equation \ref{eq:ionlow} comes from the WIM contribution which is formulated from the dispersion, scattering and distance measurements of pulsars \citep{ferriere1998}. The radial profile in  equation \ref{eq:ionup} is constructed from the mass measurement of hot gas in the halo following the analysis of OVII and OVIII emission lines \citep{mill2015}. The equation \ref{eq:ionmid}, represents the interpolation between equations \ref{eq:ionlow} and  \ref{eq:ionup}.

Figure \ref{fig:gasdensity} shows the 3D plots of  $n_{\rm{H_{2}}}$ (top), $n_{\rm{HI}}$ (middle) and $n_{\rm{HII}}$ (bottom) showing the radial and vertical distribution. It is not straightforward to compare these plots due to their complicated natures. But we can say that neutral and ionized contributions are dominant over molecular contribution at large radial distances. Another significant fact is the presence of a gap in both $n_{\rm{H_{2}}}$ and $n_{\rm{HI}}$ profiles in the range of $r \sim$1-3 kpc \citep{ferriere2007}. The Galactic bar effect is believed to be the reason behind the appearance of gap \citep{ferriere2007}.

\subsection{Methodology to obtain the proton distribution in the Galaxy}\label{sec:method}
Here, we discuss the procedure followed in this work for calculating the CR proton distribution in space and energy in the Galaxy.
 
Our calculated CR spectra near the Earth is fitted to the observed spectra to validate our models.
We consider our Galaxy to be a cylinder with maximum galactocentric radius, $R_{\rm{max}} = 40$~kpc and maximum half-height, $L = 3z_{t}$; $z_{t}$ is the halo height. For the purpose of simulations, we take into account a subset of benchmark  models, namely     PD (plain diffusion model), KRA (model including Kraichnan turbulence spectrum) \citep{kra1967,kra1980}, CON (convection model) and KOL (model including Kolmogorov turbulence spectrum) \citep{kol1941}. Each model is different from the other one on the basis of $\eta$, $\delta$, $D_{0}$ and $v_{Alf}$. Only CON model has an extra component in terms of convective velocity. In the \texttt{DRAGON} code, CON model takes an extra input i.e. $\frac{dv_{w}}{dz}$ and we set it fixed at $\frac{dv_{w}}{dz}~=~50$~$\rm{km~s^{-1}~kpc^{-1}}$. For each of the model, we follow the fitting procedure, given below,  to obtain proton distribution.

a)  It is to be noted that  $z_{t}$ and  $B_0^{\rm{turbulent}}$ are related to each other and the relation is obtained by reproducing the the observed synchrotron spectrum at 408 MHz with  the CR electron flux \citep{dibernardo2013} obtained from different CR electron propagation models. We, now, fit $^{10}$Be/$^{9}$Be to get an estimate of $z_{t}$. The $B_0^{\rm{turbulent}}$ corresponding to $z_{t}$ is obtained from the relation provided in \citet{dibernardo2013}. For fixed values of  $z_{t}$ and $B_0^{\rm{turbulent}}$, we derive $\eta$, $\delta$, $D_{0}$, and $v_{Alf}$ by minimizing $\chi^{2}/d.o.f$ \footnote{$\chi^{2} = \sum_{l} \rm{\frac{(Measured~value~ -~Simulated~value)^{2}}{(Error~in~measured~ value)^{2}}}$, $l$ denotes the  number of measured values or number of observational data points. The d.o.f = total number of data points - number of variables used for fitting.} (d.o.f means degrees of freedom) of B/C data without taking into account of solar modulation. We use Voyager \citep{cum2016}, PAMELA \citep{pambc} and CREAM \citep{creambc} data for our work. During the fitting of $^{10}$Be/$^{9}$Be  and B/C, we use a test injection spectrum for proton that roughly fits the observed proton spectra.

b) In this step, we fix  $\rho^{p}_{0,2} = 330$~GV and we tune the spectral indices  $\alpha^{p}_{1}, \alpha^{p}_{2}, \alpha^{p}_{3}$, and spectral break $\rho^{p}_{0,1}$ of the injection spectrum of proton (same for all other heavy nuclei) by minimizing $\chi^{2}/d.o.f$  of PAMELA \citep{adriani2013proton} and CREAM \citep{yoon2011} data. In this case, we also include solar modulation which is estimated by fitting PAMELA \citep{adriani2013proton} and AMS 02 \citep{aguilar2015} data as at lower energies the deviation  between data points of these two observations may indicate the difference in solar activity at the epochs of these two observations. Moreover, we want to add that Voyager proton data \citep{stone2013, cum2016} is supposed to represent proton flux in the ISM which is not affected by solar modulation and fitting of Voyager data provides a rough estimate of  $\alpha^{p}_{1}$. During minimization of  $\chi^{2}/d.o.f$, we use a range of values of  $\alpha^{p}_{1}$ around such rough estimated value. During this fitting procedure, we keep fixed all the parameter values obtained in the previous step and simultaneously check the fitted spectra of $^{10}$Be/$^{9}$Be  and B/C.

In our simulations, we follow the above fitting procedures for each model to  obtain the best fit values of the parameters such that for each  fit of B/C and  proton flux data the $\chi^{2}/d.o.f \leq 1 \sigma$ ($\sigma$ denotes the usual standard deviation). We, then, use those best fit values in the \texttt{DRAGON} code to obtain proton flux ($J_{p}^{Gal}(E_{k},r,z)$) in all the position of the Galaxy which is constrained by the local measurements of CRs. In the next section, we will use that proton flux to obtain the diffuse gamma-ray flux.

\section{Calculation of diffuse gamma-ray flux}\label{sec:gamflux}

In the preceding section, we already discussed the CR proton distribution and the total gas density profile in the Milky Way Galaxy. Now, we can calculate the gamma-ray emissivity ($J_{\gamma}(E_{\gamma}, r, z) $) at  any $r$ and $z$ using our code which is based on the semi-analytic method  presented in \citet{kelner2006}. The gamma-ray emissivity (in units of $\rm{GeV}^{-1}\rm{cm}^{-3}\rm{s}^{-1}\rm{sr}^{-1}$) is denoted by

\begin{widetext}
\begin{eqnarray}
\rm{for}~ E_{\gamma}\geq 100~\rm{GeV},\nonumber \\
J_{\gamma}(E_{\gamma}, r, z) =&& n_{g}(r,z) \int_{E_{\gamma}}^{\infty} \sigma(E_{p}) J^{Gal}_{p}(E_{p}, r, z) F_{\gamma}\Big(\frac{E_{\gamma}}{E_{p}}, E_{p}\Big) \frac{dE_{p}}{E_{p}},  \label{eq:hgam} \\
\rm{and}~\rm{for}~ 1~\rm{GeV}\leqslant E_{\gamma } \lesssim 100~\rm{GeV}, \nonumber \\
J_{\gamma}(E_{\gamma}, r, z) =&& 2\times \tilde{n}~ \frac{n_{g}(r,z)}{K_{\pi}}  \nonumber \\ && \times \int_{E_{\gamma} + \frac{m_{\pi}^{2}}{4E_{\gamma}}}^{\infty} \sigma\Big(m_{p} + \frac{E_{\pi}}{K_{\pi}}\Big) J^{Gal}_{p}\Big((m_{p} + \frac{E_{\pi}}{K_{\pi}}), r, z\Big)  \frac{dE_{\pi}}{\sqrt{E_{\pi}^{2} - m_{\pi}^{2}}},   \label{eq:lgam} 
\end{eqnarray}
\end{widetext}

where, $E_{p}$, $E_{\gamma}$ and $F_{\gamma}$ are the energy of the incident proton, gamma-ray  energy and the spectrum of the secondary gamma-ray in a single collision \footnote{see section IV.A of \citet{kelner2006} for the expression of $F_{\gamma}$ which is obtained on the basis of SIBYLL code \citep{fletch1994}.} respectively. In equation \ref{eq:lgam}, $m_{p}$, $E_{\pi}$ and $m_{\pi}$ denote the mass of the proton, energy and mass of the pion respectively. $K_{\pi}$ and $\tilde{n}$ are the free parameters.

\begin{widetext}
\begin{eqnarray}\label{eq:gamflux}
\Phi_{\gamma}(E_{\gamma}) =&&\frac{2\times 3.08\times 10^{21}}{(1- \rm{cos}~70^{\degree})}  \times \int_{\theta_{\rm{min}}=~0^{\degree}}^{\theta_{\rm{max}}=~360^{\degree}} \int_{z_{\rm{min}}=~0}^{z_{\rm{max}}=~3z_{t}~\rm{kpc}}\int_{r_{\rm{min}}=~0}^{r_{\rm{max}}=~40~\rm{kpc}}  drdzd\theta   \nonumber \\
&& ~ \times \frac{r~J_{\gamma}(E_{\gamma}, r, z)}{4\pi(r^{2} + r_{E}^{2} + z^{2} - 2rr_{E}\rm{cos}\theta)} \times \Big[ 1 - \Theta(3~{\rm{kpc}} - |z|)\Theta(15~{\rm{kpc}} - r) \Big] \nonumber \\
&& \times~ \Theta \Bigg( {\rm{sin^{-1}}}\Big(\frac{|z|}{\sqrt{(r^{2} + r_{E}^{2} + z^{2} - 2rr_{E}\rm{cos}\theta)}}\Big) - 20^{\degree}  \Bigg),
 \end{eqnarray}
\end{widetext}

\begin{figure}[!h]

\includegraphics[width=0.5\textwidth,clip,angle=0]{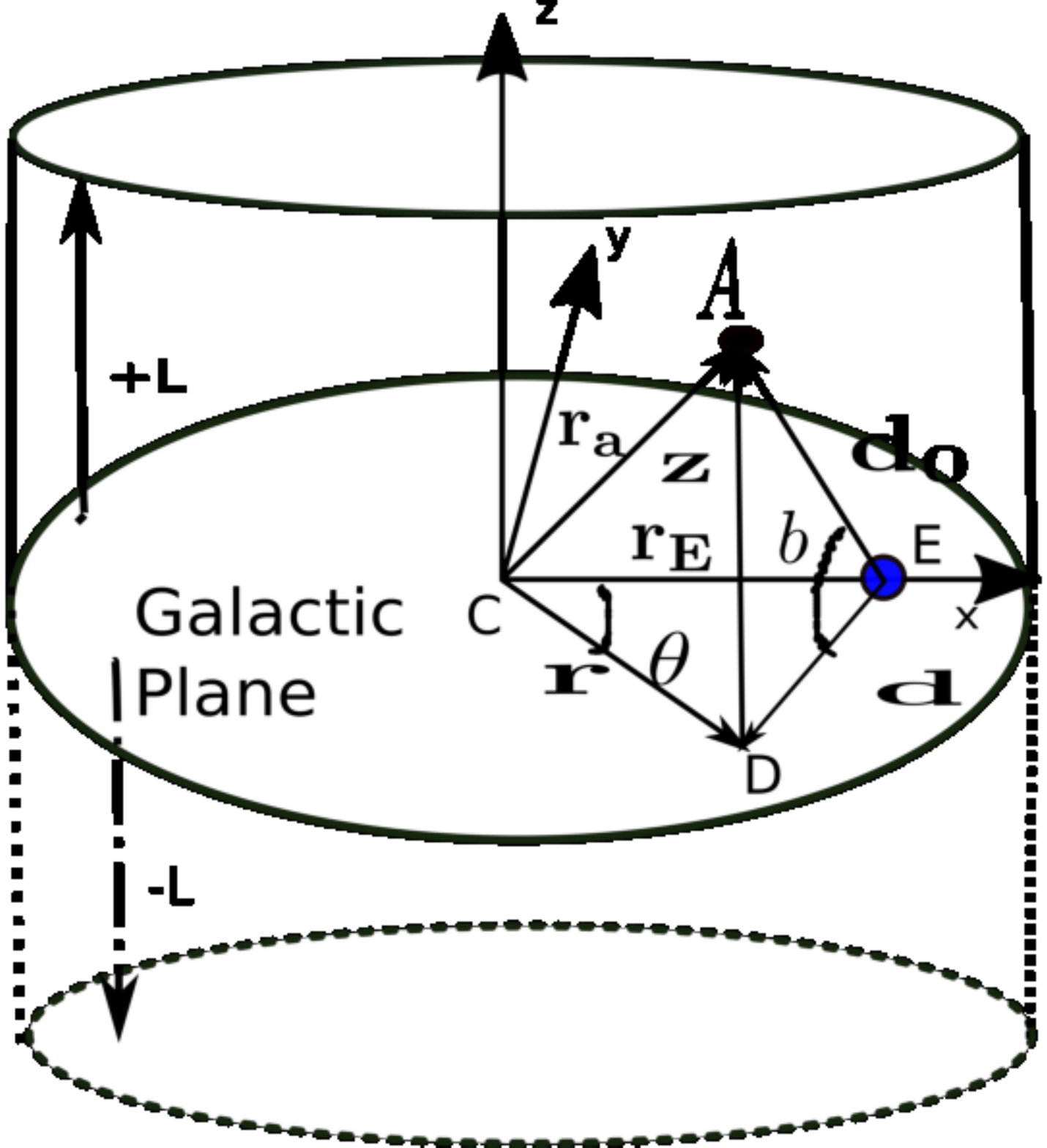}

\caption{\label{fig:diagram}  The schematic diagram shows the geometry used to model our Galaxy and calculate the total diffuse gamma-ray flux ($\phi_{\gamma}(E_{\gamma})$) (in this section) and luminosity of CR (see Sec \ref{sec:compdfluxlum}) Here, C and E represent the Galactic centre and position of the Earth respectively. The point D is the vertical projection of an arbitrary point A, where the proton distribution is calculated.} ${\bf{d_{0}}}= {\bf{d}}+{\bf{z} }= ({\bf{r} }- {\bf{r_{E}}}) + {\bf{z} }$ and $|{\bf{d_{0}}}|=\sqrt{(r^{2} + r_{E}^{2} + z^{2} - 2rr_{E}\rm{cos}\theta)}$.
\end{figure} 

 In our calculation, we consider $K_{\pi} = 0.17$ \citep{kelner2006} whereas $\tilde{n}$ can vary in the range of 0.67 - 1.10 depending on the spectral indices of proton and electrons \citep{kelner2006}. At $E_{\gamma} = 100$~GeV,  we calculate the diffuse gamma-ray flux (see equation \ref{eq:gamflux}) using each $J_{\gamma}$ expression (see equations \ref{eq:hgam} and \ref{eq:lgam} and tune $\tilde{n}$  such that both fluxes can be matched. The total diffuse gamma-ray flux ($\phi_{\gamma}(E_{\gamma})$ in units of $\rm{GeV}^{-1}\rm{cm}^{-2}\rm{s}^{-1}\rm{sr}^{-1}$) averaged over the solid angle at the Earth can be represented by equation \ref{eq:gamflux}. 
    $\Theta(x)$ denotes the Heaviside function. The square bracket term in equation \ref{eq:gamflux} represents the exclusion of gamma-ray emission from the inner Galaxy which is modeled as a cylinder with radius 15~kpc and a half-height of 3~kpc above and below the Galactic plane. The last Heaviside function in equation \ref{eq:gamflux} denotes the exclusion of low-latitude ($|b| < 20^{\degree}$) gamma-ray emission following the measurement of IGRB by \textit{Fermi} -LAT \citep{ackermann2015}. In the present calculation, we consider contribution of gamma-ray emission above the Galactic plane (+z direction) and multiply it by 2 to include the contribution coming from below the Galactic plane. Hence, the factor 2 comes in the numerator of the prefactor of the integration. The other factor, $3.08\times 10^{21}$, in the numerator represents the conversion factor from kpc to cm unit. The term $(1-\rm{cos}70^{\degree})$ in the denominator of the prefactor of the integration is due to averaging of the total flux over the solid-angle. In equation \ref{eq:gamflux} , we do not take into account the gamma-ray production from secondary electrons, produced in p-p interaction through decay of charged pions, via inverse Compton scattering of CMB photons. We also ignore the contribution of electrons produced in the electromagnetic cascades initiated due to interaction of high energy photon with CMB and infrared photon field. The study of energetics and mean free path indicate that those  processes are less favorable than the gamma-ray emission due to decay of neutral pions which are produced in p-p collisions \citep[e.g.][]{ coppi1997, liu2019}.

\section{Results}\label{sec:result}

In this section, we present the fitted CR spectra and parameters needed for fitting in different models like PD, KRA, CON and KOL. We also display the diffuse gamma-ray fluxes obtained from these models and compare with the IGRB data presented by \textit{Fermi}-collaboration \citep{ackermann2015}. The comparison of diffuse gamma-ray fluxes calculated  from KRA model by imposing different constraints on $r$, $z$ and Galactic latitude is also presented here.

\subsection{PD model}\label{sec:pd}

\begin{table}[h!]
\caption{\label{tab:pd} Models and best fitted parameter values of PD model to fit CR spectra, shown in figure \ref{fig:pdgam}, using \texttt{DRAGON} code are listed here.}

\begin{tabular}{p{4cm} c }
\hline
\hline
Model/Parameter  &   Option/Value \\
 \hline
  $R_{\rm{max}}$												&     40.0 ~ kpc \\  
   $L$                      										&     18.0~ kpc  \\
   Source Distribution          							&   Ferriere \\
  Diffusion type												&   Exp  (see equation \ref{eq:diff})   \\
  $D_0$       													&   $2.02 \times 10^{29}$~ $\rm{cm^{2}/s}$  \\
  $\rho_0$														&     3.0~GV            \\
  $\delta$														 &       0.53         	\\
  $z_t$ 															&	 6.0~kpc			\\
  $\eta$   														 &		-0.30		    \\
  $v_{Alf}$   													     &			0.0		        \\
  $\frac{dv_{w}}{dz}$                                       & 0.0                    \\
  Magnetic field type										  &	    Pshirkov			      \\
  $B_0^{\rm{disc}}$ 										  &      $2.0 \times 10^{-6}$ ~ Gauss      \\
  $B_0^{\rm{halo}}$            							  &    $4.0 \times 10^{-6}$ ~ Gauss                \\
  $B_0^{\rm{turbulent}}$   									   &      $ 6.62 \times 10^{-6}$ ~Gauss      \\
  $\alpha^{p}_{1}/ \alpha^{p}_{2}/ \alpha^{p}_{3}$		& 1.90/2.27/2.22          \\
  $\rho^{p}_{0,1}/ \rho^{p}_{0,2}$	                  		 &    4.80/330 ~GV      \\
  $\chi^{2}/(d.o.f = 22)$ (for B/C)                           &    0.51  \\
  $\chi^{2}/(d.o.f = 26)$ (for proton)                           &    0.31 \\  
  \hline
\end{tabular}

\end{table}

 \begin{figure*}[ht]
\centering
\mbox{

\includegraphics[width=0.40\textwidth,clip,angle=0]{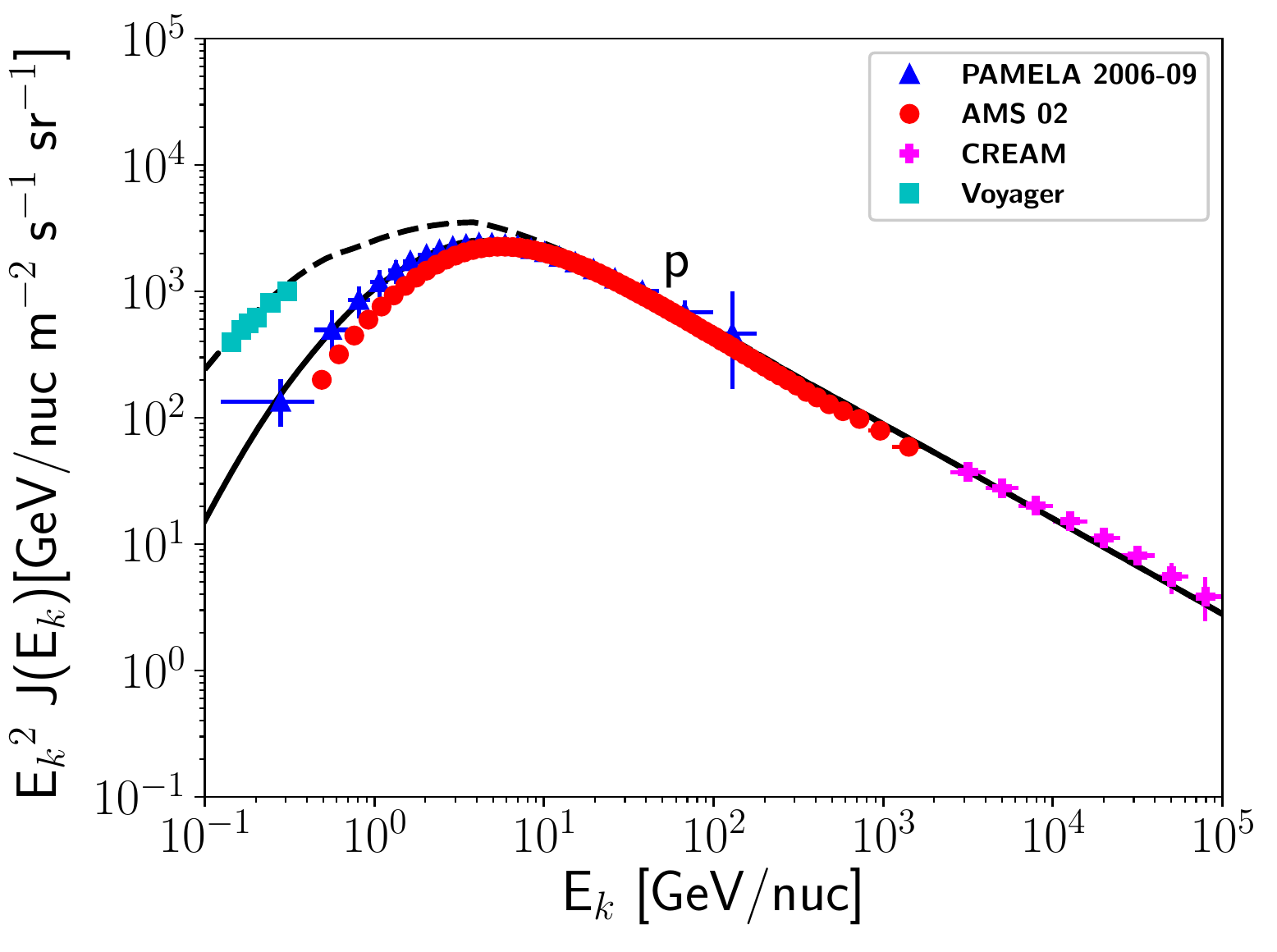}

\includegraphics[width=0.40\textwidth,clip,angle=0]{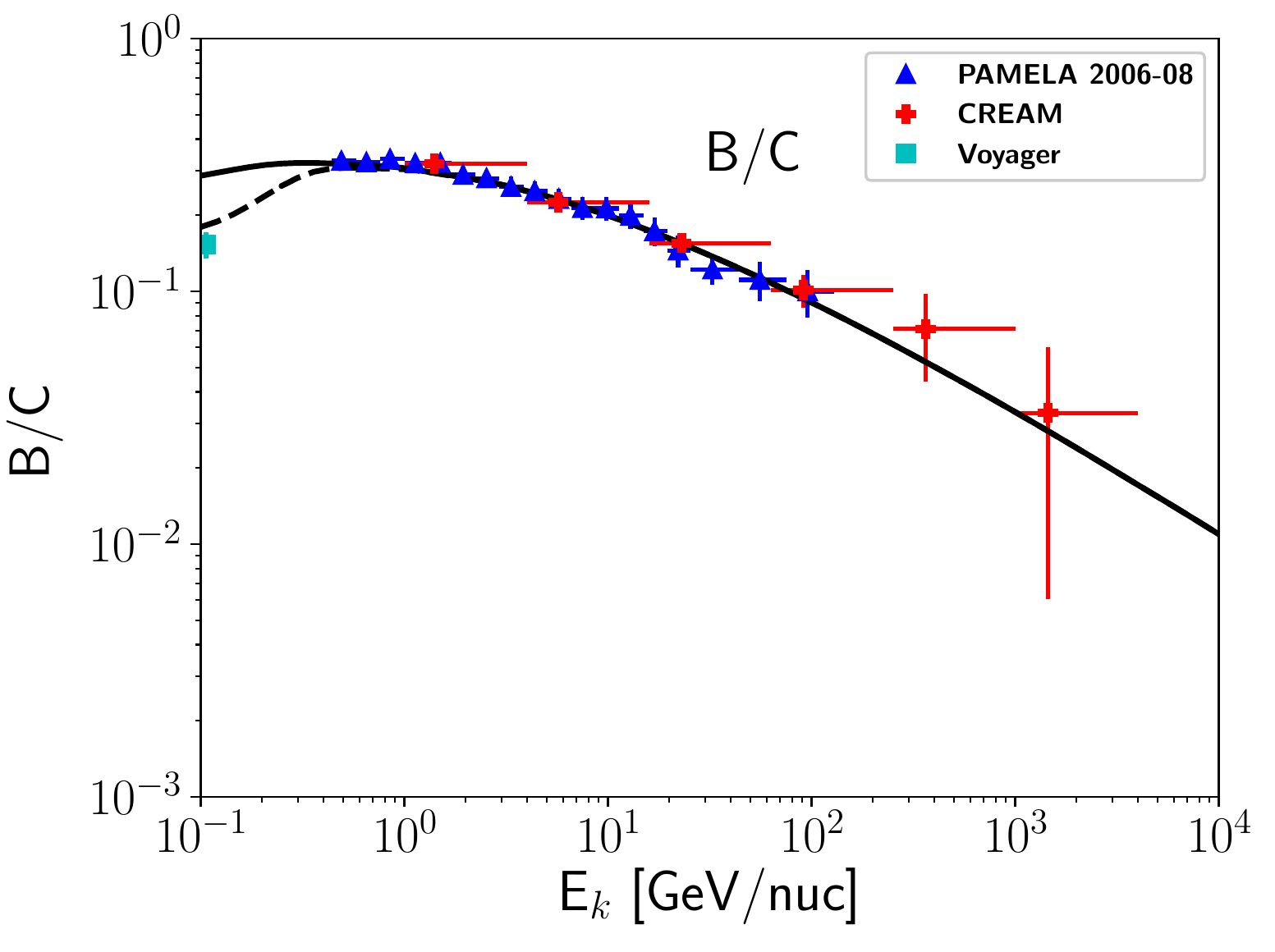}
}

\mbox{
\includegraphics[width=0.40\textwidth,clip,angle=0]{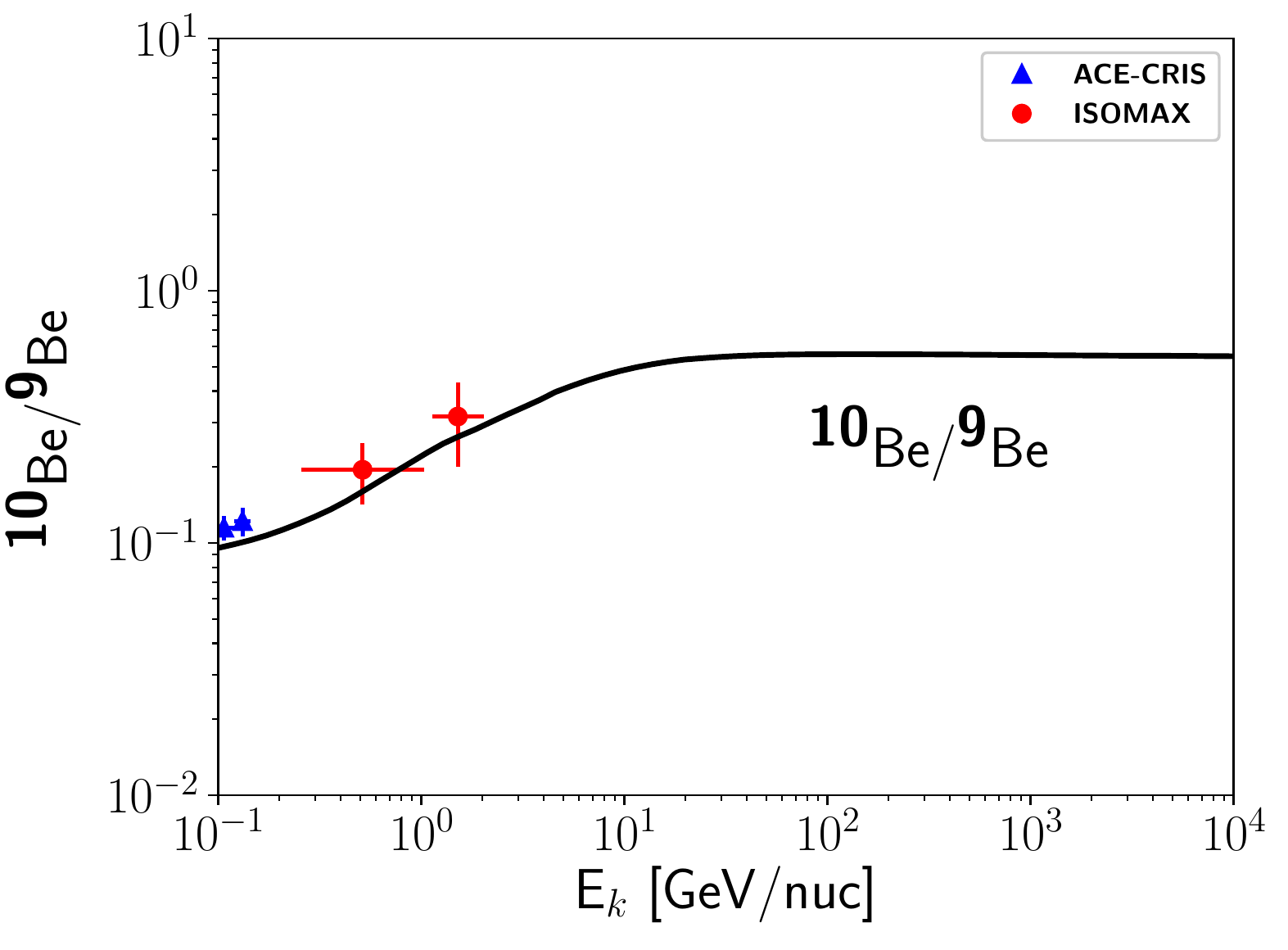}

\includegraphics[width=0.45\textwidth,clip,angle=0]{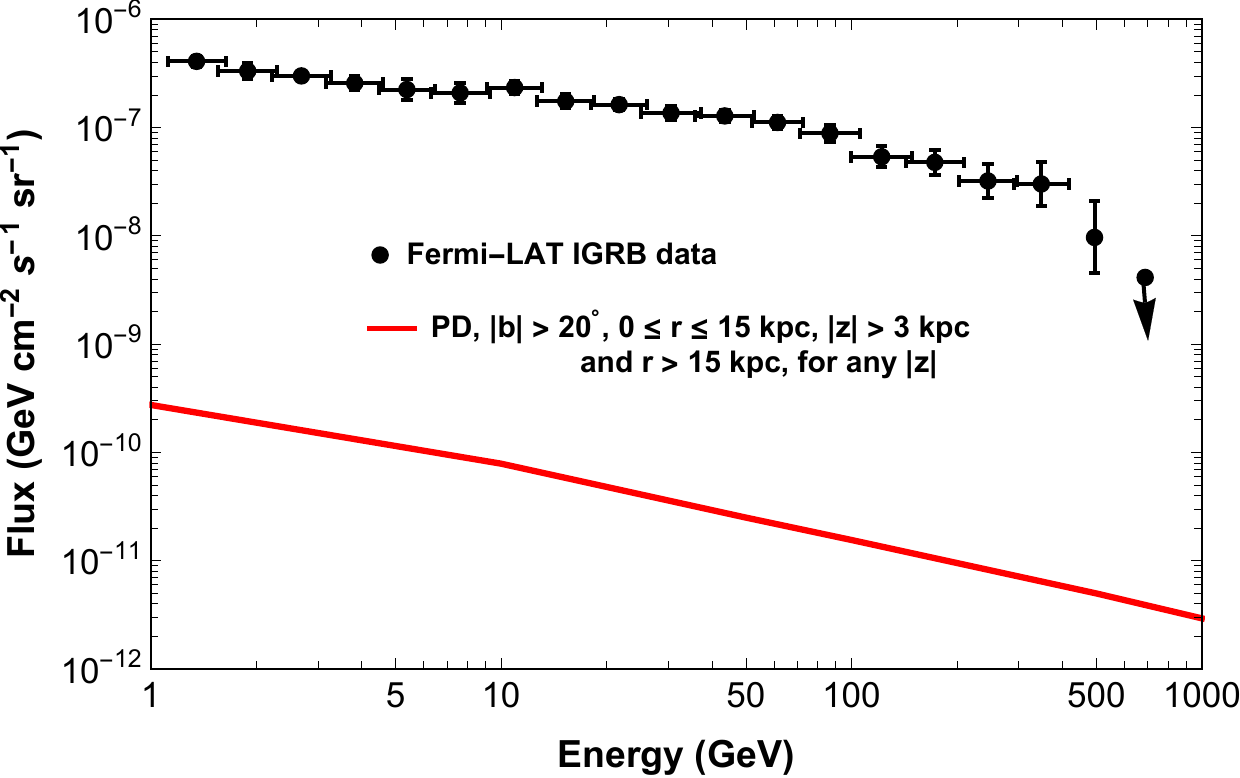}
}
\caption{\label{fig:pdgam} Energy dependence of primary CR flux, secondary to primary ratios, obtained from \texttt{DRAGON} code using PD model, are plotted with the locally measured CR fluxes. Proton flux (upper left panel) is plotted with Voyager \citep{stone2013, cum2016}, PAMELA \citep{adriani2013proton}, AMS 02 \citep{aguilar2015} and CREAM \citep{yoon2011} data. B/C (upper right panel) flux ratio is plotted with PAMELA \citep{pambc}, Voyager \citep{cum2016} and CREAM \citep{creambc} data. In case of proton and B/C, the dashed and solid lines represent spectra without and with the solar modulation ($\phi = 0.35$~GV) respectively. $^{10}$Be/$^{9}$Be  (bottom left panel) flux ratio is  plotted with ACE-CRIS \citep{ace2001} and ISOMAX \citep{isomax2004} data. Diffuse gamma-ray flux (bottom right panel) obtained from the PD model is compared with the IGRB data measured by \textit{Fermi}-LAT \citep{ackermann2015}. The downward arrow at highest energy bin (580-820) GeV represents the upper limit of flux. }
\end{figure*} 

\begin{figure*}[ht]
\centering
\mbox{
\includegraphics[width=0.38\textwidth,clip,angle=0]{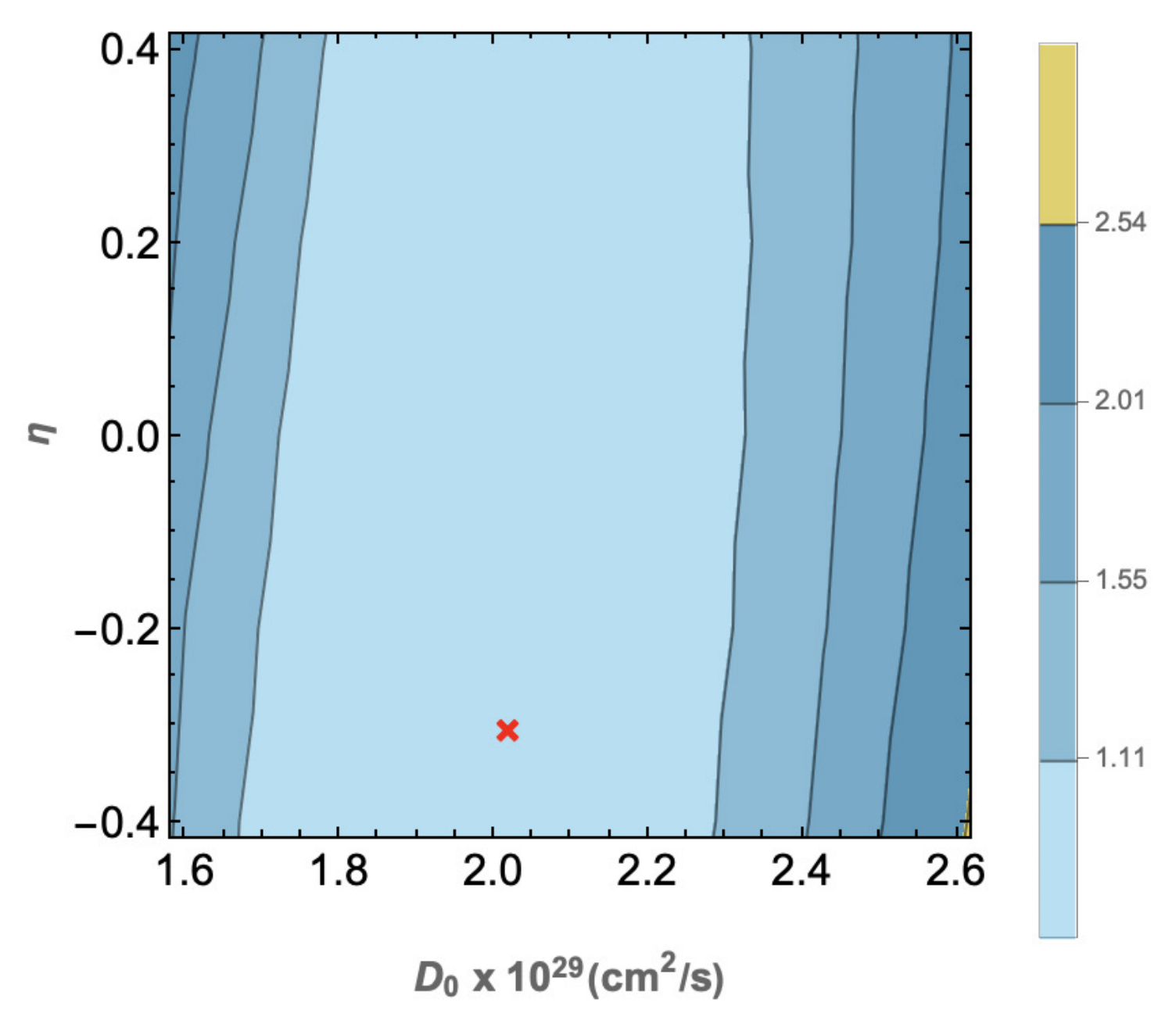}
\includegraphics[width=0.38\textwidth,clip,angle=0]{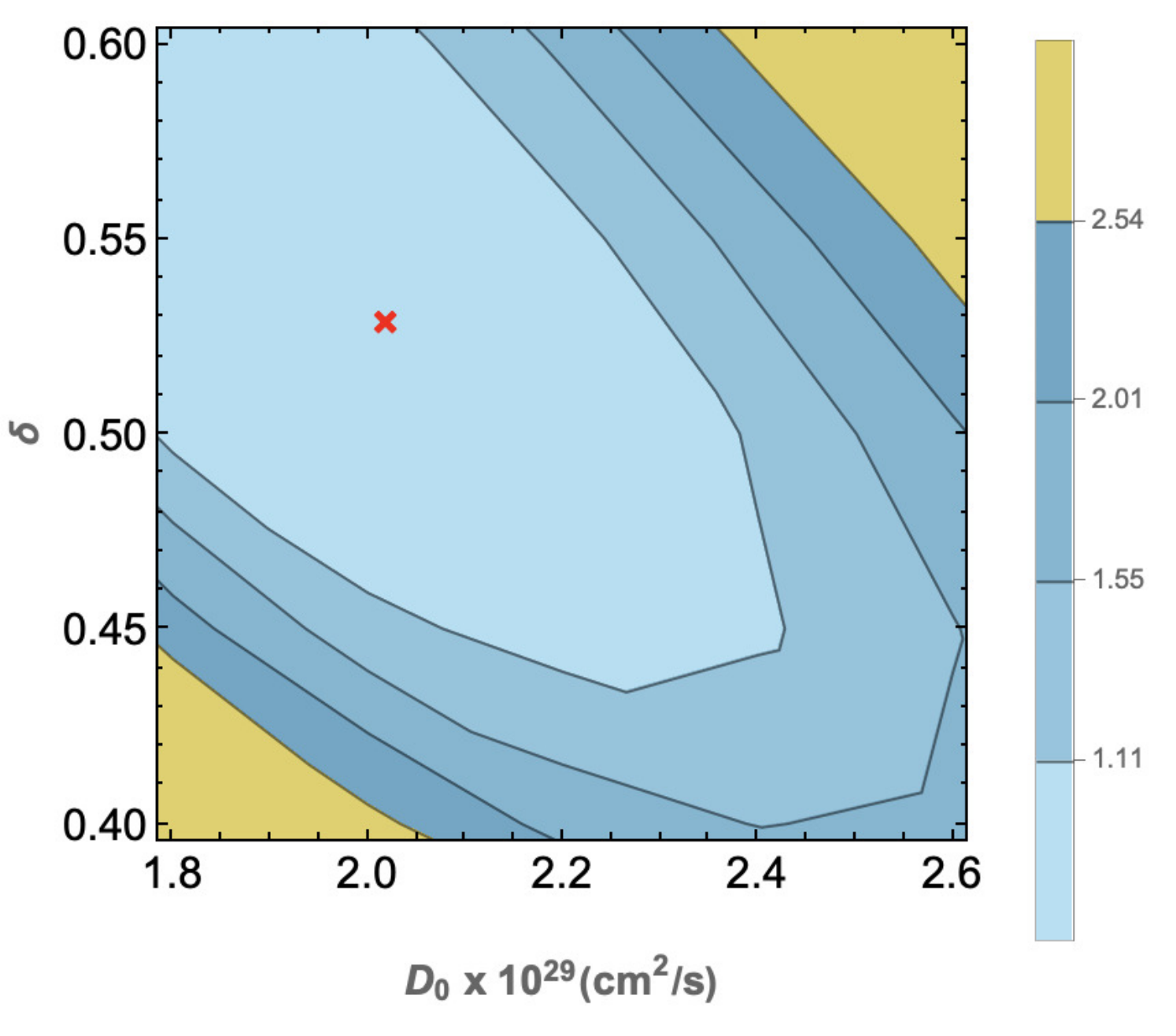}
}
\caption{\label{fig:pdcontour}  In case of PD model, the contour plots for $\eta$ against $D_{0}$ (left) and  $\delta$ against $D_{0}$ (right) are shown. The $1\sigma$, $2\sigma$, $3\sigma$, and $4\sigma$ of $\chi^{2}/d.o.f$ are represented by 1.11, 1.55, 2.01 and 2.54 respectively. The red cross marks in the plots correspond to the best fit values of the parameters shown in those plots.  }
\end{figure*}

PD model is the only model, among all the other models considered here, where re-acceleration term is absent (i.e. $v_{Alf} = 0$). We use the PD model in the \texttt{DRAGON} code and follow the procedure mentioned in section 2.4 to fit the observed CR data in the energy range of $0.1-10^{5}$~GeV/nuc. From the fitted CR spectra, we obtain the best fitted parameter values for the PD model. The parameter values and $\chi^{2}/d.o.f$ are listed in table \ref{tab:pd}.  We, finally, calculate diffuse gamma-ray flux following the procedure discussed in section \ref{sec:gamflux} and compare with IGRB data presented by \textit{Fermi} collaboration \citep{ackermann2015}.

In figure \ref{fig:pdgam}, we plot the fitted spectra of $^{10}$Be/$^{9}$Be,  B/C and proton with  locally observed CR spectra\footnote{All the CR data are  obtained from the cosmic ray database \citep{maurin2014}. Link of cosmic ray database \url{https://lpsc.in2p3.fr/cosmic-rays-db/}}. Proton spectra ($J(E_{k})$) with no solar modulation and with solar modulation ($\phi=0.35$~GV) are shown by dashed and solid lines (see the upper left panel of figure \ref{fig:pdgam}) respectively.  Diffuse gamma-ray flux obtained from PD model is also plotted with IGRB data measured by \textit{Fermi}-LAT. The results, presented here, show that PD model is consistent with both local and global observables.

We also study the correlations between $D_{0}$ and $\eta$, and $D_{0}$ and $\delta$ to understand the uncertainties lie in the range of their values. In figure \ref{fig:pdcontour},     we show the correlations through the contour plots of $D_{0}$ and $\eta$ (left),  and $D_{0}$ and $\delta$ (right) keeping other parameters fixed at the values tabulated in table \ref{tab:pd}. The different color gradients indicate different contour regions and the boundary values i.e., 1.11, 1.55, 2.01 and 2.54 indicate the $1\sigma$, $2\sigma$, $3\sigma$, and $4\sigma$ values of $\chi^{2}/d.o.f$ respectively. In the present work, we focus only to the values those lie $\leq~ 1\sigma$. The best fit values are marked by red  crosses in the figure \ref{fig:pdcontour}. The ranges of $D_{0}$ within $1 \sigma$ contour are shown in the figure \ref{fig:pdcontour}. Within $1 \sigma$ contour, $\eta$ value varies in a wide range of permissible negative and positive values, whereas $\delta$ can have lower limit of $\sim 0.45$.

\subsection{KRA model}\label{sec:kra}

\begin{table}[!h]
\caption{\label{tab:kra} Models and best fitted parameter values of KRA model to fit CR spectra, shown in figure \ref{fig:kragam}, using \texttt{DRAGON} code are listed here.}

\begin{tabular}{ p {4 cm}  c }
\hline
\hline
Model/Parameter  &   Option/Value \\
 \hline
  $R_{\rm{max}}$																		  &     40.0 ~ kpc \\  
   $L$                      																&     18.0~ kpc  \\
   Source Distribution          													&   Ferriere \\
  Diffusion type															&   Exp  (see equation \ref{eq:diff})   \\
  $D_0$       													&   $2.05 \times 10^{29}$~ $\rm{cm^{2}/s}$  \\
  $\rho_0$																						&     3.0~GV            \\
  $\delta$																		                &       0.50         	\\
  $z_t$ 																							&	 6.0~kpc			\\
  $\eta$   																						&		-0.30		    \\
  $v_{Alf}$   															               &  25.0 $\rm{km~s^{-1}}$		   \\
  $\frac{dv_{w}}{dz}$                                                                &  0.0                    \\
  Magnetic field type																&	    Pshirkov			      \\
  $B_0^{\rm{disc}}$ 												&      $2.0 \times 10^{-6}$ ~ Gauss      \\
  $B_0^{\rm{halo}}$            								&    $4.0 \times 10^{-6}$ ~ Gauss                \\
  $B_0^{\rm{turbulent}}$   									   &      $ 6.62 \times 10^{-6}$ ~Gauss      \\
  $\alpha^{p}_{1}/ \alpha^{p}_{2}/ \alpha^{p}_{3}$		& 1.92/2.30/2.26          \\
  $\rho^{p}_{0,1}/ \rho^{p}_{0,2}$	                  					 &    5.2/330 ~GV      \\
   $\chi^{2}/(d.o.f = 22)$ (for B/C)                           &    0.51  \\
  $\chi^{2}/(d.o.f = 26)$ (for proton)                           &    0.45 \\  
  \hline
  \end{tabular}

\end{table}

\begin{figure*}[ht]
\centering
\mbox{

\includegraphics[width=0.40\textwidth,clip,angle=0]{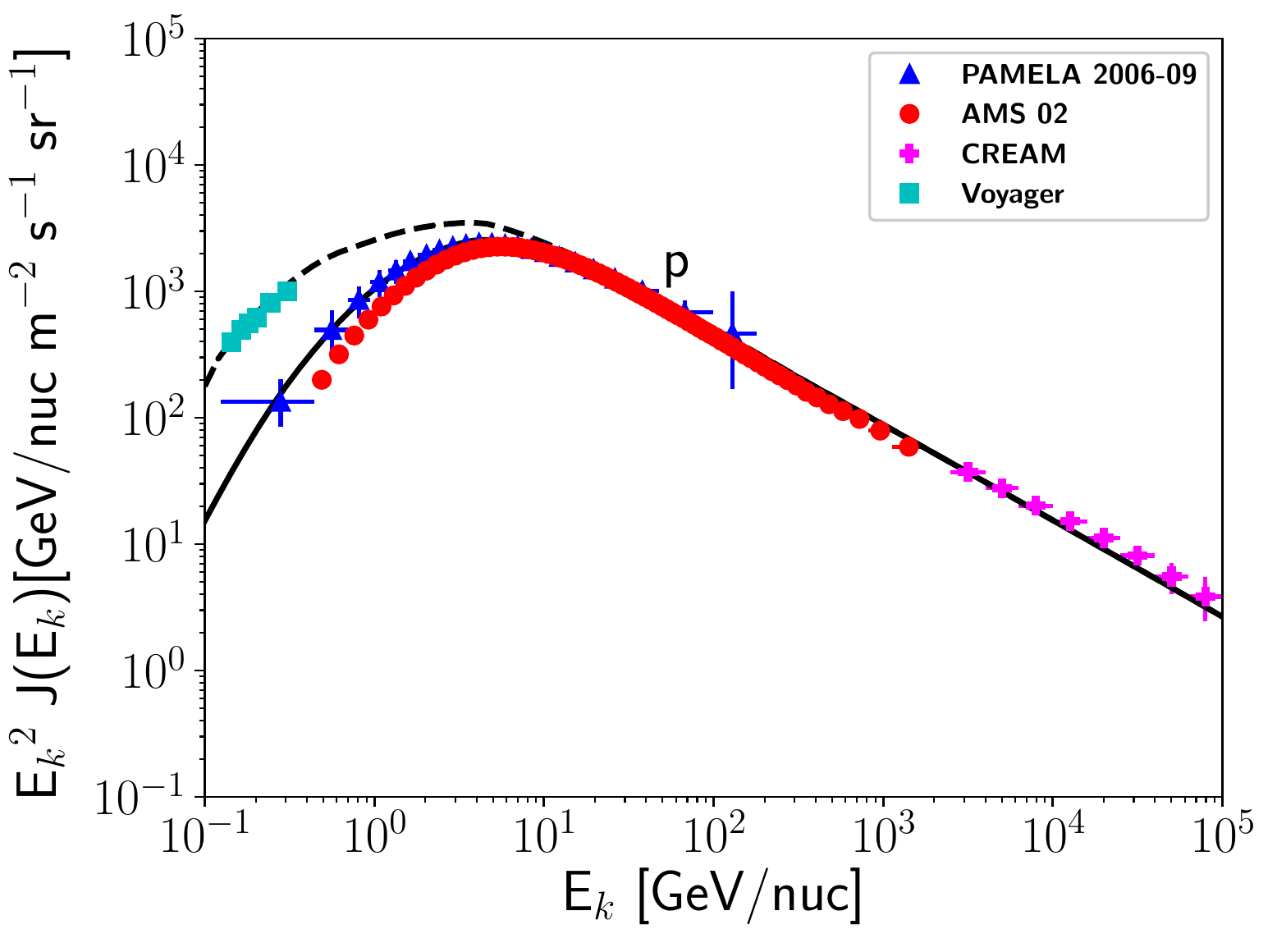}

\includegraphics[width=0.40\textwidth,clip,angle=0]{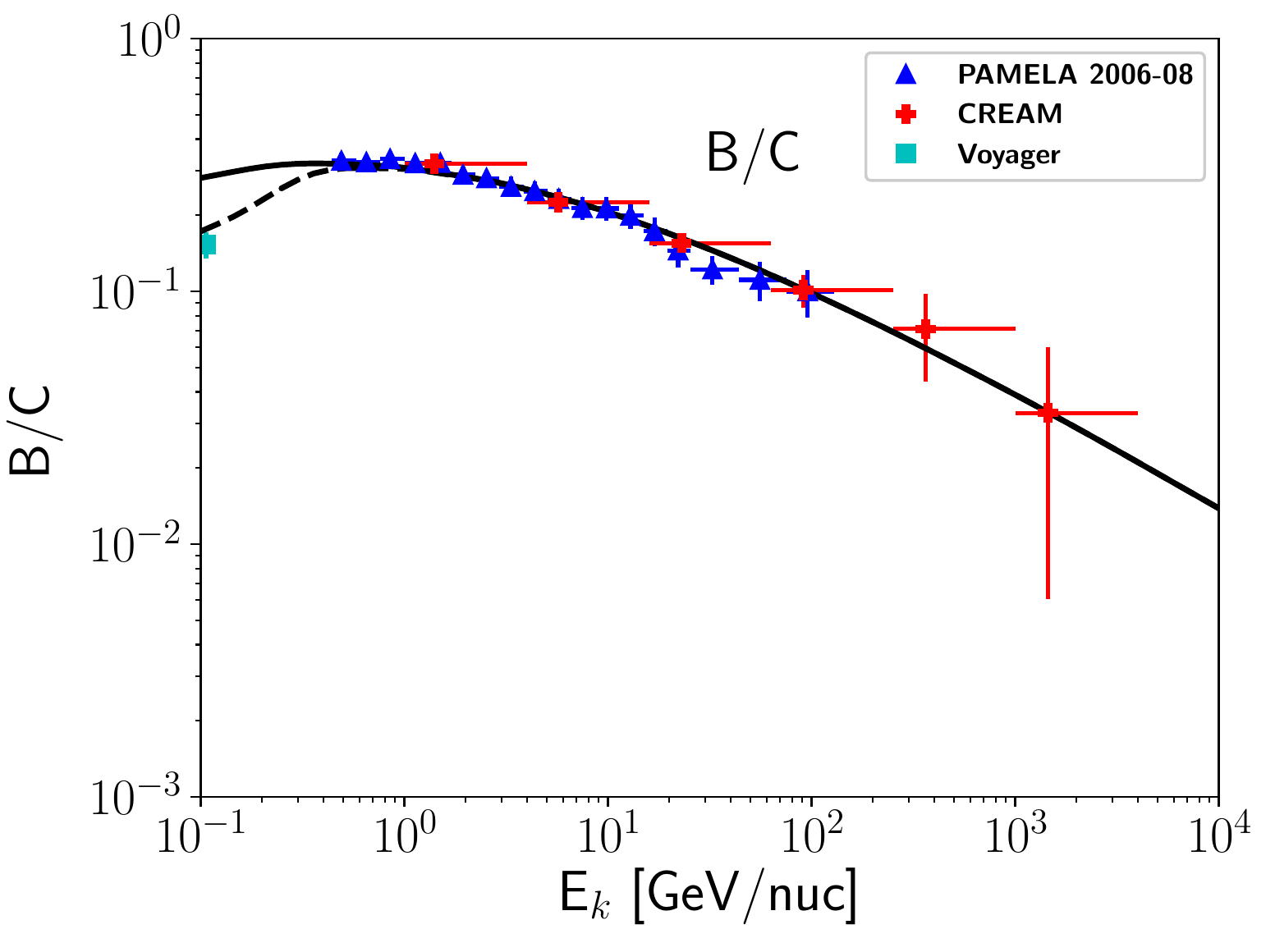}
}

\mbox{
\includegraphics[width=0.40\textwidth,clip,angle=0]{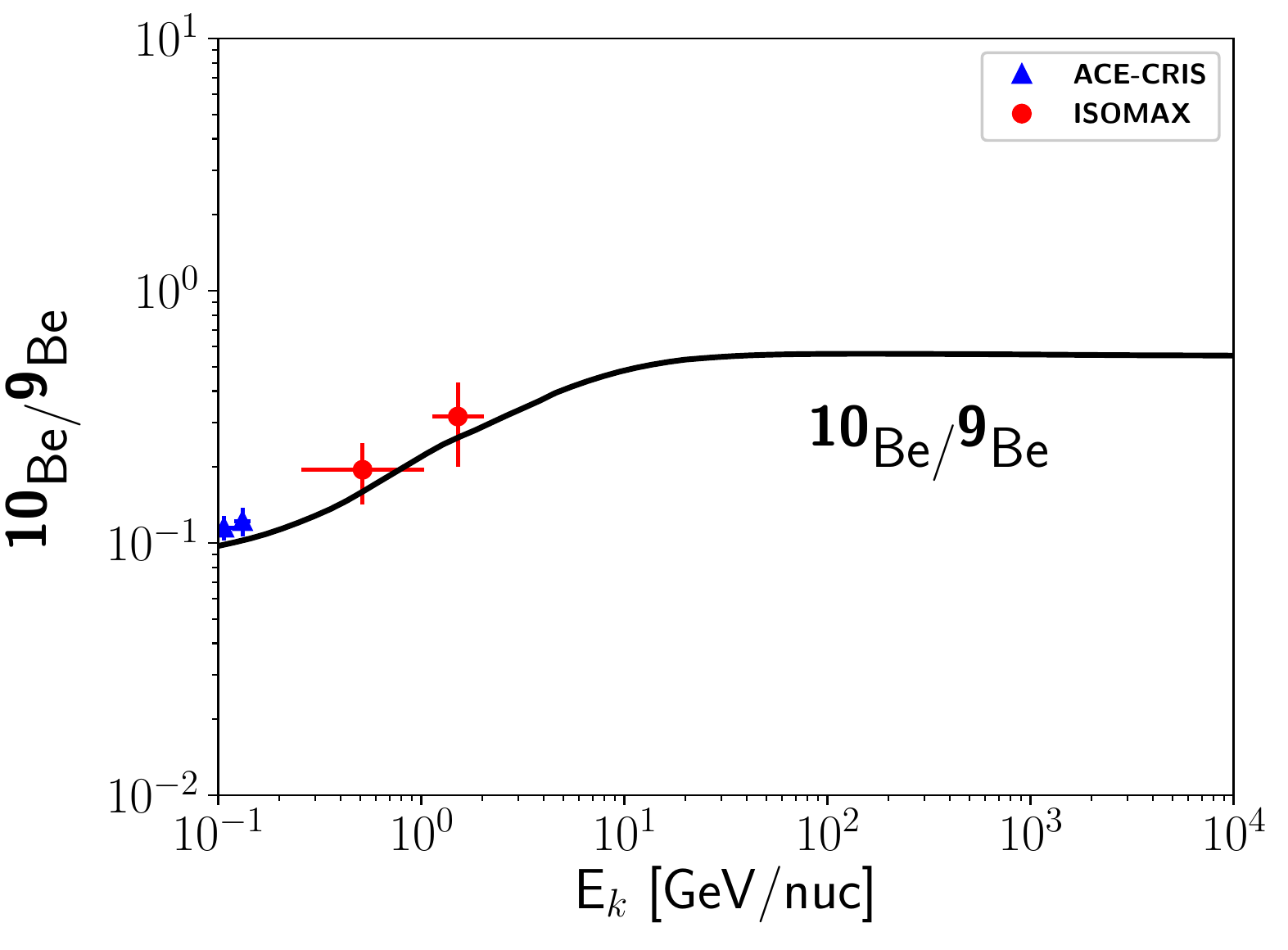}

\includegraphics[width=0.45\textwidth,clip,angle=0]{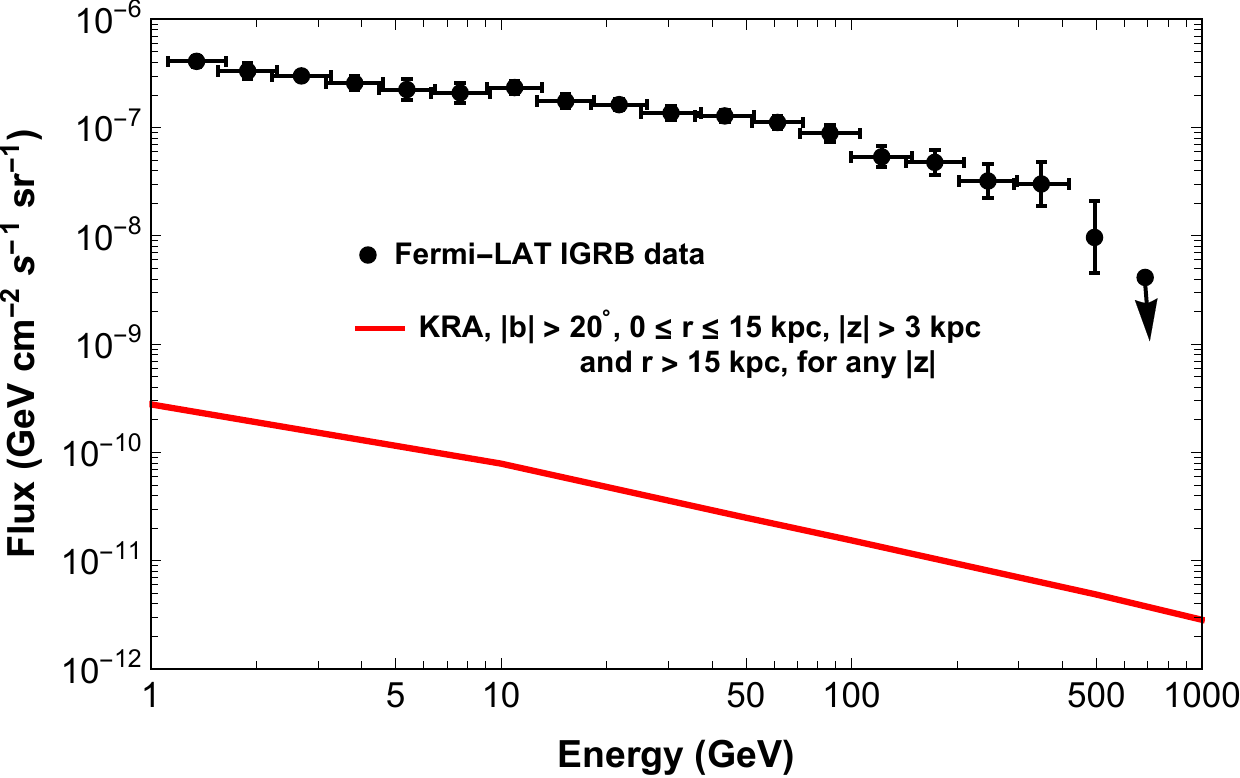}
}
\caption{\label{fig:kragam} Energy dependence of primary CR flux, secondary to primary ratios, obtained from \texttt{DRAGON} code using KRA model, are plotted with the locally measured CR fluxes. Proton flux (upper left panel) is plotted with Voyager \citep{stone2013, cum2016}, PAMELA \citep{adriani2013proton}, AMS 02 \citep{aguilar2015} and CREAM \citep{yoon2011} data. B/C (upper right panel) flux ratio is plotted with PAMELA \citep{pambc}, Voyager \citep{cum2016} and CREAM \citep{creambc} data.  In case of proton and B/C, the dashed and solid lines represent spectra without and with the solar modulation ($\phi = 0.35$~GV) respectively. $^{10}$Be/$^{9}$Be  (bottom left panel) flux ratio is  plotted with ACE-CRIS \citep{ace2001} and ISOMAX \citep{isomax2004} data. Diffuse gamma-ray flux (bottom right panel) obtained from the KRA model is compared with the IGRB data measured by \textit{Fermi}-LAT \citep{ackermann2015}. The downward arrow at highest energy bin (580-820) GeV represents the upper limit of flux. }
\end{figure*} 

\begin{figure*}[ht]
\centering
\mbox{
\includegraphics[width=0.38\textwidth,clip,angle=0]{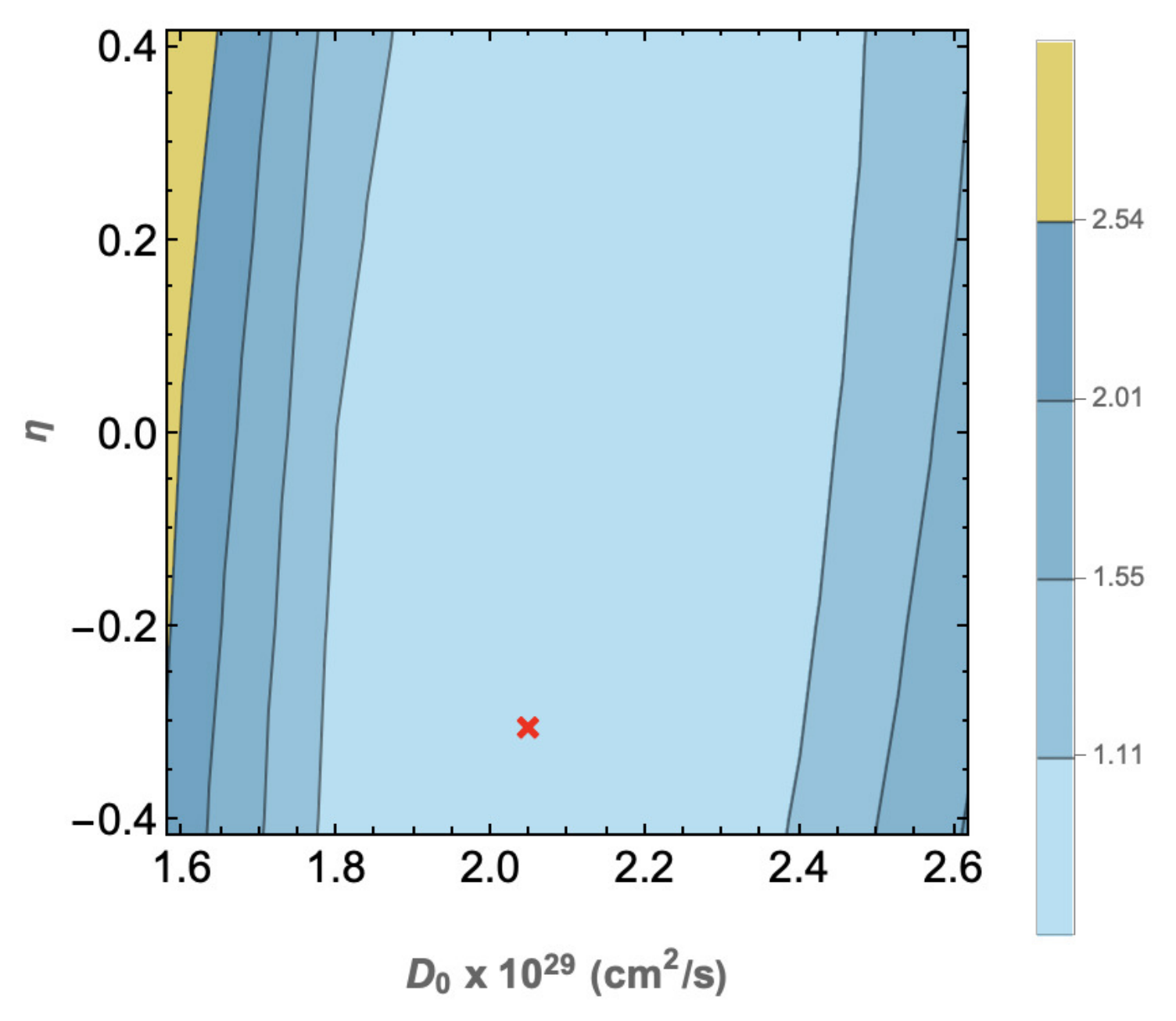}

\includegraphics[width=0.38\textwidth,clip,angle=0]{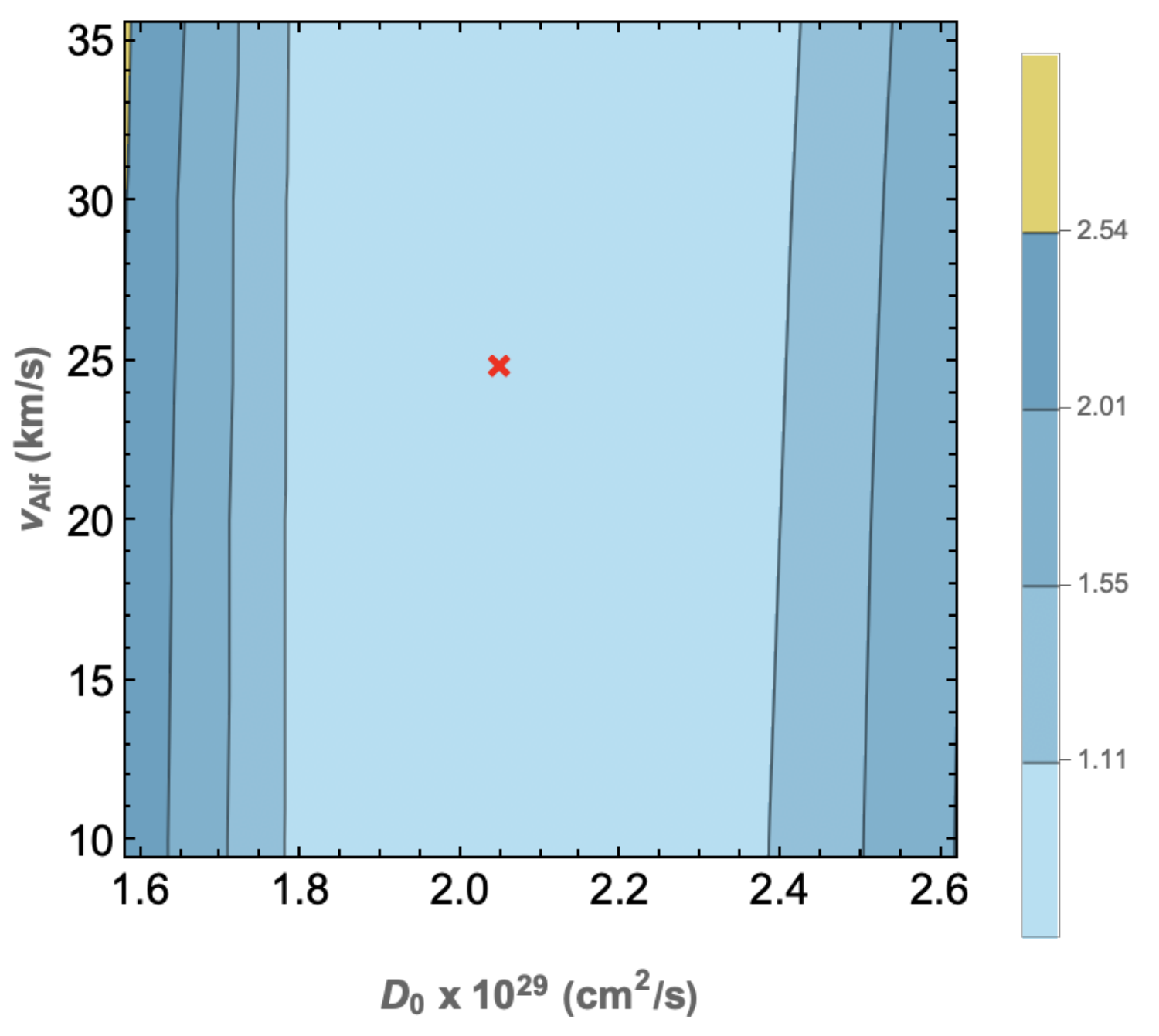}
}

\caption{\label{fig:kracontour}   In case of KRA model, the contour plots for $\eta$ against $D_{0}$ (left) and  $v_{Alf}$ against $D_{0}$ (right) are shown. The $1\sigma$, $2\sigma$, $3\sigma$, and $4\sigma$ of $\chi^{2}/d.o.f$ are represented by 1.11, 1.55, 2.01 and 2.54 respectively. The red cross marks in the plots correspond to the best fit values of the parameters shown in those plots. }
\end{figure*}

KRA model is characterized by its fixed $\delta$ value; $\delta = 0.50$. Although the fitted $\delta$ value of PD (see the table \ref{tab:pd}) is close to the $\delta$ value of KRA, $v_{Alf}$ parameter exists in KRA which makes the difference between KRA and PD. It is also noted that $\delta$ value in KRA model has a fixed value. In PD model, the $\delta$ can be varied to tune the observed CR spectra. The best fitted parameter values and $\chi^{2}/d.o.f$  of KRA model which are needed to fit the locally measured CR spectra are listed in table \ref{tab:kra}. The fitted CR spectra and the diffuse gamma-ray flux obtained from KRA model are shown in the figure \ref{fig:kragam}. The diffuse gamma-ray fluxes of KRA model over the energy range of 1 -1000 GeV are consistent with the IGRB data.

In figure \ref{fig:kracontour}, we show the contour plots of $\eta$ against $D_{0}$ (left) and $v_{Alf}$ against $D_{0}$ (right). Within $1 \sigma$ contour, the range of $D_{0}$ of KRA model is almost similar to PD model. Like PD, $\eta$ values of KRA model also vary in the similar range having negative to positive values within $1 \sigma$ contour. Unlike PD, KRA model contains non zero value of $v_{Alf}$ and its possible range is shown in figure \ref{fig:kracontour}.

\subsection{CON model}\label{sec:con}

\begin{table}[!h]
\caption{\label{tab:con} Models and best fitted parameter values of CON model to fit CR spectra, shown in figure \ref{fig:congam}, using \texttt{DRAGON} code are listed here.}

\begin{tabular}{ p {4 cm}  c }
\hline
\hline
Model/Parameter  &   Option/Value \\
 \hline
  $R_{\rm{max}}$																		  &     40.0 ~ kpc \\  
   $L$                      																&     18.0~ kpc  \\
   Source Distribution          													&   Ferriere \\
  Diffusion type															&   Exp  (see equation \ref{eq:diff})   \\
  $D_0$       													&   $1.22 \times 10^{29}$~ $\rm{cm^{2}/s}$  \\
  $\rho_0$																						&     3.0~GV            \\
  $\delta$																		                &       0.63         	\\
  $z_t$ 																							&	 6.0~kpc			\\
  $\eta$   																						&		-0.30		    \\
  $v_{Alf}$   																			&  55.0 $\rm{km~s^{-1}}$	\\
  $\frac{dv_{w}}{dz}$                                                         &  50.0 $\rm{km~s^{-1}kpc^{-1}}$	      \\
  Magnetic field type																&	    Pshirkov			      \\
  $B_0^{\rm{disc}}$ 												&      $2.0 \times 10^{-6}$ ~ Gauss      \\
  $B_0^{\rm{halo}}$            								&    $4.0 \times 10^{-6}$ ~ Gauss            \\
  $B_0^{\rm{turbulent}}$   									   &      $ 6.62 \times 10^{-6}$ ~Gauss      \\
  $\alpha^{p}_{1}/ \alpha^{p}_{2}/ \alpha^{p}_{3}$		& 2.0/2.33/2.12          \\
  $\rho^{p}_{0,1}/ \rho^{p}_{0,2}$	                  				&    5.8/330 ~GV      \\
   $\chi^{2}/(d.o.f = 21)$ (for B/C)                           &    0.86  \\
  $\chi^{2}/(d.o.f = 26)$ (for proton)                           &    0.43 \\  
  \hline
  \end{tabular}
\end{table}

\begin{figure*}[ht]
\centering
\mbox{
\includegraphics[width=0.40\textwidth,clip,angle=0]{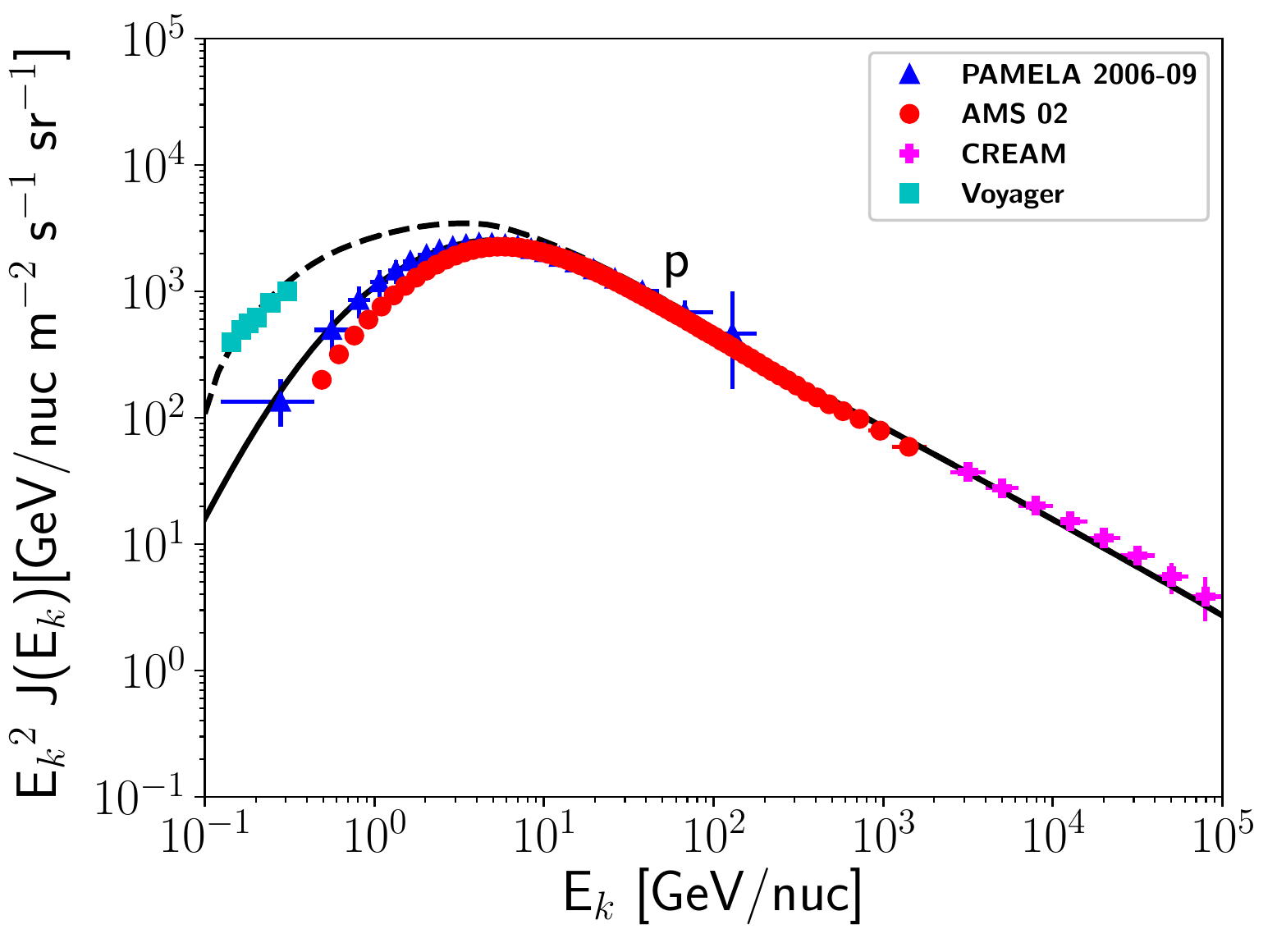}

\includegraphics[width=0.40\textwidth,clip,angle=0]{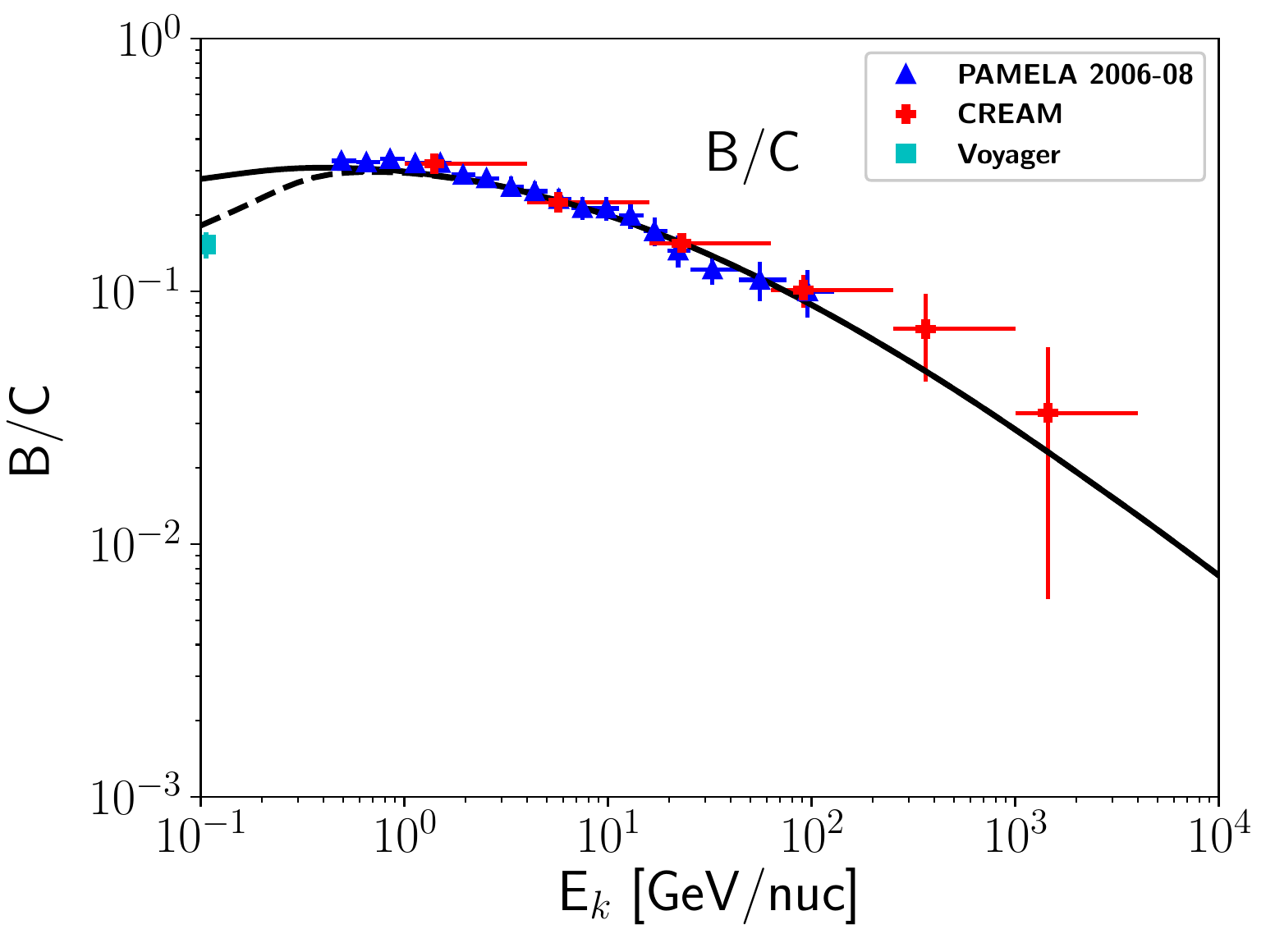}
}

\mbox{
\includegraphics[width=0.40\textwidth,clip,angle=0]{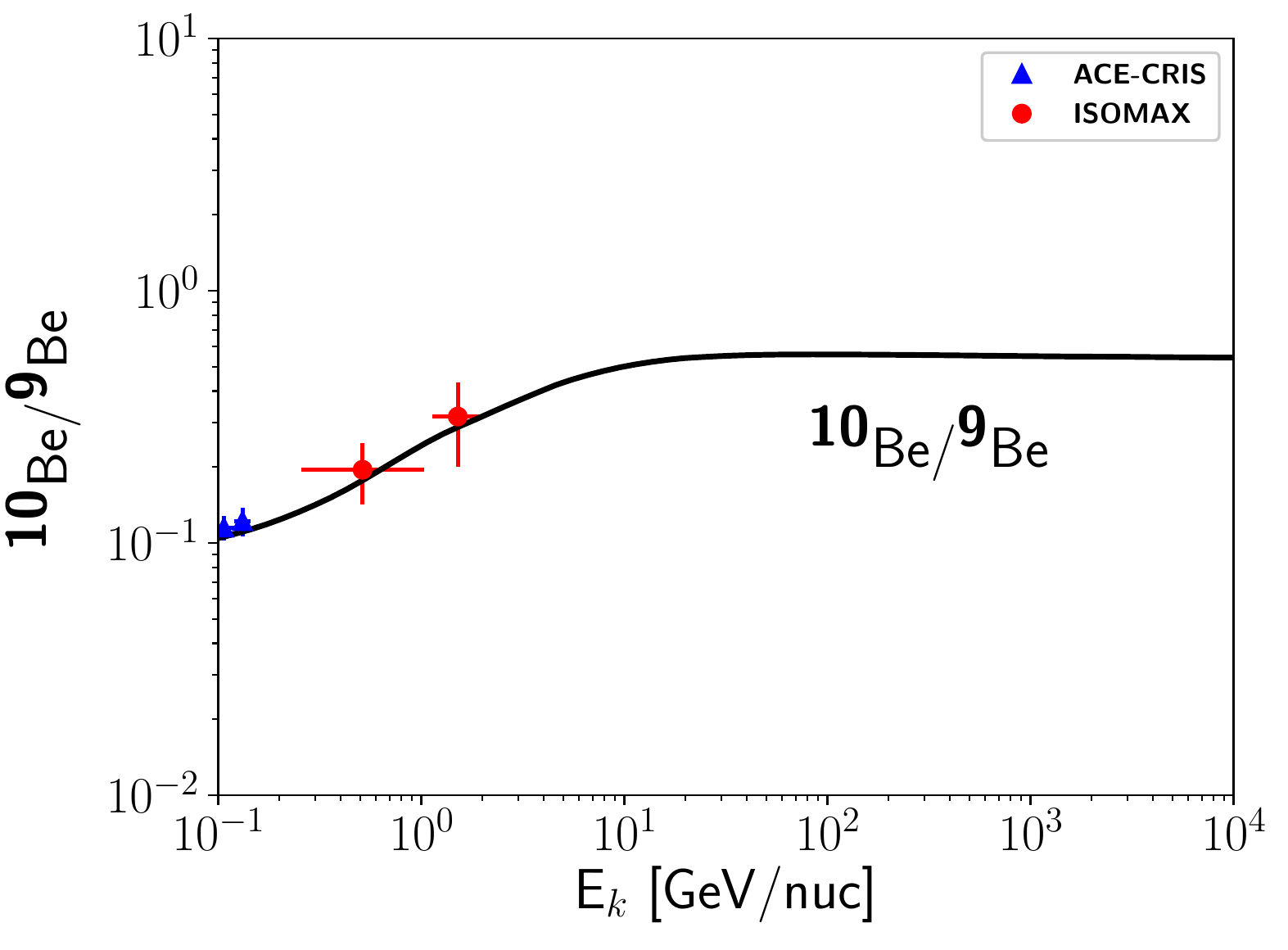}

\includegraphics[width=0.45\textwidth,clip,angle=0]{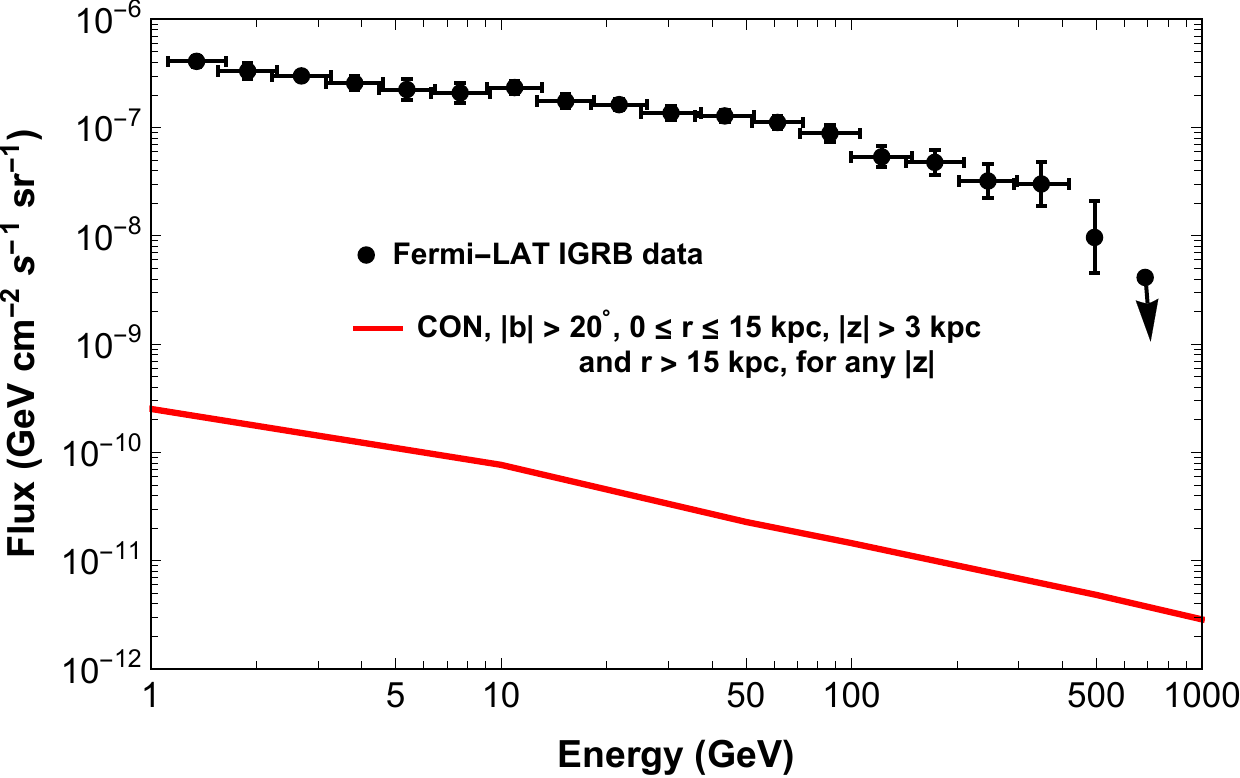}
}
\caption{\label{fig:congam} Energy dependence of primary CR flux, secondary to primary ratios, obtained from \texttt{DRAGON} code using CON model, are plotted with the locally measured CR fluxes. Proton flux (upper left panel) is plotted with Voyager \citep{stone2013, cum2016}, PAMELA \citep{adriani2013proton}, AMS 02 \citep{aguilar2015} and CREAM \citep{yoon2011} data. B/C (upper right panel) flux ratio is plotted with PAMELA \citep{pambc}, Voyager \citep{cum2016} and CREAM \citep{creambc} data.  In case of proton and B/C, the dashed and solid lines represent spectra without and with the solar modulation ($\phi = 0.35$~GV) respectively. $^{10}$Be/$^{9}$Be  (bottom left panel) flux ratio is  plotted with ACE-CRIS \citep{ace2001} and ISOMAX \citep{isomax2004} data. Diffuse gamma-ray flux (bottom right panel) obtained from the CON model is compared with the IGRB data measured by \textit{Fermi}-LAT \citep{ackermann2015}. The downward arrow at highest energy bin (580-820) GeV represents the upper limit of flux. }
\end{figure*}

\begin{figure*}[h!]
\centering
\mbox{
\includegraphics[width=0.38\textwidth,clip,angle=0]{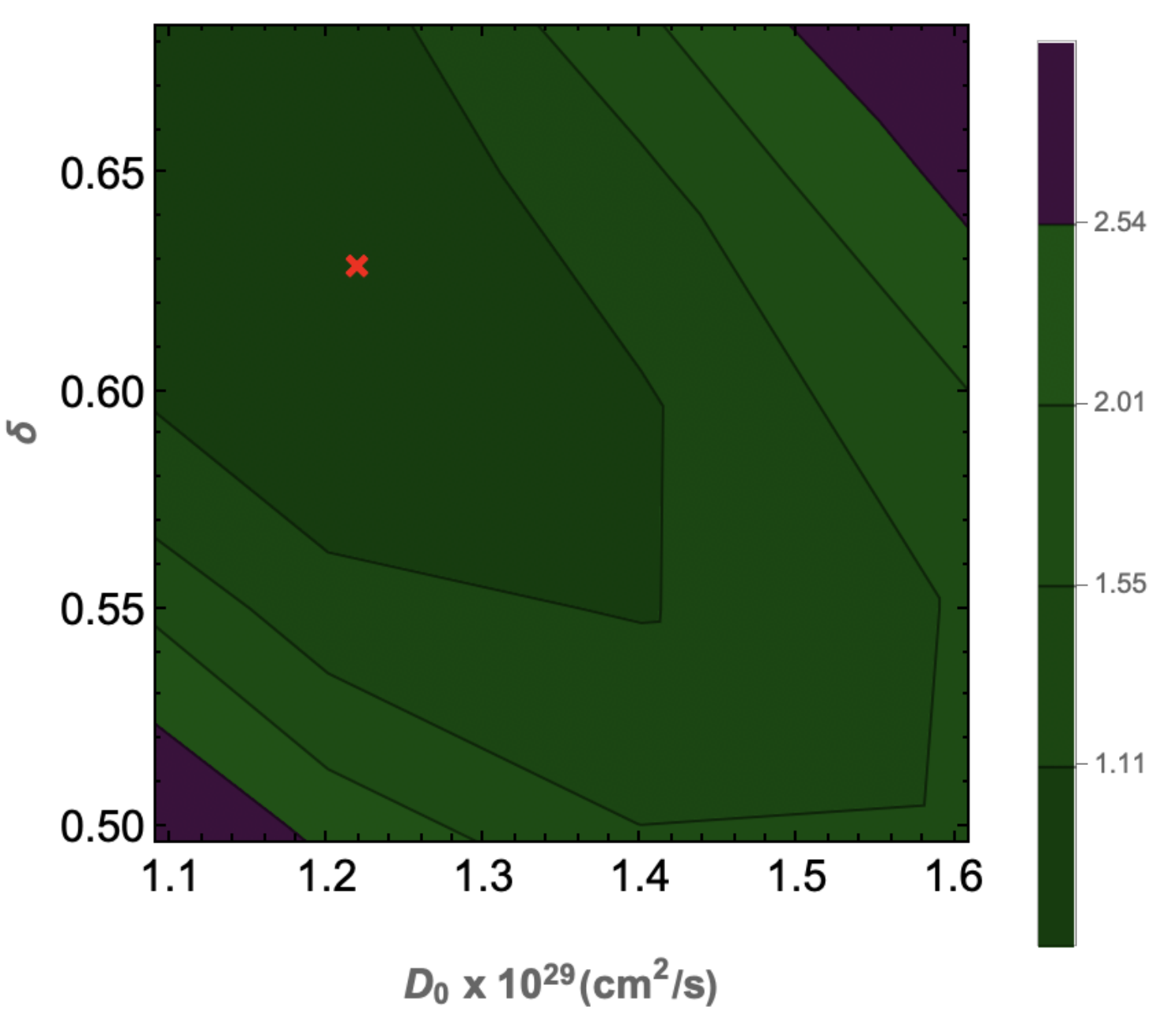}
\includegraphics[width=0.38\textwidth,clip,angle=0]{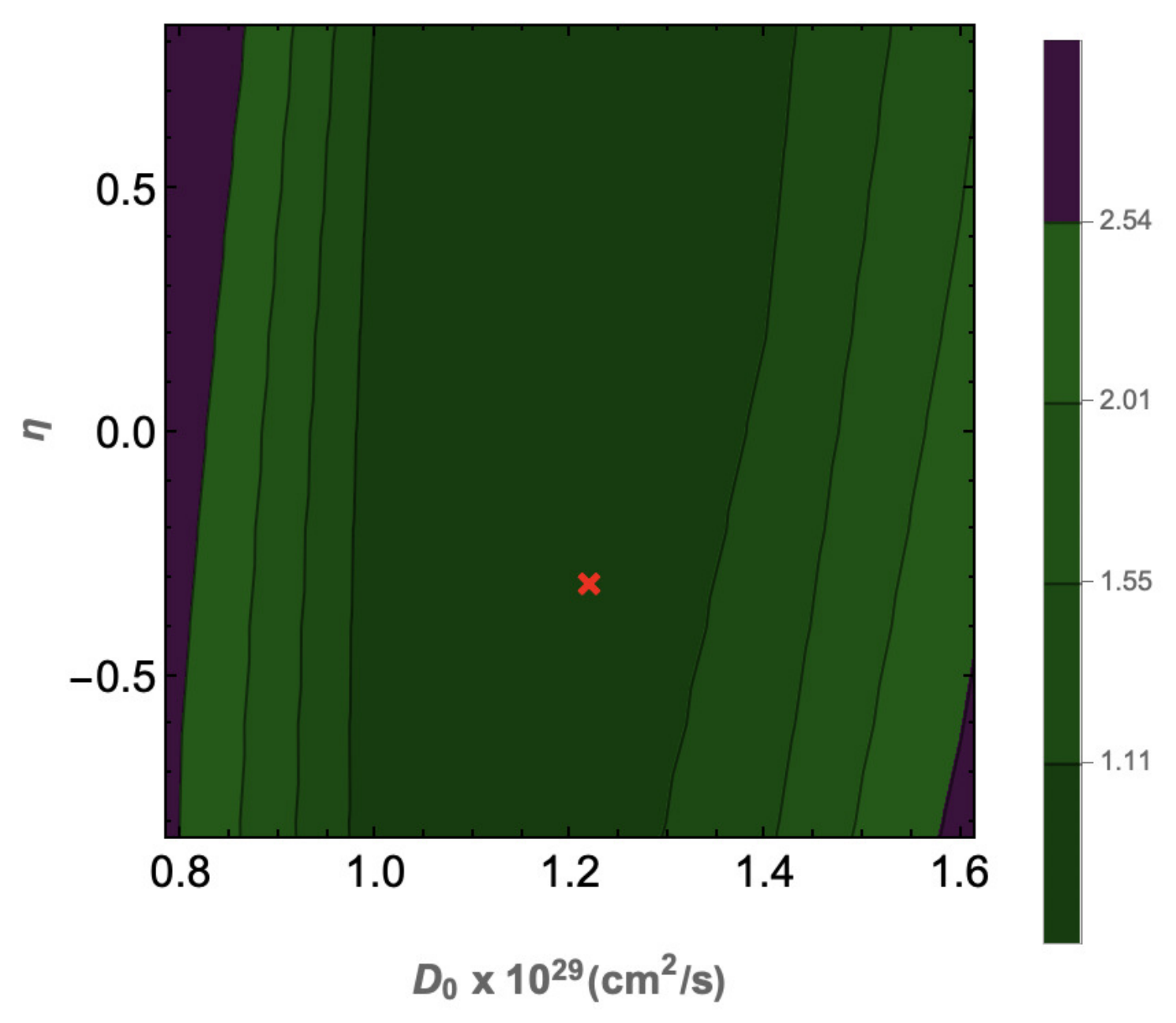}
}
\subfigure{}
\includegraphics[width=0.38\textwidth,clip,angle=0]{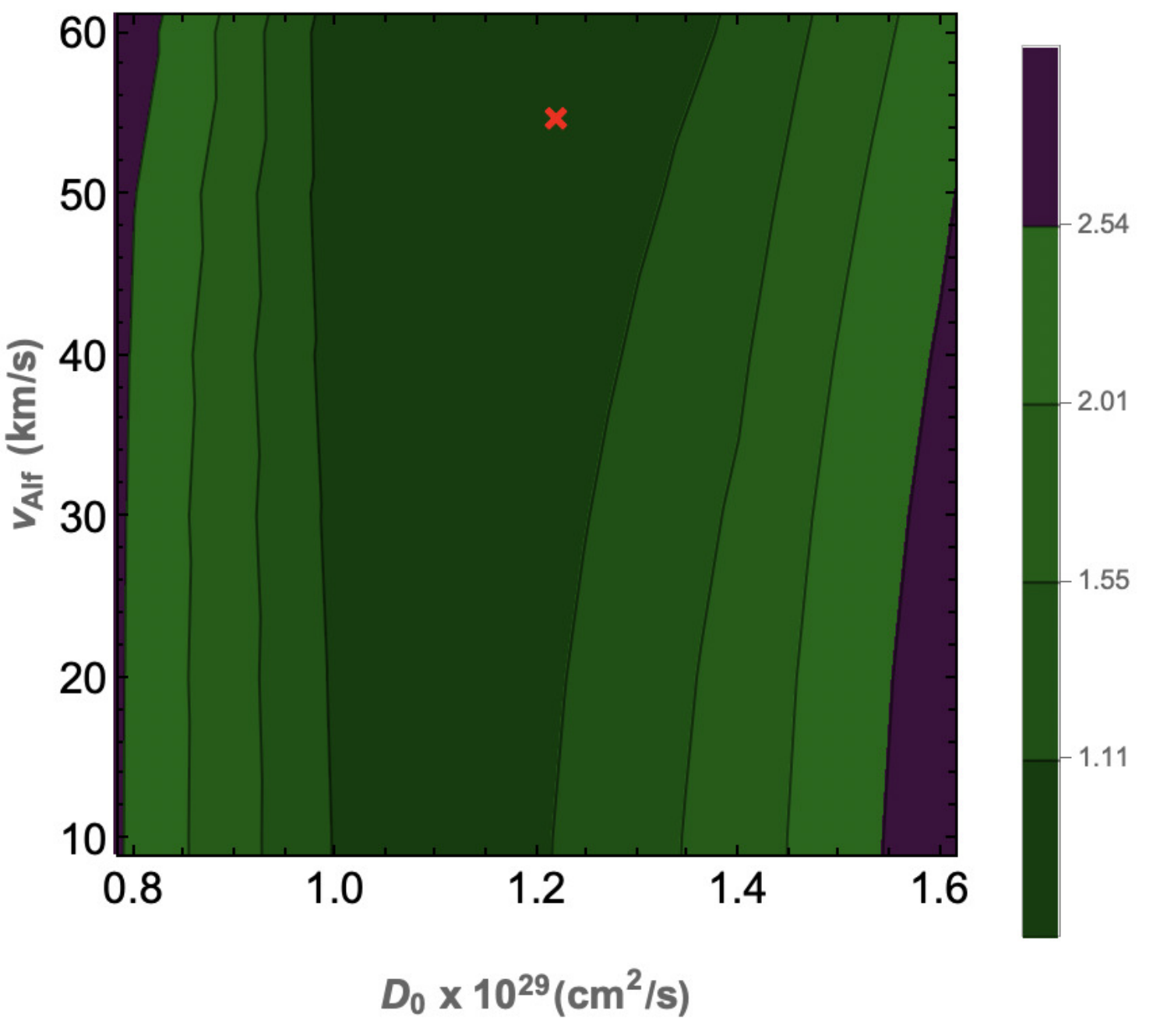}

\caption{\label{fig:concontour} In case of CON model, the contour plots for $\delta$ against $D_{0}$ (upper left), $\eta$ against $D_{0}$ (upper right) and  $v_{Alf}$ against $D_{0}$ (bottom) are shown. The $1\sigma$, $2\sigma$, $3\sigma$, and $4\sigma$ of $\chi^{2}/d.o.f$ are represented by 1.11, 1.55, 2.01 and 2.54 respectively. The red cross marks in the plots correspond to the best fit values of the parameters shown in those plots. }
\end{figure*}

CON model incorporates the convection wind speed via the parameter $\frac{dv_{w}}{dz}$ (= 50.0 $\rm{km~s^{-1}kpc^{-1}}$, fixed value chosen for the model) which is a major difference from the other models used here. Unlike KRA model, the $\delta$ value of CON model is not fixed. The best fitted parameter values and $\chi^{2}/d.o.f$ of CON model are listed in the table \ref{tab:con}.  Figure \ref{fig:congam} shows the results obtained from CON model. The results are consistent with the observed data.

In figure \ref{fig:concontour}, we show the contour plots of $\delta$ against $D_{0}$ (upper left), $\eta$ against $D_{0}$ (upper right), and $v_{Alf}$ against $D_{0}$ (bottom). For $D_{0}$ values of CON model, the $1\sigma$ contour shifts towards lower range compared to PD and KRA models. The  $1 \sigma$ ranges of $\eta$ and $v_{Alf}$ are almost similar to KRA model.

\subsection{KOL model}\label{sec:kol}

\begin{table}[!h]
\caption{\label{tab:kol} Models and best fitted parameter values of KOL model to fit CR spectra, shown in figure \ref{fig:kolgam}, using \texttt{DRAGON} code are listed here.}

\begin{tabular}{ p {4 cm} c }
\hline
\hline
Model/Parameter  &   Option/Value \\
 \hline
  $R_{\rm{max}}$																		  &     40.0 ~ kpc \\  
   $L$                      																&     15.0~ kpc  \\
   Source Distribution          													&   Ferriere \\
  Diffusion type															&   Exp  (see equation \ref{eq:diff})   \\
  $D_0$       													&   $2.98 \times 10^{29}$~ $\rm{cm^{2}/s}$  \\
  $\rho_0$																						&     3.0~GV            \\
  $\delta$																		                &       0.33         	\\
  $z_t$ 																							&	 5.0~kpc			\\
  $\eta$   																						&		2.5		    \\
  $v_{Alf}$   																			&  66.0 $\rm{km~s^{-1}}$	\\
  $\frac{dv_{w}}{dz}$                                                         &  0.0 	      \\
  Magnetic field type																&	    Pshirkov			      \\
  $B_0^{\rm{disc}}$ 												&      $2.0 \times 10^{-6}$ ~ Gauss      \\
  $B_0^{\rm{halo}}$            								&    $4.0 \times 10^{-6}$ ~ Gauss            \\
  $B_0^{\rm{turbulent}}$   									   &      $ 6.98 \times 10^{-6}$ ~Gauss      \\
  $\alpha^{p}_{1}/ \alpha^{p}_{2}/ \alpha^{p}_{3}$		& 1.98/2.45/2.42          \\
  $\rho^{p}_{0,1}/ \rho^{p}_{0,2}$	                  				&    9.1/330 ~GV      \\
   $\chi^{2}/(d.o.f = 22)$ (for B/C)                           &    0.96  \\
  $\chi^{2}/(d.o.f = 26)$ (for proton)                           &    0.73 \\  
\hline  
\end{tabular}

\end{table}

\begin{figure*}[ht]
\centering
\mbox{

\includegraphics[width=0.40\textwidth,clip,angle=0]{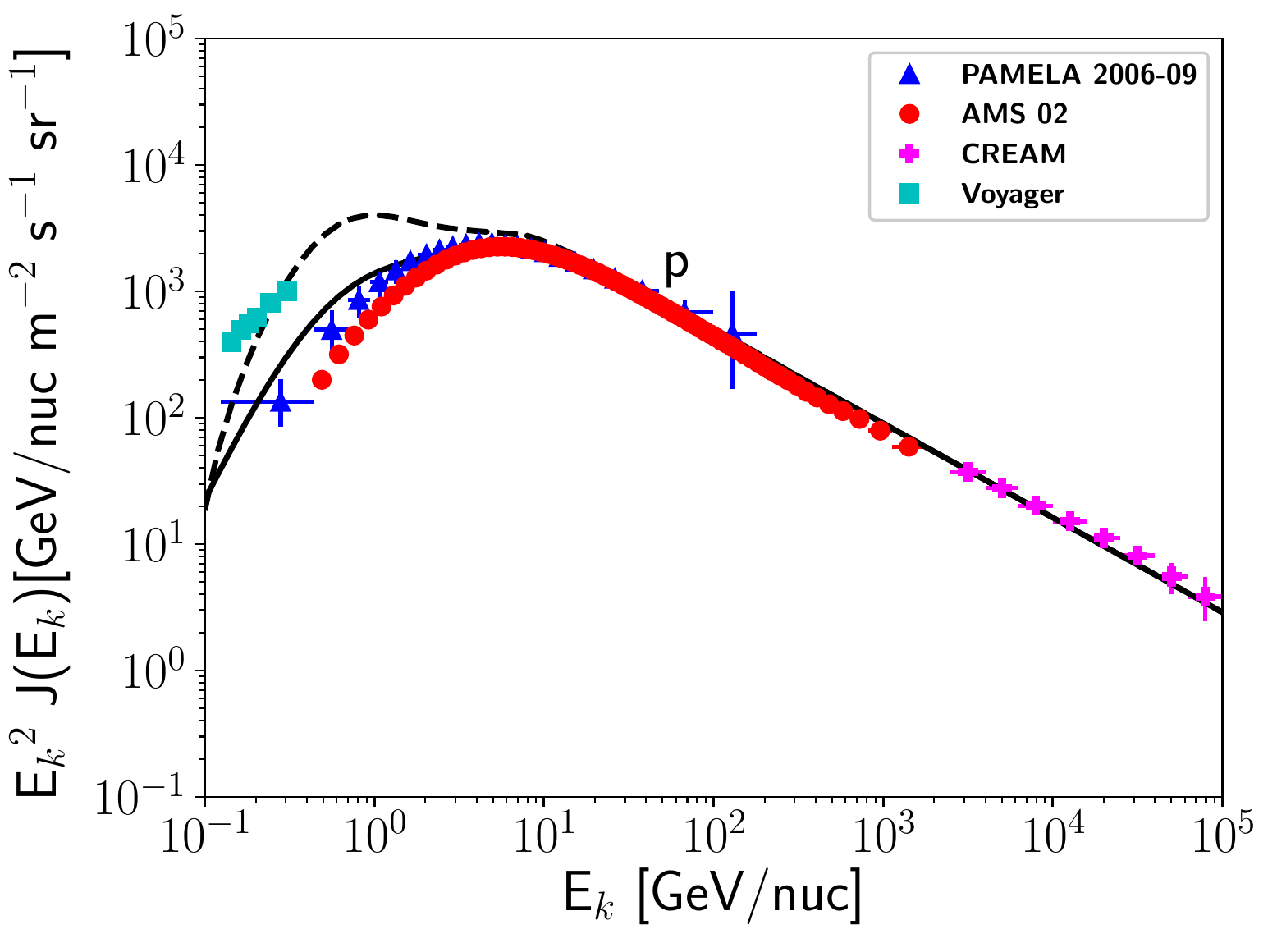}

\includegraphics[width=0.40\textwidth,clip,angle=0]{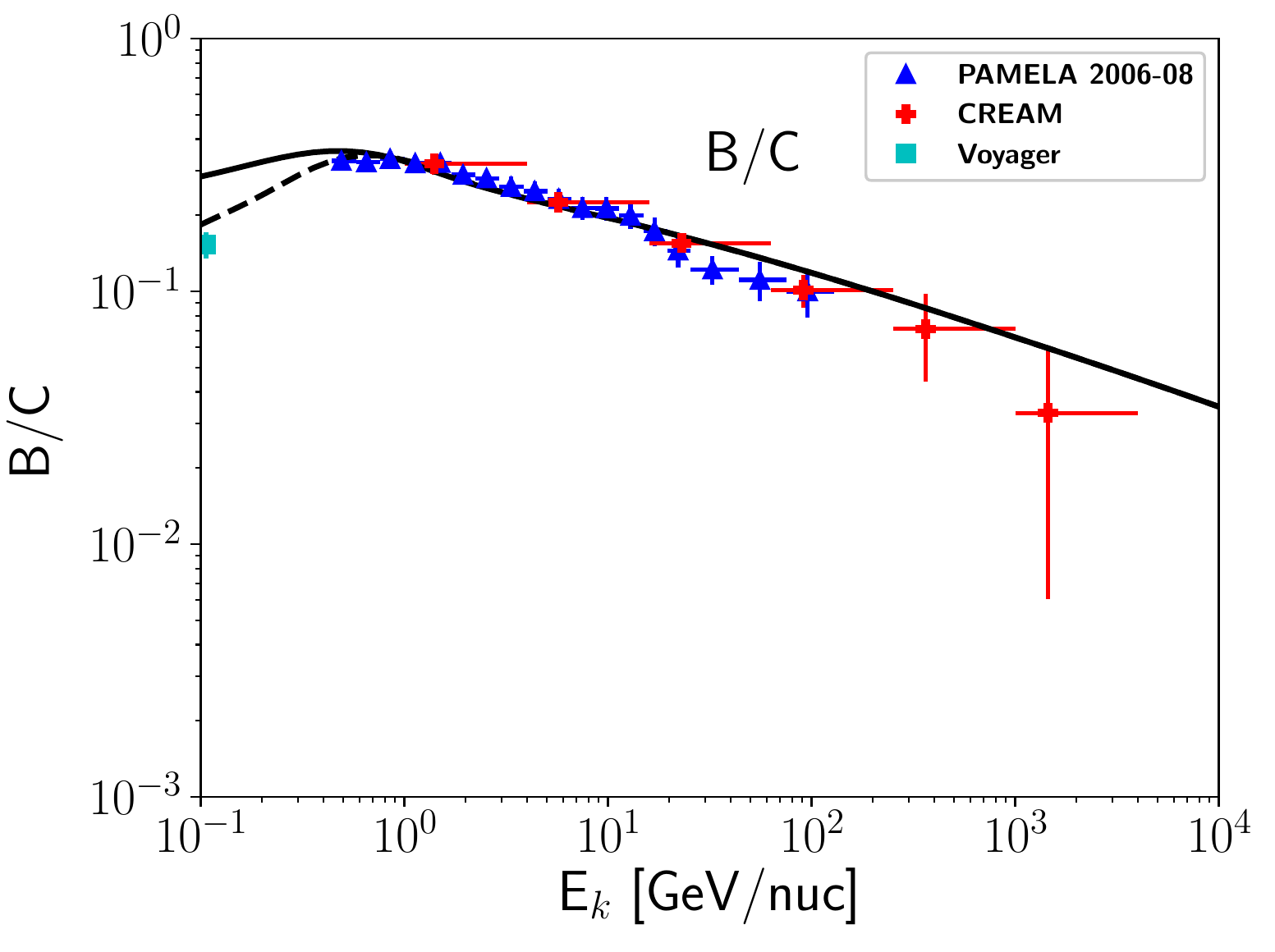}
}

\mbox{
\includegraphics[width=0.40\textwidth,clip,angle=0]{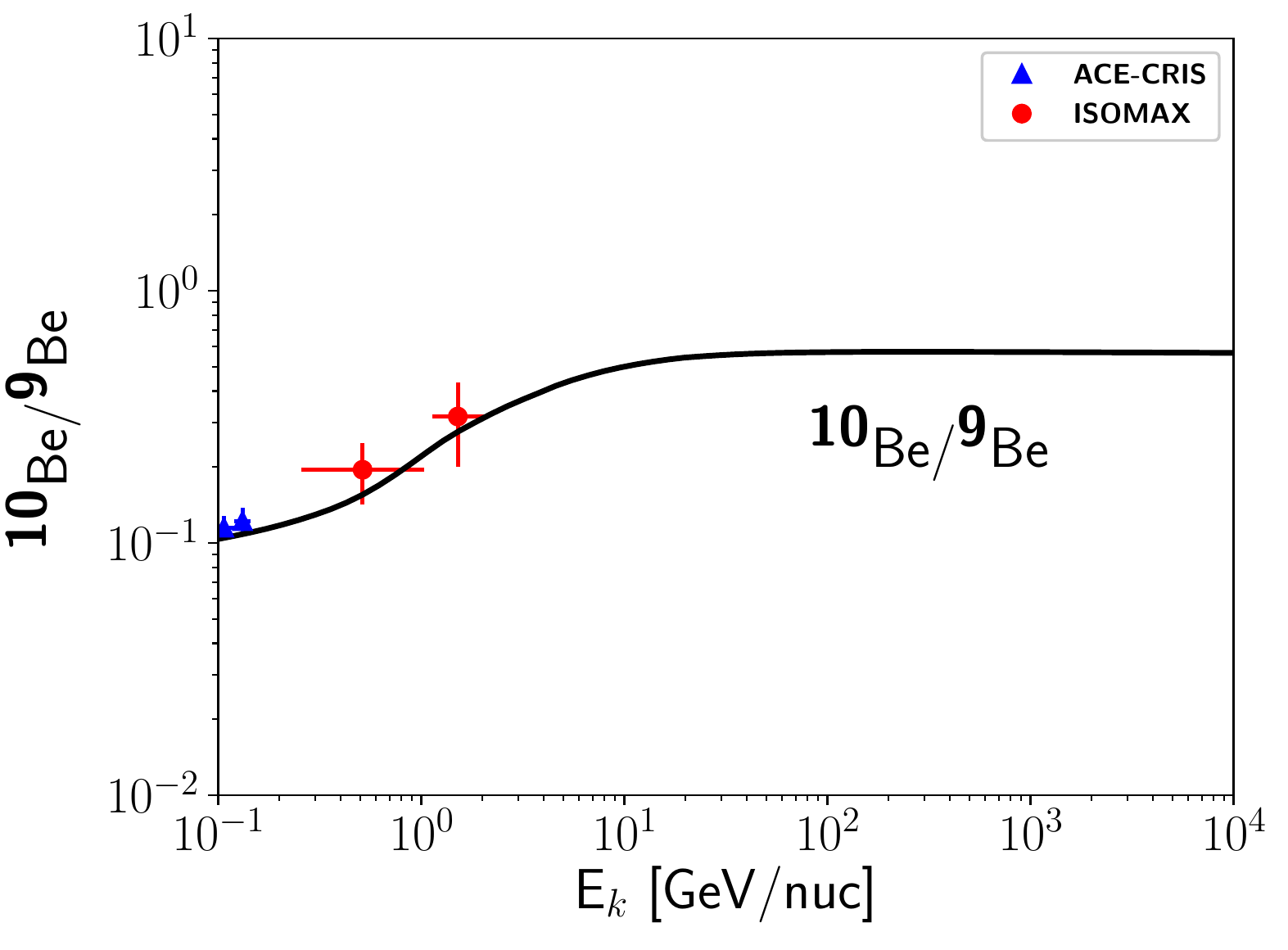}

\includegraphics[width=0.45\textwidth,clip,angle=0]{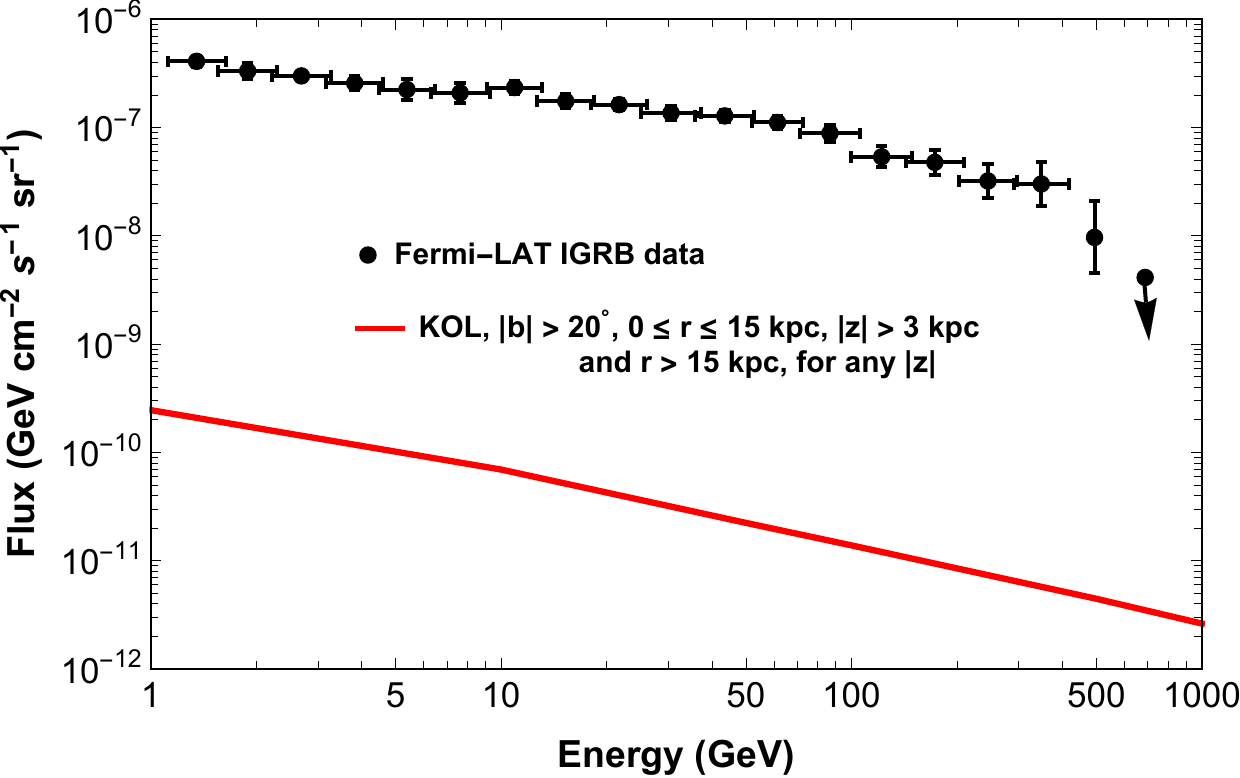}
}
\caption{\label{fig:kolgam}Energy dependence of primary CR flux, secondary to primary ratios, obtained from \texttt{DRAGON} code using KOL model, are plotted with the locally measured CR fluxes. Proton flux (upper left panel) is plotted with Voyager \citep{stone2013, cum2016}, PAMELA \citep{adriani2013proton}, AMS 02 \citep{aguilar2015} and CREAM \citep{yoon2011} data. B/C (upper right panel) flux ratio is plotted with PAMELA \citep{pambc}, Voyager \citep{cum2016} and CREAM \citep{creambc} data.  In case of proton and B/C, the dashed and solid lines represent spectra without and with the solar modulation ($\phi = 0.35$~GV) respectively. $^{10}$Be/$^{9}$Be  (bottom left panel) flux ratio is  plotted with ACE-CRIS \citep{ace2001} and ISOMAX \citep{isomax2004} data. Diffuse gamma-ray flux (bottom right panel) obtained from the KOL model is compared with the IGRB data measured by \textit{Fermi}-LAT \citep{ackermann2015}. The downward arrow at highest energy bin (580-820) GeV represents the upper limit of flux. }
\end{figure*} 

\begin{figure*}[ht]
\centering
\mbox{
\includegraphics[width=0.38\textwidth,clip,angle=0]{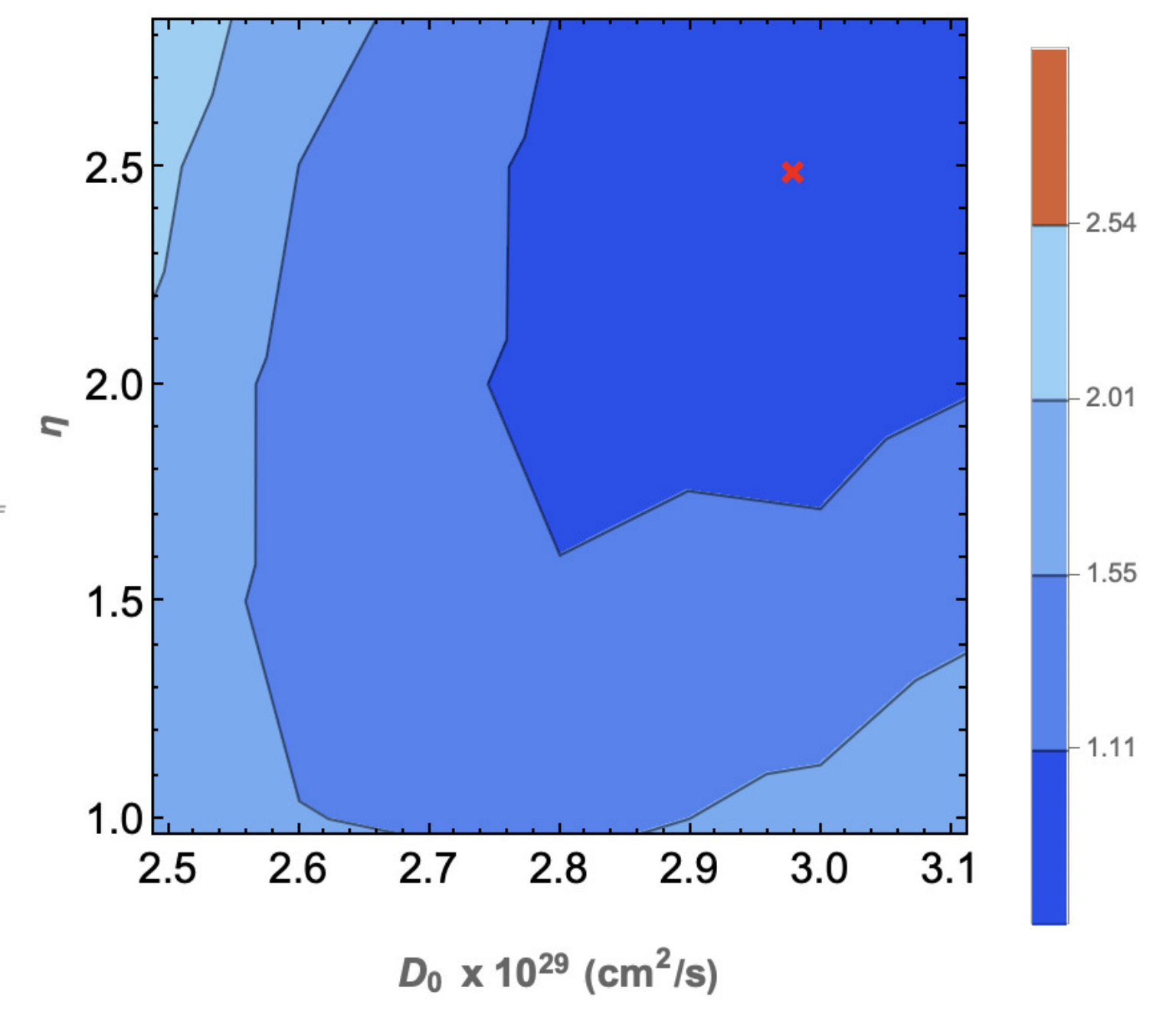}

\includegraphics[width=0.38\textwidth,clip,angle=0]{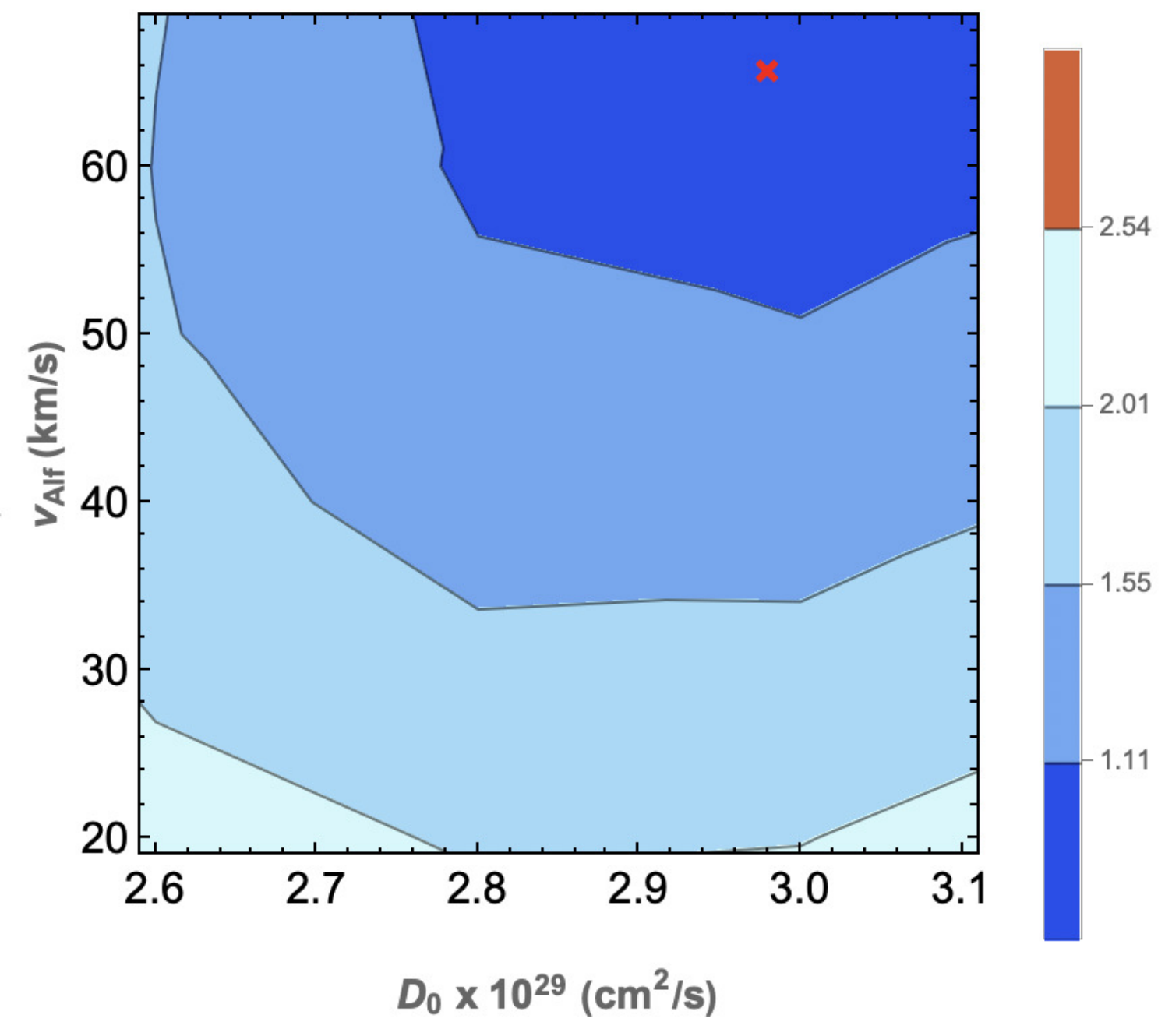}
}

\caption{\label{fig:kolcontour}  In case of KOL model, the contour plots for  $\eta$ against $D_{0}$ (left) and  $v_{Alf}$ against $D_{0}$ (right) are shown. The $1\sigma$, $2\sigma$, $3\sigma$, and $4\sigma$ of $\chi^{2}/d.o.f$ are represented by 1.11, 1.55, 2.01 and 2.54 respectively. The red cross marks in the plots correspond to the best fit values of the parameters shown in those plots.}
\end{figure*}

Similar to KRA model, KOL model is characterized by its fixed $\delta$ value; $\delta = 0.33$.  The best fit parameter values and $\chi^{2}/d.o.f$ of KOL model which are needed to fit the locally measured CR spectra are listed in table \ref{tab:kol}. The fitted CR spectra and the diffuse gamma-ray flux obtained from KOL model are shown in the figure \ref{fig:kolgam}.

We show the contour plots of $\eta$ against $D_{0}$ (left) and $v_{Alf}$ against $D_{0}$ (right) in  figure \ref{fig:kolcontour}.  In case of KOL model, the $1\sigma$ contour for $D_{0}$ shifts towards the higher range of values compared to PD and KRA models. Similarly, lower limit of $1 \sigma$ contour for $v_{Alf}$ also shifts towards very higher values in comparison of KRA and CON models. In comparison with other three models, KOL model has significantly different range of  $\eta$ values within $1\sigma$ contour. Here, the $1 \sigma$ contour excludes the negative values of $\eta$.

\subsection{Comparison of diffuse gamma-ray fluxes and luminosity calculation}\label{sec:compdfluxlum}

In this section, we compare the diffuse gamma-ray fluxes obtained in two different cases using KRA model. In the first case, we calculate the diffuse gamma-ray flux excluding the diffuse emission from the inner cylindrical region of the Galaxy with radius 15 kpc and half-height 3 kpc (see the equation \ref{eq:gamflux} and discussion in section \ref{sec:gamflux}). In another case, we also calculate same diffuse gamma-ray flux without putting any constraint on $r$ and $z$. In both the cases, we consider $|b|> 20^{\degree}$ and use the parameter values obtained in KRA model (see the section \ref{sec:kra} and table \ref{tab:kra}). Comparison of the diffuse fluxes in two cases reveals that if we do not put any constraint on $r$ and $z$ then the diffuse gamma-ray flux increases by $\sim$~2 orders of magnitude than the other case where we exclude the diffuse emission from the inner Galactic region.  We, finally, compare our results with the IGRB data presented by \textit{Fermi} collaboration \citep{ackermann2015}. We find our results are well below the IGRB data.

\begin{figure}[h!]
\includegraphics[width=0.5\textwidth,clip,angle=0]{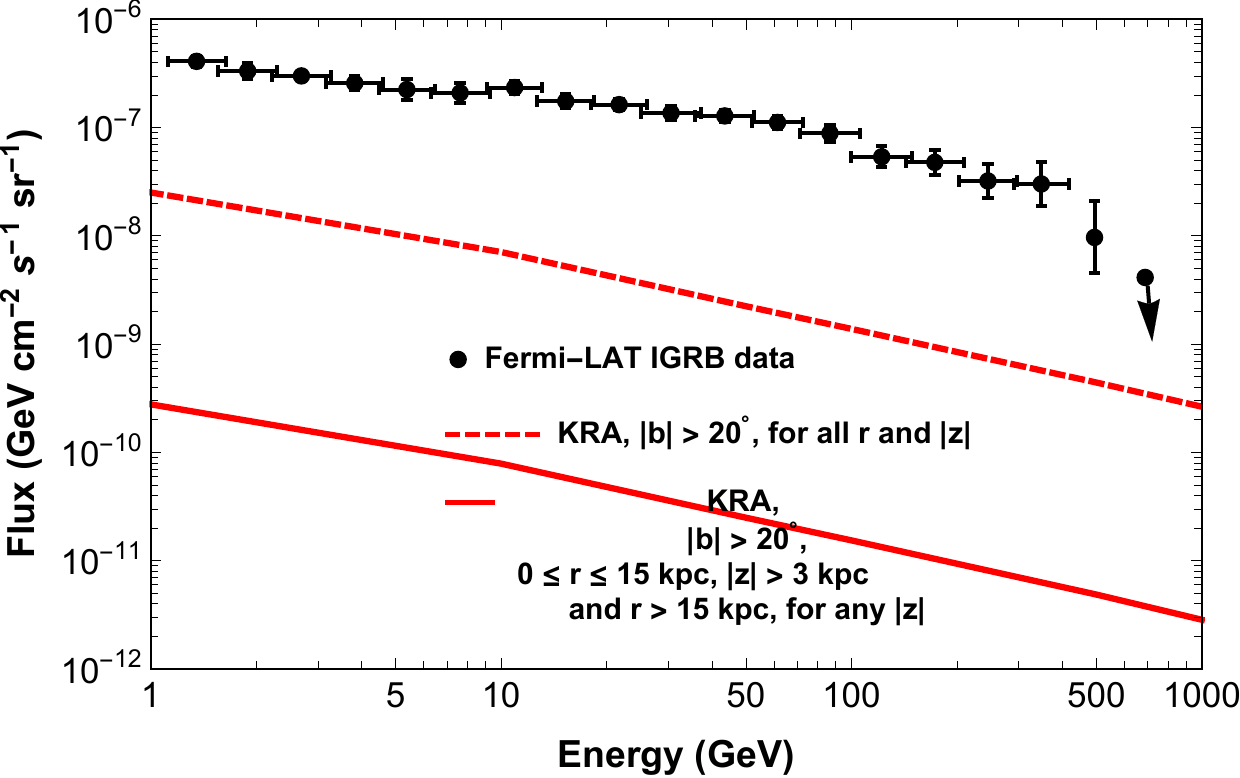}
  \caption{\label{fig:partfull} Diffuse gamma-ray fluxes obtained from KRA model is compared with the IGRB data measured by \textit{Fermi}-LAT \citep{ackermann2015}. The downward arrow at highest energy bin (580-820) GeV represents the upper limit of flux. The dashed line represents the diffuse gamma-ray flux for no restriction on $r$ and $z$, and $|b|> 20^{\degree}$. The solid line represents the diffuse gamma-ray flux with constraints on $r$ and $z$, and $|b|> 20^{\degree}$. }  
\end{figure}

Figure \ref{fig:partfull} shows the plots of diffuse gamma-ray fluxes obtained in two cases, as discussed here, with KRA model. IGRB data is also plotted in the same plot for comparison.

In the following, we have calculated the CR luminosity ($L_{\rm{CR}}$) for the KRA model. The total CR luminosity  is defined as, 

\begin{widetext}
\begin{eqnarray}
L_{{\rm{CR}}}=&& 2 \times \frac{4 \pi}{c~t_{g}} \times \int_{1~{\rm{GeV}}}^{10^{5}~\rm{GeV}}\int_{0}^{2\pi}\int_{z_{\rm{min}}=~0}^{z_{\rm{max}}=~3z_{t}~\rm{kpc}}\int_{r_{\rm{min}}=~0}^{r_{\rm{max}}=~40~\rm{kpc}} 
 \frac{r~E_{p}~ J^{Gal}_{p}(E_{p}, r, z)} {\beta(E_{p})}    drdz d\theta dE_{p} \nonumber \\
=&&\frac{2\times 2\pi\times 4\pi\times (3.08\times10^{19})^{3}\times (1.6\times 10^{-3})}{ t_{g} \times 3.0\times 10^{8}} \nonumber \\
&& \times \int_{1~{\rm{GeV}}}^{10^{5}~\rm{GeV}} \int_{z_{\rm{min}}=~0}^{z_{\rm{max}}=~3z_{t}~\rm{kpc}}\int_{r_{\rm{min}}=~0}^{r_{\rm{max}}=~40~\rm{kpc}} 
 \frac{r~E_{p}~ J^{Gal}_{p}(E_{p}, r, z)} {\beta(E_{p})}    drdzdE_{p} ~\rm{erg~s^{-1}},  \label{eq:lum}
\end{eqnarray}
\end{widetext}
where, $c$ is the speed of light and $\beta(E_{p}) = \frac{\sqrt{E_{p}^{2} + 2 E_{p}m_{p}}}{E_{p}+m_{p}}$. Here, $ t_{g}$ is the time at which the CR injection have started.  The prefactors $3.08\times10^{19}$ and $1.6\times 10^{-3}$ represent the conversion factor from kpc to meter and GeV to erg respectively. If we put $t_{g} = 12~{\rm{Gyr}} = 12 \times 3.15\times 10^{16}$~s \citep{liu2019}, then $L_{{\rm{CR}}} = 4.75\times10^{38}~\rm{erg~s^{-1}}$.  

\section{Summary and discussion}\label{sec:sandd}

In this work, we combine both the local CR measurements and diffuse gamma-ray fluxes. We have used few benchmark models with diverse parameter settings for modeling of CR propagation in the ISM. For each model, we choose parameter values by checking the goodness of the fit to the CR data and study the correlation between different parameters to understand the possible ranges of uncertainties in those parameters.  Moreover, we find that diffuse gamma-ray data do not strongly constrain the $\delta$ parameter and therefore diffuse gamma-ray data can not discriminate among the CR propagation models considered in this work. We discuss our findings below.

\subsection{Outcomes of the correlation studies between different parameters used for CR propagation}
\begin{itemize}
\item  Within $1\sigma$ contour,  the $D_{0}$ has almost a similar range of variation  in PD ($D_0 \sim~1.7 \times 10^{29} - 2.4\times 10^{29}~\rm{cm^{2}/s}$) and KRA ($D_0 \sim~1.8 \times 10^{29} - 2.35\times 10^{29}~\rm{cm^{2}/s}$) models, but in  CON ($D_0 \sim~1.0 \times 10^{29} - 1.4\times 10^{29}~\rm{cm^{2}/s}$) and KOL ($D_0 \sim~2.8 \times 10^{29} - 3.1\times 10^{29}~\rm{cm^{2}/s}$)  models $D_{0}$ varies in lower and upper ranges compared to PD and KRA models respectively.

\item  KRA and KOL models have fixed $\delta$ values 0.50 and 0.33 respectively.  Within $1 \sigma$ region, PD and CON models have $\delta \gtrsim 0.45$ and $\delta \gtrsim 0.55$ respectively. 

\item  Within $1 \sigma$ region, $\eta$ can only have  positive values  ($\eta > 1.6$) in KOL model. In rest of the models, $\eta$ can have both negative and positive values. 

\item  In PD model, $v_{Alf} = 0$. Almost similar range of preferred non zero values of $v_{Alf}$ are found in the $1\sigma$ region in  KRA and CON models. In case of KOL model, $1\sigma$ contour shifts towards more higher values ($v_{Alf} \gtrsim 52~\rm{km/s}$) of $v_{Alf}$.

\item  From the B/C data fitting, $\chi^{2}/d.o.f$ values for PD, KRA, CON and KOL models  are 0.51, 0.51, 0.86 and 0.96 respectively.  The $\chi^{2}/d.o.f$ are obtained as  0.31, 0.45, 0.43 and 0.73 during the proton data fitting with PD, KRA, CON and KOL models respectively.

\end{itemize}

\subsection{Comparison with previous works}

\begin{figure*}[ht]

\centering
\mbox{

\includegraphics[width=0.40\textwidth,clip,angle=0]{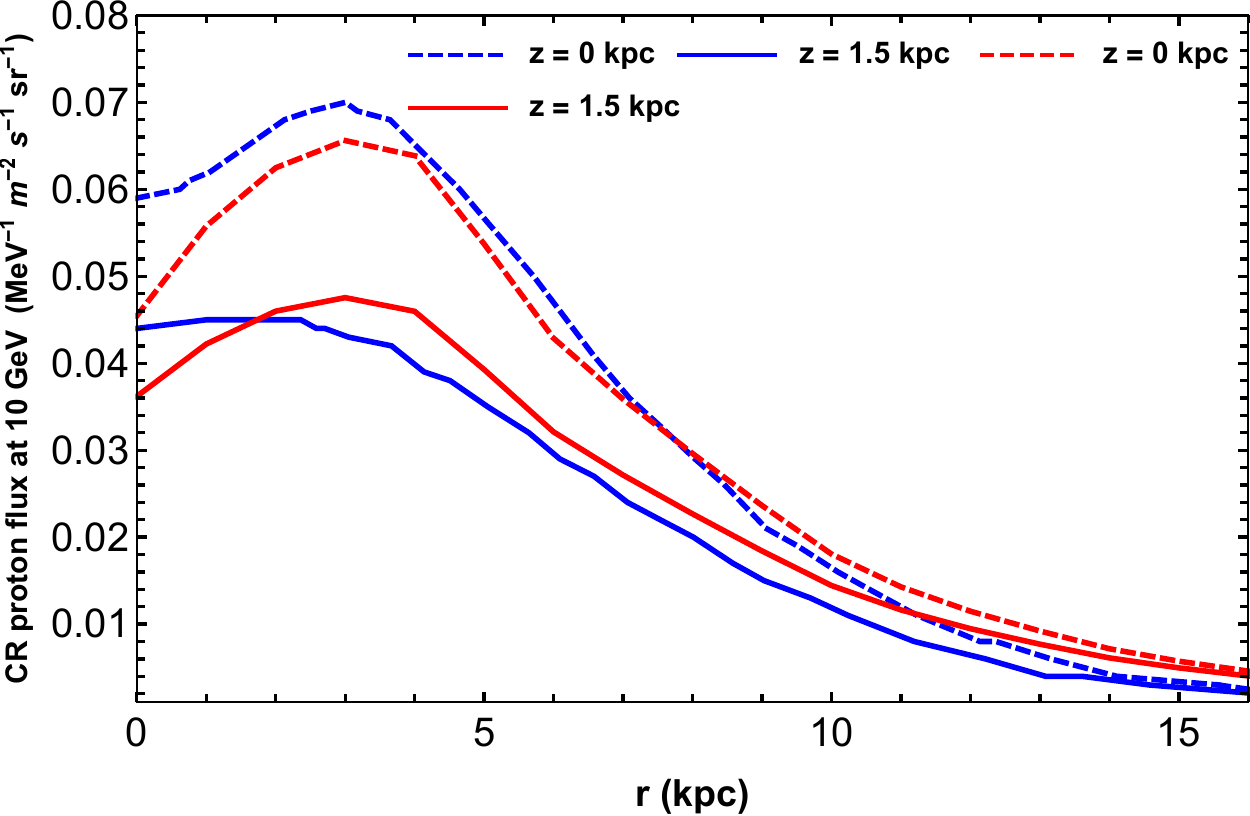}

\includegraphics[width=0.40\textwidth,clip,angle=0]{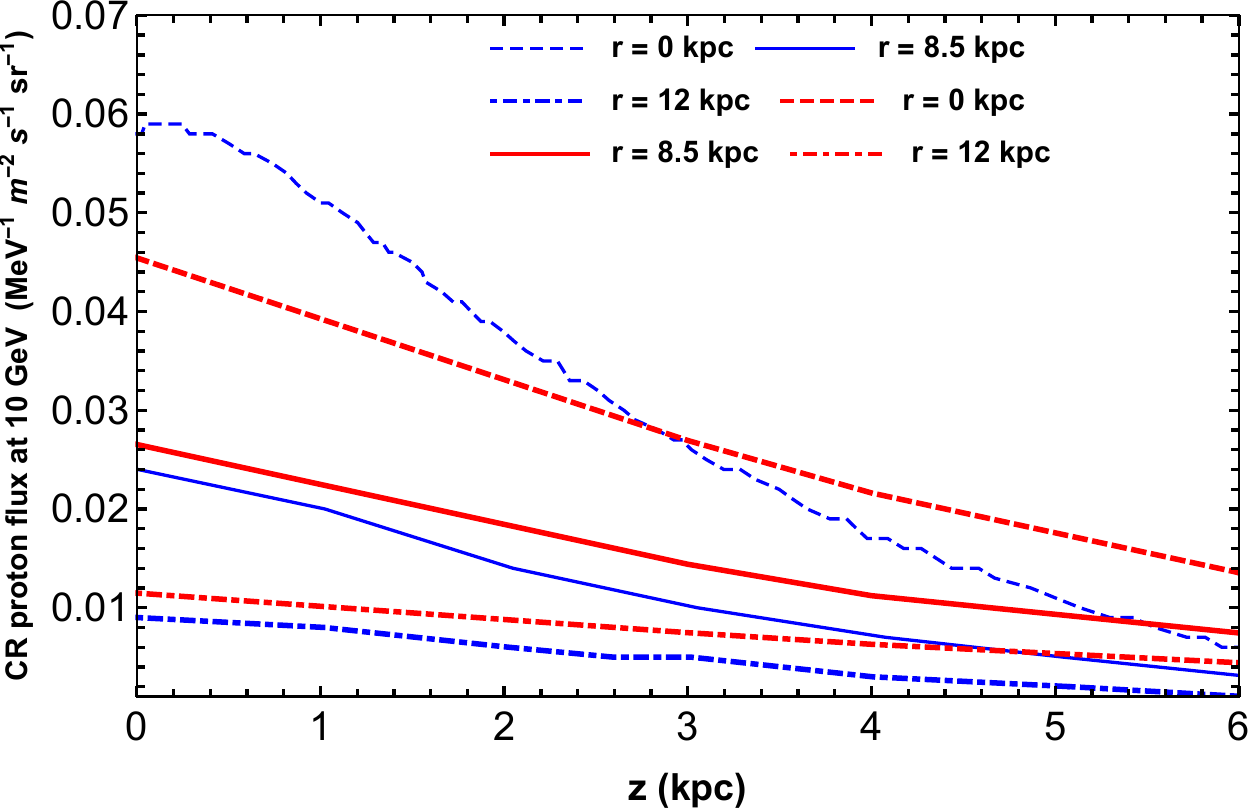}
}

\mbox{
\includegraphics[width=0.40\textwidth,clip,angle=0]{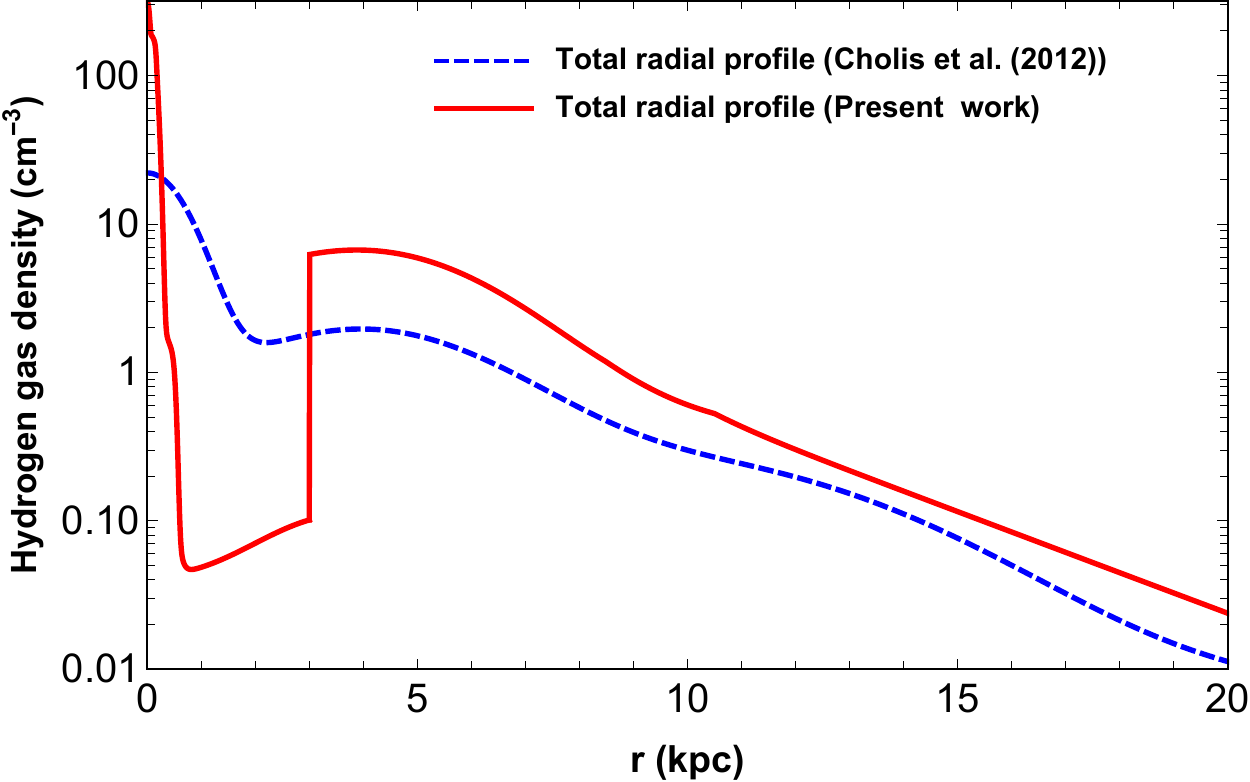}

\includegraphics[width=0.45\textwidth,clip,angle=0]{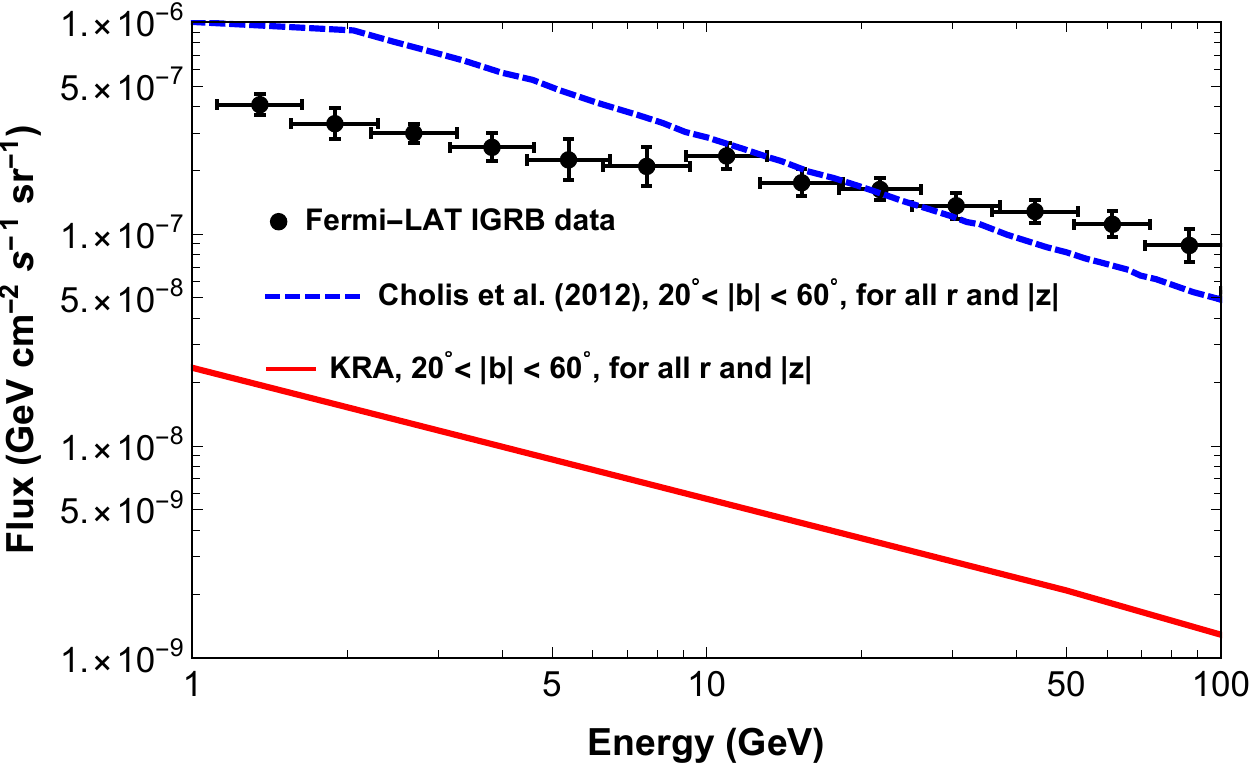}
}
\caption{\label{fig:cholvsme} The radial variations (upper left panel) of CR proton flux at different r and the vertcal varitions (upper right panel) of CR proton flux at different z are shown here. The total gas densities (bottom left) of both the works are also plotted. Comparison of diffuse gamma-ray fluxes (bottom right panel) obtained from KRA model and \cite{cholis2012} are compared with the IGRB data measured by \textit{Fermi}-LAT \citep{ackermann2015}. In all the figures, blue lines indicate the results of \cite{cholis2012} and the red lines indicate our results.  }
\end{figure*}

Previously, a similar kind of analysis has been carried out by \citet{cholis2012}. In that work, diffusion coefficient contains an extra radially dependent term which is absent in our work. The gas density profiles used in their work are significantly different from those used by us. Our gas density profiles are more updated and based on recent observations and hydrodynamical simulations. Most of the CR data used by us are more updated than the previous data sets used in their work. The \texttt{DRAGON} code has been used in both of the works to obtain CR proton flux in the Galaxy. The best fit values of the model parameters in both the cases have been obtained by minimizing $\chi^{2}/d.o.f$ of the  observed B/C and proton data. In the work of \citet{cholis2012}, the parametrization in p-p interaction is based on \citet{kam2006}, whereas we use \citet{kelner2006} for the parametrization.  The IGRB data \citep{abdo2010} used in the previous work is upto 100 GeV, but in our work we have extended our gamma-ray flux calculation upto 1 TeV and we have used the recent IGRB data \citep{ackermann2015} with an extension upto $~ 800$~GeV. In figure \ref{fig:cholvsme}, we compare few of our results with the results obtained by \citet{cholis2012}.  We find that the radial and vertical distributions of our CR proton flux at 10 GeV (see upper two panels of figure \ref{fig:cholvsme}) do not have significant numerical deviation from the results of \citet{cholis2012}. The gaseous components such as ${\rm{H_{2}}}$ \citep{nak2006}, HI \citep{nak2003} and HII \citep{cord1991} are used in \citet{cholis2012}. We have used  ${\rm{H_{2}}}$ \citep{ferriere1998, ferriere2007, feld2013, biswas2018}, HI \citep{ferriere1998, ferriere2007, kalberla2008, feld2013, biswas2018}  and HII \citep{ferriere1998, ferriere2007, mill2015, feld2013,biswas2018} gas profiles in our work. The radial profiles of the total hydrogen gas density in the two cases show significant discrepancy between 1-3 kpc (see bottom left panel of figure \ref{fig:cholvsme}) where our estimated values are $\sim$1-2 orders of magnitude lower than the values obtained by \citet{cholis2012}. Such discrepancy may trigger the deviation in the diffuse gamma-ray fluxes (see bottom right panel of figure \ref{fig:cholvsme}) obtained in these two works. Our calculated diffuse gamma-ray flux values are $\sim$ 1-2 orders of magnitude less than the values obtained by \citet{cholis2012} over the energy range of 1-100 GeV. Another important finding is that if we compare the gamma-ray fluxes of \citet{cholis2012}  with the new IGRB data \citep{ackermann2015}, the fluxes below 10 GeV are higher than the IGRB data and rest of them are comparable with IGRB data.  Such result excludes the region below 10 GeV and indicates that IGRB data mostly comprises of the diffuse gamma-rays produced due to p-p interactions in our Galaxy. The major difference between \citet{cholis2012} and our present work is that our results with most updated gas density profile show that diffuse gamma-ray fluxes, originated through p-p interactions and subsequent decay of neutral pions, are quite well below the IGRB data \citep{ackermann2015} over the entire energy range of 1-1000 GeV. 

\begin{figure}[!h]

\includegraphics[width=0.45\textwidth,clip,angle=0]{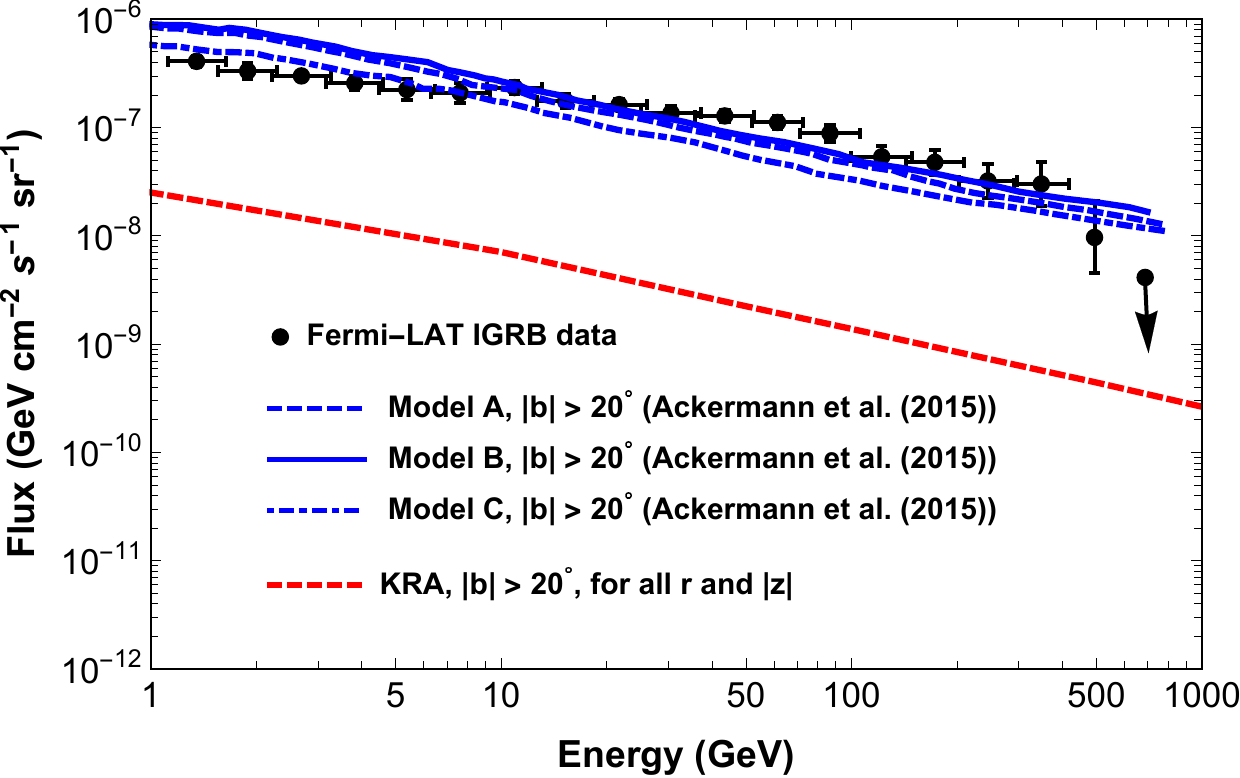}

\caption{\label{fig:ackvsme}  Comparison of diffuse gamma-ray fluxes obtained from KRA model and \cite{ackermann2015} (for different foreground models) with $|b|> 20^{\degree}$ are shown here. The IGRB data measured by \textit{Fermi}-LAT \citep{ackermann2015} are also plotted. In the figure, blue lines indicate the diffuse gamma-ray fluxes  of \cite{ackermann2015} and the red dashed line indicates our results.}
\end{figure}

In figure \ref{fig:ackvsme}, we have shown the diffuse gamma-ray fluxes of \citet{ackermann2015} for different Galactic foreground models namely Model A, Model B and Model C with $|b|> 20^{\degree}$. These models are defined in terms of sources of CR nuclei and electrons, and the nature of diffusion coefficient and reacceleration strength through Galaxy \citep{ackermann2015}. Different components of hydrogen gas such as  ${\rm{H_{2}}}$ \citep{dame2001,ackermann2012}, HI \citep{kalberla2005, ackermann2012} and HII \citep{gaen2008, ackermann2012} are used in those calculations. For each of the Galactic foreground models, the proton distribution in the Galaxy was obtained by using the \texttt{GALPROP} code \citep{strong2000,vlad2011} and by fitting the observed proton, B/C and $^{10}{\rm{Be}}/^{9}{\rm{Be }}$ data. The diffuse gamma-ray emission has been obtained by using the parametrization of \citet{kam2006} and the procedure followed in \citet{ackermann2012}. In \texttt{GALPROP} code, the diffusion coefficient has a power law dependence on rigidity with power law index $\delta = 0.33$. For our calculations, we use \texttt{DRAGON} code in which the diffusion coefficient includes the power law dependence on rigidity and exponential variation of z. In the figure \ref{fig:ackvsme}, we have shown  our result for KRA ($\delta = 0.5$)  model with $|b|> 20^{\degree}$. The fluxes are nearly same even for KOL ($\delta = 0.33$) model in our case. The diffuse gamma-ray fluxes of \citet{ackermann2015} are comparable with the IGRB data. Our  obtained diffuse gamma-ray fluxes, however, are lower by less than 2 orders of magnitude than the IGRB data over the energy range considered here.

\begin{figure*}[ht]
\centering
\mbox{
\includegraphics[width=0.45\textwidth,clip,angle=0]{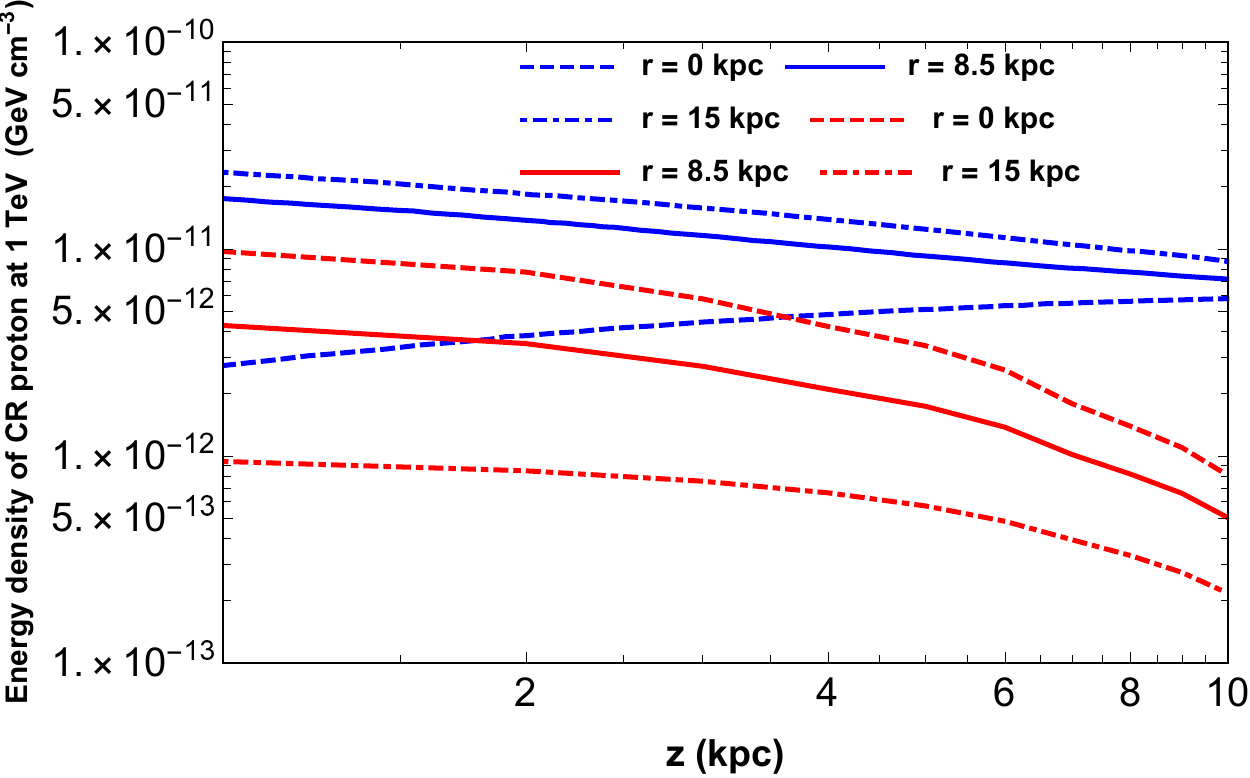}

\includegraphics[width=0.45\textwidth,clip,angle=0]{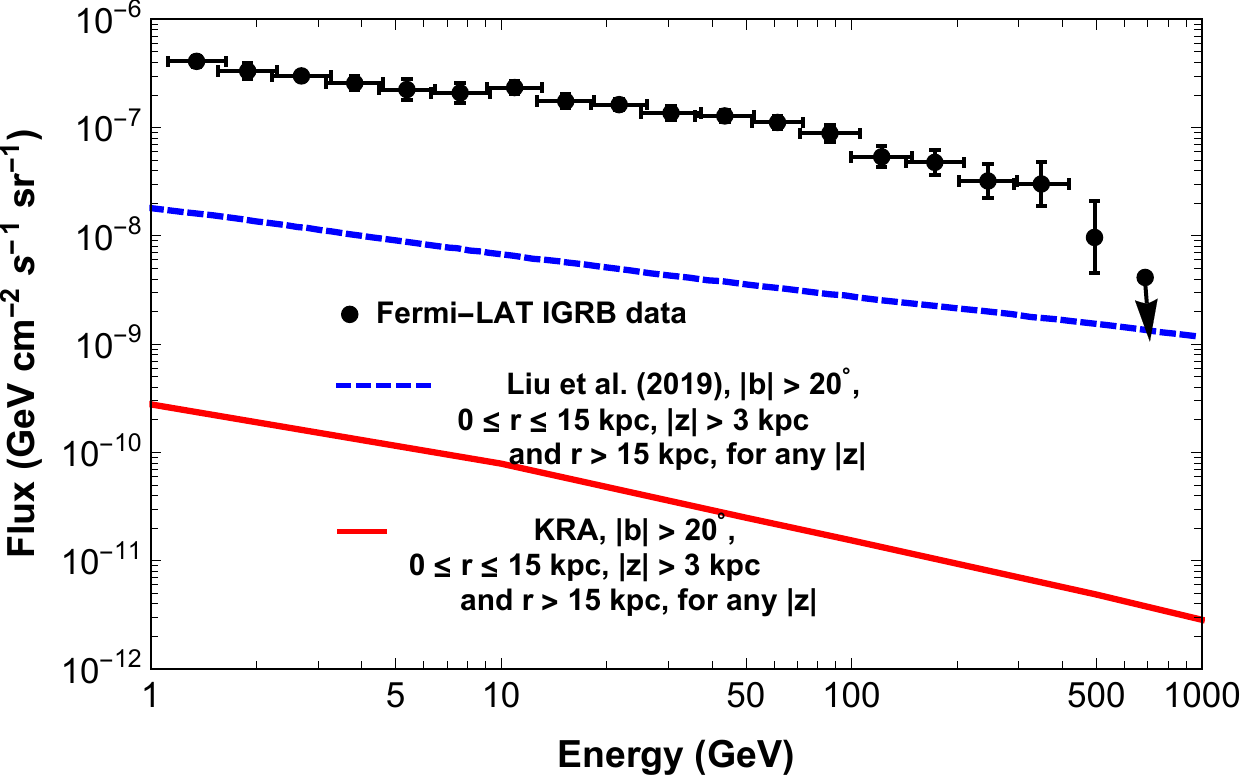}
}
\caption{\label{fig:liucompare}  In the figures, the vertical variations (left) of CR densities at different radial distance are shown and  diffuse gamma-ray fluxes (right) obtained from KRA model and \cite{liu2019} (with wind velocity $\sim 300~\rm{km/s}$) are compared with the IGRB data measured by \textit{Fermi}-LAT \citep{ackermann2015}. In both the figures, blue lines indicate the results of \cite{liu2019} and the red lines indicate our results.}
\end{figure*} 

Recently, similar problem has been addressed by \citet{liu2019}. The radial profile of equation \ref{eq:ionup} was used as the total gas density profile and was marked as ``Model A" in their work. The total CR luminosity in their work is taken as $10^{41}~\rm{erg~s^{-1}}$ which is three orders of magnitude higher than our calculated luminosity for KRA model (see section \ref{sec:compdfluxlum}). In \citet{liu2019}, the analytical solution for CR proton density at a point was obtained by following the solving procedure of a CR transport equation as provided in \citet{berez2006}. In \citet{liu2019}, the diffusion coeffiecient is energy dependent and $\delta = 0.33$. In our case, we use the \texttt{DRAGON} code to solve the CR transport equation to obtain proton flux at any point in the Galaxy. In addition to that, the proton distribution in the Galaxy is obtained by us by fitting the locally observed CR spectra, which is a major difference from the procedure followed in \cite{liu2019}. In the previous work, the diffuse gamma-ray flux was calculated by following the same method  (see section \ref{sec:gamflux}) used by us in the present work. The diffuse fluxes of Model A without Galactic wind overshoot the upper limit of IGRB flux measured by \textit{Fermi}-LAT \citep{ackermann2015}, whereas the diffuse gamma-ray fluxes predicted by Model A with Galactic wind (velocity $\sim 300~\rm{km/s}$) are below the measured IGRB fluxes. In our case, the diffuse gamma-ray fluxes obtained from all the models are well below the measured fluxes of IGRB. Our obtained diffuse-gamma  ray fluxes are lower by $\sim$2-3 orders of magnitude (depending on the energy) than the results obtained by \citet{liu2019} for Model A with wind (see the figure  \ref{fig:liucompare}). The left plot of the figure \ref{fig:liucompare}  shows the vertical distributions of CR density at 1 TeV for different radial distances obtained by us and \citet{liu2019}. For $3~\rm{kpc}<z \lesssim 10~\rm{kpc}$ with $r = 8.5$~kpc and $r = 15$~kpc , our derived CR densities at 1 TeV  are lower by $\sim$ 4-40 factor than the results of \citet{liu2019} at different z. For $3~\rm{kpc} < z \lesssim10~\rm{kpc}$ with $r = 0$, the two results differ by few factor ($<10$). In our calculations, we take maximum half height upto 18 kpc and the CR densities fall rapidly with increase in vertical distance. On the other hand, the maximum vertical distance is taken as 100 kpc in  \citet{liu2019} with  CR densities at 1 TeV  $\sim 10^{-12}~\rm{GeV~cm^{-3}}$ at higher z values upto 100 kpc. The difference in the diffuse gamma-ray fluxes in both the cases may be due to the cumulative effect of the  variation of CR densities at different z and r.

\subsection{Probing Galactic Molecular Clouds  with CR protons}
\begin{figure}[!h]

\includegraphics[width=0.45\textwidth,clip,angle=0]{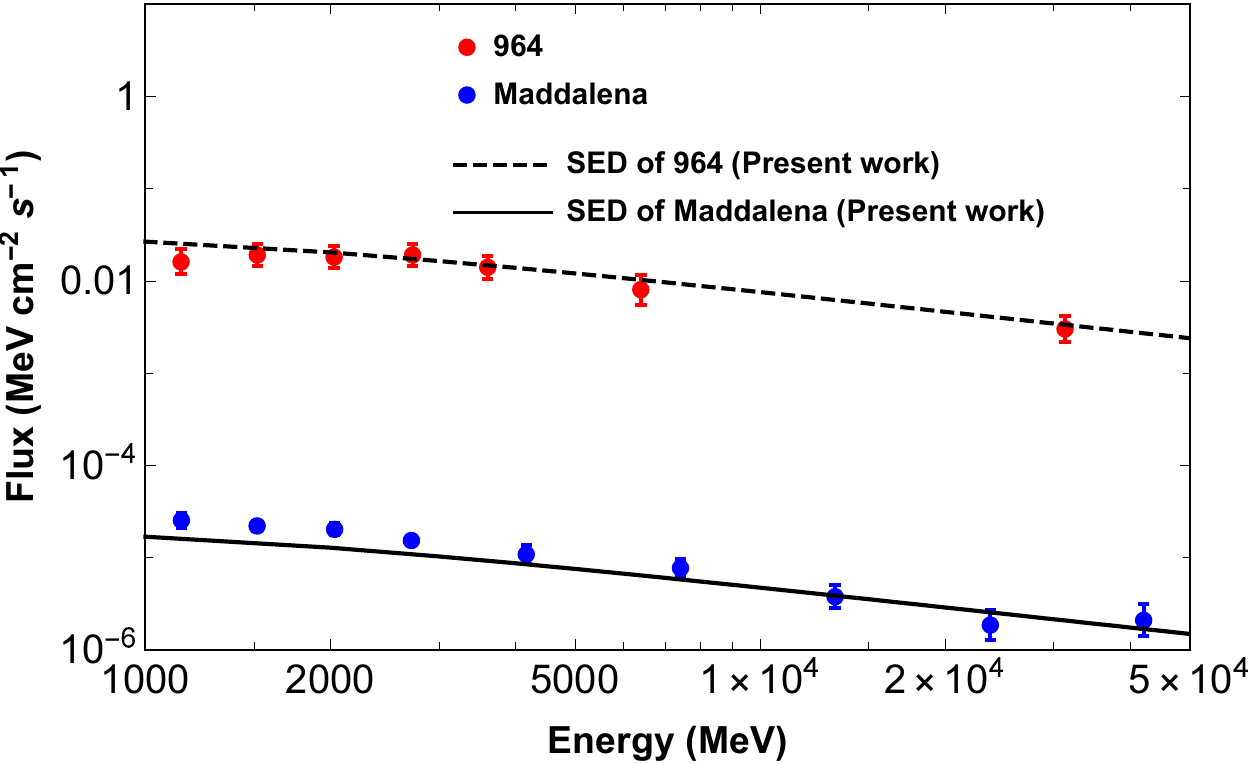}

\caption{\label{fig:gmc}  The SEDs, obtained from \textit{Fermi}-LAT analysis, of the GMCs named 964 and Maddalena are shown here. We also applied our CR proton distribution, obtained in  KRA model, to calculate the gamma-ray flux produced in the  interactions of CR protons with those GMCs. A factor of $10^{3}$ is multiplied with the fluxes (both observed and calculated) of GMC 964 to separate the SED of GMC 964 from the SED of GMC Maddalena. The SEDs of 964 (dashed) and  Maddalena (solid line), obtained from our calculations, are shown in this figure.}
\end{figure}

The CR protons can penetrate Galactic Molecular Clouds (GMCs) and produce gamma-rays through the p-p interactions. Thus gamma-rays from individual GMCs can provide a straightforward and localized information of CR protons. We already have  calculated the energy and spatial distribution of CR protons in our Galaxy for different propagation models. We, here, use the Galactic CR proton fluxes obtained in the KRA model and apply them to calculate the gamma-ray fluxes from GMCs at different energies. Finally, we compare our  results with the spectral energy distributions (SEDs)  of those selected GMCs obtained from \textit{Fermi}-LAT analysis. The gamma-ray flux (in units of $\rm{GeV}^{-1}\rm{cm}^{-2}\rm{s}^{-1}$) originated in p-p interactions from a single GMC  is denoted as \citep{aharon2018} 

\begin{equation}
\phi_{{\rm{\gamma,GMC}}}(E_{\gamma}) = \zeta_{{\rm{ISM}}}\frac{(M_{{\rm{GMC}}}/m_{p})}{n_{{\rm{GMC}}}~d^{2}} ~J_{\gamma}(E_{\gamma}, r, z).
 \label{eq:gmcflux}
\end{equation}   
In equation \ref{eq:gmcflux}, $J_{\gamma}(E_{\gamma}, r, z)$ takes the expressions denoted in equations \ref{eq:hgam} and  \ref{eq:lgam} depending on the energies and the total gas densities ($n_{g}(r, z)$) in those equations are replaced by  gas density of GMC, i.e. $n_{{\rm{GMC}}} =1~\rm{cm}^{-3}$. We set $\zeta_{{\rm{ISM}}} \approx 1.8$ which denotes the composition of ISM and CRs \citep{kaf2014}. Here, $M_{{\rm{GMC}}} $, $m_{p}$, $r$ and $d$ denote the mass of the GMC, proton mass, galactocentric distance and distance of the GMC from the Earth respectively. For the present work, we have chosen two GMCs named as 964 and Maddalena with masses $(3 \pm 2)\times 10^{5}~M_{\odot}$ ($M_{\odot}$ is the solar mass) and $5.29 \times 10^{5}~M_{\odot}$ respectively \citep{aharon2018}. The 964 and Maddalena  have $r = 6.4$~kpc and  $r = 10.1$~kpc respectively, and $d =  1.9 \pm 0.6$~kpc and  $d = 2.1 \pm 0.1$~kpc respectively \citep{aharon2018}. The Galactic longitudes ($l$) and latitudes ($b$) of  964 and Maddalena are $l=345.57^{\degree}$, $b=0.79^{\degree}$, and $l=216.5^{\degree}$, $b=-2.5^{\degree}$ respectively \citep{aharon2018}.  In figure \ref{fig:gmc}, we plotted the SEDs of 964 and Maddalena obtained from \textit{Fermi}-LAT analysis \citep{aharon2018}. In that plot we also plotted the SEDs of those two GMCs obtained by us by using the equation \ref{eq:gmcflux}. We found that our results fitted well with the observed values. The comparison of our results with the observed ones indicates that at the positions of the two GMCs the gamma-ray production is dominated by p-p interactions   and the CR proton densities at those two positions can be traced back solely from the analysis of SEDs of those two GMCs. We can also conclude that the results presented in this work are consistent with the observational results. 
 
\section{Conclusion}\label{sec:end}
In this work, we use the \texttt{DRAGON} code to study the CR propagation in the Galaxy. We take into account the updated gas density profiles and minimize $\chi^{2}/d.o.f$ of B/C and proton data to obtain the best fit values of the parameters for each of the benchmark models used here. We, then, calculate the diffuse gamma-ray fluxes by considering only  p-p interactions in the ISM. Our obtained diffuse gamma-ray fluxes for the different models do not show any signature that can be used
 to distinguish between them.

However, we find that our calculated diffuse gamma-ray fluxes are lower by $\sim 1-3$ orders of magnitude than the previous results, where the CR propagation models were significantly different. We find that if we exclude the emission from the inner Galaxy, then our diffuse gamma-ray fluxes are lower by $\sim 3$ orders of magnitude (see figure \ref{fig:liucompare}) than the IGRB data. If we include the emission from inner Galaxy, then our diffuse gamma-ray fluxes are lower by less than 2 orders of magnitude (see figures \ref{fig:cholvsme}, \ref{fig:ackvsme}) than the IGRB data. We, hence, conclude that the contribution of the Galactic diffuse gamma-rays to the IGRB  flux is quite less, which suggests IGRB is mostly of extragalactic origin. Such conclusion is also supported by  previous studies \citep{ajello2015, ackermann2016,lisanti2016}, where it was reported that the IGRB is  likely to be dominated by unresolved extragalactic sources. With future gamma-ray missions it could be possible to resolve more extragalactic GeV-TeV gamma-ray sources, which may
 further restrict the possible contribution from Galactic diffuse gamma-rays to the IGRB. 
Our results may provide useful clues to the origin of Galactic CRs. In our case, we find $L_{{\rm{CR}}} \sim 10^{38}~\rm{erg~s^{-1}}$. A smaller CR luminosity ($< 10^{41}~\rm{erg~s^{-1}}$) would indicate the required acceleration efficiency of SNRs would be lower, and the other potential CR accelerators such as Galactic Center \citep{abram2016}, OB associations \citep{ces1983,bykov2001,aharonian2018a} and pulsars \citep{bedna2004} could be major sources of Galactic CRs  in terms of energy budget.

 The analysis framework developed by us can be used to obtain diffuse gamma-ray fluxes in p-p interactions at different latitudes and longitudes in GeV to TeV energy band. Furthermore, our obtained CR proton fluxes are applied to probe GMCs and we infer that individual GMCs can trace the local CR proton density.





\bibliography{gcr.bib}

\end{document}